\documentclass[11pt, a4paper]{article}
\pdfoutput=1
\usepackage{jcappub}
\usepackage[english]{babel}
\usepackage{amsfonts,amsmath,amssymb,amsthm,mathtools}
\usepackage{comment,enumerate,footnote,graphicx,subfloat,relsize}
\usepackage{array,tabularx,tabu,multirow,framed,afterpage}
\usepackage[usenames,dvipsnames]{xcolor}
\numberwithin{equation}{section}
\usepackage{cleveref}

\usepackage{amsmath}
\usepackage{amsfonts}
\usepackage{amssymb}
\usepackage{graphicx, rotating}
\usepackage{epstopdf}
\usepackage{epsfig}
\usepackage{latexsym}
\usepackage{graphicx}
\usepackage{caption}
\usepackage{subcaption}
\usepackage{color}
\usepackage{amsmath,amssymb}
\usepackage{slashed,cancel}
\usepackage{hyperref}
\usepackage{multirow}
\hypersetup{colorlinks, citecolor=bluscuro, linkcolor=black, urlcolor=bluscuro}
\definecolor{rossos}{cmyk}{0,1,1,0.55}
\definecolor{bluscuro}{rgb}{0.15, 0.2, .85}
\definecolor{bluchiaro}{cmyk}{1,.3,0.,0.1}

\newcommand{\lag}{\mathcal{L}}

\newcommand{\nn}{\nonumber}

\newcommand{\be}{\begin{equation}}
\newcommand{\ee}{\end{equation}}
\newcommand{\bea}{\begin{eqnarray}}
\newcommand{\eea}{\end{eqnarray}}
\newcommand{\bc}{\begin{center}}
\newcommand{\ec}{\end{center}}

\def\Im{{\rm Im\,}}
\def\Tr{{\rm Tr\,}}

\def\dd {{\rm d}}

\def\ora#1{\overset{\text{\scriptsize$\rightarrow$}}{#1}}
\def\ola#1{\overset{\text{\scriptsize$\leftarrow$}}{#1}}

\title{CP-violation for Electroweak Baryogenesis from Dynamical CKM Matrix}

\author[a]{ Sebastian Bruggisser,}
\author[a]{Thomas Konstandin,} 
\author[a,b]{G\'eraldine  Servant}

\affiliation[a]{DESY, Notkestra{\ss}e 85, D-22607 Hamburg, Germany}
\affiliation[b]{II. Institute of Theoretical Physics, University of Hamburg, D-22761 Hamburg}

\emailAdd{sebastian.bruggisser@desy.de}
\emailAdd{thomas.konstandin@desy.de}
\emailAdd{geraldine.servant@desy.de}

\date{\today}

\arxivnumber{1706.08534}

\abstract{{We show that the CKM matrix can be the source of CP violation for electroweak baryogenesis if Yukawa couplings vary at the same time as the Higgs acquires its vacuum expectation value. This offers new avenues for explaining the baryon asymmetry of the universe.} These ideas  apply if  the mechanism explaining
the flavour structure of the Standard Model  is connected to electroweak symmetry breaking. 
We compute the resulting baryon asymmetry for various low-scale flavour models and different configurations  of the Yukawa coupling variation across the bubble wall, and show that it can naturally be of the right order.
}

\begin{document}
\maketitle
\flushbottom


\section{Introduction}

Electroweak baryogenesis (EWBG) is a mechanism to explain the matter antimatter asymmetry of the universe  using Standard Model baryon number violation \cite{Kuzmin:1985mm}. It relies on a charge transport mechanism in the vicinity of bubble walls during a first-order electroweak phase transition \cite{Cohen:1990it}. It is particularly attractive as it relies on electroweak scale physics only and is therefore testable experimentally. It requires an extension of the Higgs sector leading to a first-order electroweak phase transition. This is minimally achieved by adding a singlet scalar to the Standard Model. In electroweak baryogenesis, CP violation  comes into play when chiral fermions scatter off the Higgs at the phase interface. A chiral asymmetry is created in front of the bubble wall that is converted by sphalerons into a baryon number.  CP violation in the Standard Model is too suppressed to explain the baryon asymmetry \cite{Gavela:1993ts}.
New sources that have been commonly  studied in the literature are either in the chargino/neutralino mass matrix \cite{Carena:1997gx, Carena:2000id, Carena:2002ss,Konstandin:2005cd, Cirigliano:2009yd, Li:2008ez} {or the sfermion sector in supersymmetric models  \cite{Carena:1997gx,Carena:2000id,Chung:2008aya,Kozaczuk:2012xv} }or coming from a varying top quark Yukawa coupling \cite{Fromme:2006wx,Bodeker:2004ws} as motivated in composite Higgs models \cite{Espinosa:2011eu} or  in Two-Higgs doublet models \cite{Cline:1995dg,Cline:1996mga,Fromme:2006cm,Cline:2011mm,Dorsch:2016nrg,Alanne:2016wtx} where the CP violating source comes from the changing phase in the Higgs VEV  during the electroweak phase transition. A new possibility was studied recently where CP violation is coming from the dark matter particle \cite{Cline:2017qpe}.
A typical constraint on EWBG comes from Electric Dipole Moments, which is particularly severe in the case of supersymmetric scenarios \cite{Cirigliano:2009yd}.

In this paper,  we are interested in  models where the source of CP violation is changing with time, which is a natural way to evade constraints. Following the same philosophy, strong CP violation from the QCD axion was studied in the context of cold baryogenesis in \cite{Servant:2014bla}. 
We are now considering the possibility that the structure of the Cabibbo-Kobayashi-Maskawa (CKM) matrix is varying during the EW phase transition such that Yukawa couplings start with natural values of order one in the electroweak symmetric phase and end up with their present values in the broken phase. This way, CP violation is no longer suppressed by small Yukawa couplings during the EW phase transition \cite{Berkooz:2004kx}.  The main motivation is to link EWBG to low-scale flavour models.  If the physics responsible for the structure of the Yukawa couplings is linked to EW symmetry breaking, we can expect the Yukawa couplings to vary at the same time as the Higgs is acquiring a VEV, in particular if the flavour structure is controlled by a new scalar field which couples to the Higgs.
In principle, one  has to compute the full scalar potential, to determine the field path in a multiple scalar field space and compute the evolution of the Yukawa coupling during the phase transition. 
Particular realisations in explicit constructions were studied in  \cite{Baldes:2016gaf,vonHarling:2016vhf}.
It is crucial to show that the Yukawa can vary with the Higgs, i.e. that the Higgs and Flavon fields vary simultaneously and not subsequently. 
For this to happen, the new scalar has to be light, i.e. not much heavier than the Higgs.
This is typically in tension with flavour bounds in Froggatt-Nielsen constructions \cite{Baldes:2016gaf}. However, in Randall-Sundrum  models, this can be achieved \cite{vonHarling:2016vhf}. The case of Composite Higgs models is particularly interesting \cite{CompositeEWPT}.

In this work, we will first provide a model-independent analysis, assuming that the Higgs and the Yukawa couplings vary at the same time. 
We will  assume that  the field trajectory is given in a specific model and we will parametrize it. 
We stress that in the framework we have in mind, the Yukawa couplings do not depend explicitly on the Higgs VEV. Instead, the Yukawas are controlled by some other scalar field. However, because of the couplings between this additional scalar and the Higgs field, the Yukawa coupling effectively varies at the same time as the Higgs during the EW phase transition. A dependence of the Yukawa coupling on the Higgs VEV is induced during the EW phase transition, but today there is no such dependence.  Our scenario is therefore very different from \cite{Bauer:2015fxa,Bauer:2015kzy}, which is very constrained experimentally.
 
 Our goal is to compute the baryon asymmetry when the CP-violating source is coming from varying Yukawa couplings.
In Section \ref{sec:baryonasymmetry}, we review the calculation of the baryon asymmetry in EWBG. In Section  \ref{sec:kinetic_equations}, we derive the kinetic equations and extract the CP-violating force induced by varying Yukawas.
Section \ref{sec:diffusion_Network}  describes the diffusion network. Section \ref{sec:showcase_oneFlavour} applies the formalism to the case where only the top quark Yukawa coupling varies. In Section \ref{sec:twoflavour}, we generalise to a two-flavour system, and study the  (Top, Charm), (Up, Charm) and ($\tau$, $\mu$) cases. We eventually also consider the full system including all quark flavours.
In Section \ref{sec:flavour_models}, we apply our results to specific models such as Froggatt-Nielsen and Randall-Sundrum. Technicalities are moved to appendices. Appendix A presents matrices for diffusion equations. Appendices B and C report some parts of the derivation of the transport equations. Appendix D is a repository of the Yukawa coupling expressions used in   Randall-Sundrum models.
 
\section{Basics of the calculation of the baryon asymmetry  during EW baryogenesis}\label{sec:baryonasymmetry}
 
 At the nucleation temperature, quantum tunnelling takes place from the symmetric vacuum towards the global minimum of the Higgs potential where electroweak symmetry is spontaneously broken. Bubbles are created and the two phases coexist until bubbles percolate and the whole universe is converted into the broken phase.
To compute the  baryon asymmetry produced during bubble expansion, it is enough to focus on one single bubble and integrate the baryonic density over the radial coordinate perpendicular to the bubble wall. The bubble expands fast with a velocity $v_w$ through the plasma and we can approximate the bubble wall to  be planar with a characteristic thickness $L_w$.
Inside the bubble the Higgs has a non-vanishing vacuum expectation value $\phi$. Outside the bubble, the electroweak symmetry is unbroken $\phi=0$. 
We define the direction perpendicular to the bubble wall as $z$ and we choose the origin of our comoving (with the bubble wall) coordinate system as the middle of the wall. The phase where the electroweak symmetry is broken is for negative $z$ and the symmetric phase is for positive $z$. As the wall is passing, particles in the plasma get reflected and transmitted depending on their interactions with the wall. If some of these interactions are CP-violating, left- and right-handed particles will have different reflection and transmission coefficients and  a chiral asymmetry will develop inside and in front of the bubble wall. The resulting excess of left-handed fermions in front of the bubble wall can be converted into a net baryon number by the weak sphalerons, which are unsuppressed in the symmetric phase in front of the bubble.

\begin{figure}[t]
\centering
\includegraphics[width=0.75\textwidth]{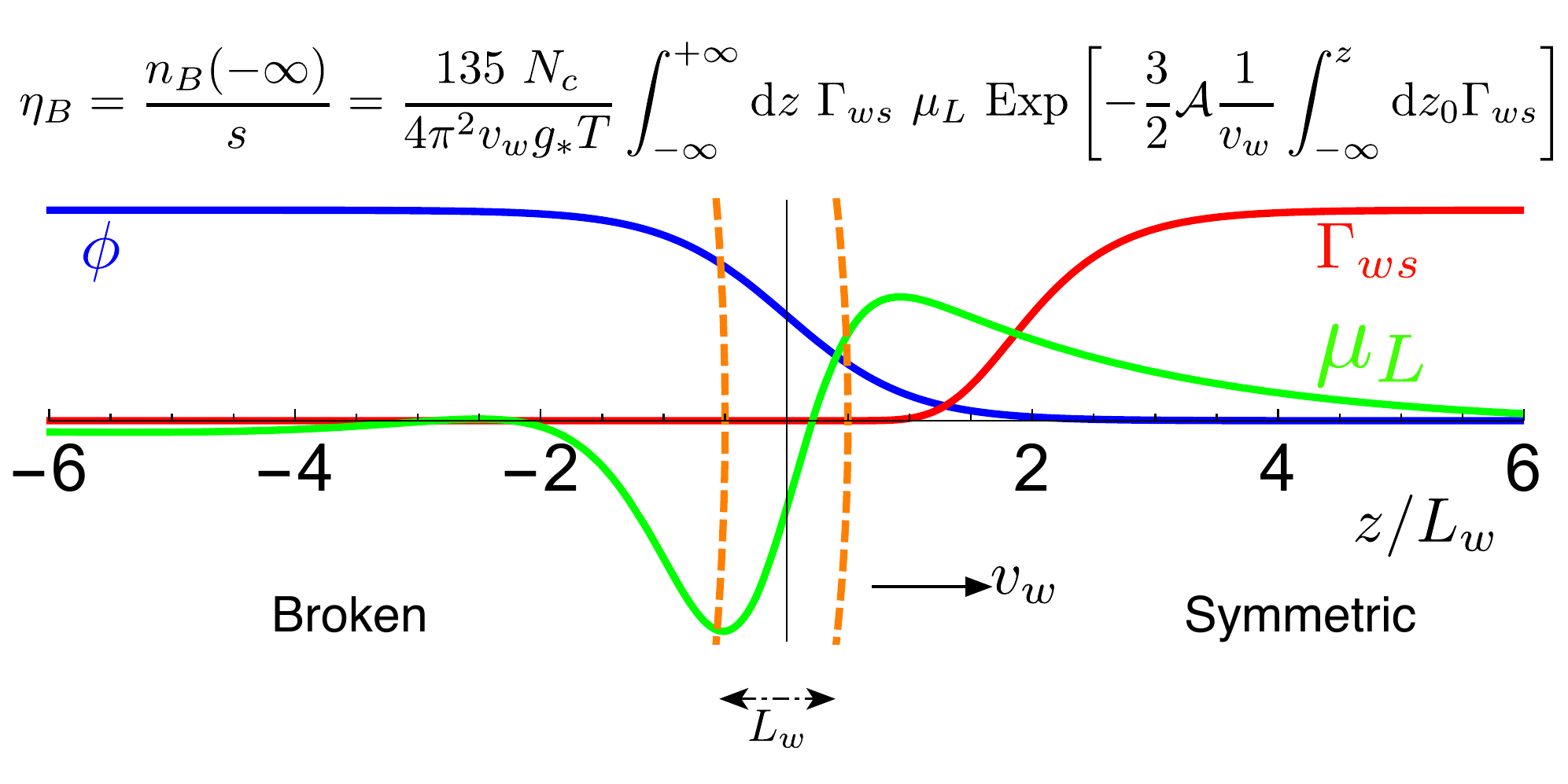}
\caption{\label{fig:SchematicBubbleWall} \em
A cut through the bubble wall, which moves from the left to the right (in the direction of positive $z$, i.e. $v_w>0$). In blue we show the profile of the Higgs VEV through the bubble wall. The rate for the sphaleron transitions (red) becomes important only in front of the bubble wall. The green line is the profile of the total chemical potential for the quark doublets. All curves are in arbitrary units, for illustration only.
}
\end{figure}

The baryon asymmetry  $n_B$ denotes the difference between the baryon and anti-baryon number densities. It is determined by solving the diffusion equation of the weak sphaleron~\cite{Cline:2000nw}
\begin{equation}\label{eqn:sphaleron_diffusion}
\partial_z \, n_B = \frac{3}{2} v_w^{-1} \Gamma_{ws} 
\left(N_c \mu_L T^2 - {\cal A} \,  \, n_B \right) \, ,
\end{equation}
with 
\be
\Gamma_{ws} = 10^{-6} \, T \, \exp(- a \phi(z)/T)
\ee
 being the $\phi$-dependent sphaleron rate and $N_c=3$ is the number of colours. The constant ${\cal A}$ is determined by the number of quark and lepton species that are in equilibrium during the sphaleron transitions. Its value in the SM is ${\cal A}= 15 /2$~\cite{Cline:2000nw}. The other constant $a \simeq 37=E_{Sph}/v$~\cite{Quiros:2007zz,Klinkhamer:1984di} is the sphaleron energy normalized to the Higgs VEV today.  The left-handed chemical potential of the quark doublets 
\begin{equation}\label{eqn:leftPotental}
	\mu_L = \frac{\mu_{t_L} + \mu_{b_L} + \mu_{c_L} + \mu_{s_L}+ \mu_{u_L} + \mu_{d_L}}{2}
\end{equation}
is a measure for the density of left handed quarks in front of the bubble wall. The first term in the parenthesis on the right hand side of equation (\ref{eqn:sphaleron_diffusion}) represents the source, i.e. the  excess of left handed quarks being converted into a net baryon number by the weak sphaleron. The second term accounts for the washout, i.e. the fact that the sphaleron tends to relax any baryon asymmetry to zero if it has enough time to do so. If the bubble wall expands at a very low speed compared to the typical diffusion time scale, the sphaleron washes-out the baryon asymmetry. If, however, the wall has a sizable velocity, a non-negligible fraction of the baryon asymmetry diffuses into the bubble, where the weak sphaleron is suppressed. This way the baryon asymmetry can be frozen inside the bubble. 
The whole mechanism is illustrated in figure~\ref{fig:SchematicBubbleWall}  which also clarifies our notations and conventions.

\subsection{Chemical potentials and local velocities}

The source term in equation (\ref{eqn:sphaleron_diffusion}) is $\mu_L$, the density of the excess of left-handed fermions in front of the bubble wall which depends on how fermions are transported through the bubble wall. This is determined by their interactions with the wall and with other particles in the plasma. We need to determine the profiles of the chemical potentials ($\mu_i$) for each of the particle species whose local velocity in the plasma ($u_i$) is influencing the diffusion through the bubble wall. We therefore have to determine $\mu_i$ and $u_i$ simultaneously. For electroweak baryogenesis, only the CP-violating contribution is of interest. The CP-violating part of the chemical potentials and the local velocities will crucially depend on the (new) source of CP-violation that is necessary to create an excess of left-handed particles. This gives rise to a system of coupled differential equations, the so-called transport equations, which, as we will see in the next section, can be brought to the form:
\begin{equation}
\label{eqn:diffusionSystemMatrix}
 A(z)\cdot r'(z) +B(z)\cdot r(z) =\bar {\cal S}(z)
\end{equation}
where
\be
r=(\mu_1,\mu_2,\dots,\mu_N,u_1,u_2,\dots,u_N)^T
\ee
 is the $2N$-dimensional vector of the solutions of the differential equations, $A$ and $B$ are $2N\times2N$ matrices that encode the dynamics and interactions of the particles and $\bar {\cal S}$ is the vector containing the CP-violating source. Here $N$ is the number of particle species that are taken into account in the diffusion system. As stated in Appendix \ref{sec:diffusionSystem}, for simplicity we will take 
 \be
 N=9,
 \ee
 corresponding to the LH and RH chiralities of the Top, Bottom, Charm, and Strange quarks as well as the Higgs, except in Section \ref{sec:fullsystem} where we take into account all quark flavours i.e. $N=13$ (it increases the computation time for our analysis  but numerical results are not radically changed).
Notice that the matrices $A$ and $B$ are space-dependent. Besides, we impose that the solution vector vanishes in both limits $z \to \pm \infty$. In general, it is not guaranteed that such a solution exists and is unique, but it does in our context as long as the wall velocity is not too large. 

To solve this system, we construct a Green's function $G$ such that 
\begin{equation}\label{eqn:solution_to_linear_system}
r(z) = \int \, dy \, G(z,y) \,  A^{-1} \, \bar{\cal S}(y) \, .
\end{equation}
$G$ is a suitably normalized linear combination of the solutions of the homogeneous equations ( $r'+A^{-1}Br=0$) multiplied with a Heaviside step function. We choose two points outside of the wall, $z_0 \ll - L_w$, $z_1 \gg L_w$. Since $A$ and $B$ are constant outside of the wall, we determine the eigenvalues ($\lambda_i$) of $A^{-1}B$ with the correct sign at the points $z_0$ and $z_1$, such that the corresponding solutions $w_i(z)=e^{-\lambda_i z}$ go to zero at $\pm\infty$. 
Typically,  half of the solutions $\lambda_i$ have either sign in both points such that in total we find  the correct number of solutions that vanish beyond the wall.

The corresponding functions $w_i(z)$ can then be numerically continued into the wall and beyond, taking the space-dependence of $A$ and $B$ into account. They blow up exponentially beyond the wall. When these functions are multiplied with the appropriate Heaviside functions, $\Theta(\pm(z - y))$, one obtains solutions to the equation of motion that vanish at $z \to \pm \infty$ and contain a discontinuity at $z=y$. An appropriate linear superposition then yields the Green's function $G(z,y)$.

\subsection{Baryon asymmetry}
The relation (\ref{eqn:sphaleron_diffusion}) can be inverted:
\begin{equation}\label{eqn:totalBaryonAsymmetry}
  \eta_B=\frac{n_B(-\infty)}{s}=\frac{135~N_c}{4\pi^2 v_w g_* T }\int_{-\infty}^{+\infty} \dd z ~ \Gamma_{ws}~\mu_L~ e^{-\frac{3}{2} {\cal A} \frac{1}{v_w}\int_{-\infty}^{z} \dd z_0 \Gamma_{ws}}\, ,
\end{equation}
where $s=\frac{2\pi^2}{45}g_*T^3$ is the entropy density, $g_*$ the number of relativistic degrees of freedom, and $\mu_L$ is the chemical potential of the left handed quark species and hence is a linear combination of the entries of the solution vector
\be
\mu_L=V^Tr(z),
\ee
 where $V$ is the vector that defines the linear combination (see equation (\ref{eqn:leftPotental})). 
 From dimensional analysis, the baryon asymmetry (\ref{eqn:totalBaryonAsymmetry}) will scale like 
\be
\eta_B \sim \frac{\Gamma_{ws} \mu_L L_w}{g_* T} \sim \frac{10^{-8} \mu_L}{T} \mbox{ for } L_w\sim \frac{1}{T}
\ee
so it is clear that for EW baryogenesis to be successful, $\mu_L/T$ has to be large and this requires new sources of CP violation beyond the Standard Model.

Using equations (\ref{eqn:solution_to_linear_system}) and (\ref{eqn:totalBaryonAsymmetry}), the total baryon asymmetry can be written as:
\begin{equation}
  \eta_B=\sum_i\int_{-\infty}^{+ \infty}\dd y ~K_i(y) ~\bar{\cal S}_i(y)
\end{equation}
with the kernel $K(z)$ given by (we have replaced ${\cal A}=15/2$):
\begin{equation}
  K_i(z)=\frac{135~N_c}{4\pi^2v_wg_*}\int_{-\infty}^{+\infty} {\rm d} z_0~ \Gamma_{ws}(z_0)~ V_j G_{jk}(z_0,z)~A^{-1}_{ki}~e^{-\frac{45}{4v_w}\int_{-\infty}^{z_0} {\rm d} y \Gamma_{ws}}
\end{equation}

We have separated the CP-violating source term $\bar{\cal S}_i(z)$ from the effects of the diffusion through the bubble wall and the weak sphaleron $K_i(z)$. To understand what controls the final prediction for the baryon asymmetry, it will be enlightening to plot separately those two physical effects, for different assumptions of the source term and bubble configuration.
In sections~\ref{sec:showcase_oneFlavour}, \ref{sec:twoflavour} and~\ref{sec:flavour_models} we will present the kernel $K(z)$ and source $\bar {\cal S{}}(z)$ in various models. Before that, in the next two sections we derive the equations leading to the master equation (\ref{eqn:diffusionSystemMatrix}).

\section{Kinetic equations and CP-violating force} 

\label{sec:kinetic_equations}

\subsection{Kinetic equations}

We now  justify  our master equation (\ref{eqn:diffusionSystemMatrix}) which is the compact form of the equations governing the diffusion of different particles through the bubble wall. 
Starting from first principles, the diffusion equations are derived from the Kadanoff-Baym equations (for a review of the formalism and comparison with other approaches see \cite{Konstandin:2013caa}). 
In the limit where we truncate the derivative expansion of $\hat{E}=e^{\frac{i}{2} \ola{\partial_z}\cdot\partial_k}$ in Kadanoff-Baym equations, 
these equations coincide with the usual Boltzmann equations, which, after using a fluid ansatz for the particle distributions, lead to the diffusion equations  (\ref{eqn:diffusionSystemMatrix}) for the chemical potential and particle velocities.
 At the end, only the interaction terms lead to mixing between the different flavours.
 
A brief derivation of this procedure is given in the appendix \ref{sec:KB_eq_spin_projection}. We can split the Kadanoff-Baym equations into a Hermitian and an anti-Hermitian part. The anti-Hermitian part is often called the constraint equations  leading to the dispersion relations for the energies of the particles, which we derive in appendix \ref{sec:constraint}. 
The Hermitian part of the Kadanoff-Baym equations gives the 
kinetic equations that define the diffusion.
Truncating the  expansion  of operator $\hat{E}=e^{\frac{i}{2} \ola{\partial_z}\cdot\partial_k}$ to second order in derivatives through manipulations similar to the ones performed in appendix \ref{sec:constraint}, Kadanov-Baym equations can be brought to the following form \cite{Konstandin:2004gy}  starting from equations (\ref{eqn:kinetic_equations_exact1}--\ref{eqn:kinetic_equations_exact4}) in appendix \ref{sec:KB_eq_spin_projection} :
\begin{equation}\label{eqn:right-density}
\begin{split}
k_z\partial_z g_R 
&+\underbrace{\frac{i}{2}\left[m^\dagger m,g_R\right]}_{\text{Term 2}}-\underbrace{\frac{1}{4}\left\{\left(m^\dagger m\right)',\partial_{k_z} g_R\right\}}_{\text{Term 3}}\\
&+\underbrace{\frac{1}{4k_z}\left({m^{\dagger}}'mg_R+g_Rm^\dagger m'\right)}_{\text{Term 4}}-\underbrace{\frac{1}{4k_z}\left({m^{\dagger}}'g_Lm+m^\dagger g_L m'\right)}_{\text{Term 5}}\\
&-\underbrace{\frac{i}{16}\left[\left(m^\dagger m\right)'',\partial_{k_z}^2g_R\right]}_{\text{Term 6}}+\underbrace{\frac{i}{8k_z}\left[{m^\dagger}' m',\partial_{k_z}g_R\right]}_{\text{Term 7}}\\
&+\underbrace{\frac{i}{8}\left({m^\dagger}'' m\partial_{k_z}\left(\frac{g_R}{k_z}\right)-\partial_{k_z}\left(\frac{g_R}{k_z}\right)m^\dagger m''\right)}_{\text{Term 8}}\\
&-\underbrace{\frac{i}{8}\left({m^\dagger}''\partial_{k_z}\left(\frac{g_L}{k_z}\right) m-m^\dagger \partial_{k_z}\left(\frac{g_L}{k_z}\right) m''\right)}_{\text{Term 9}}=coll.\, ,
\end{split}
\end{equation}
where $m$ is the fermion mass matrix in the flavour basis, $g_L$ and $g_R$ are the left and right-handed density matrices that will eventually be related to the fermion distribution functions. We have the same Equation (\ref{eqn:right-density}) for the up and down quarks so the  $m$ and $g$ matrices both  have dimension $(N-1)/2$. The  term $coll.$ in the right-hand side stands for the collision terms of equation~(\ref{eqn:collisionInKB}) that will be discussed in section~\ref{sec:diffusion_Network}. For now we neglect these collision terms and reintroduce them in the final expression. 
We remark that the equation for the left-handed density can be obtained by the replacements:
\begin{equation}
g_R\leftrightarrow g_L \qquad m^\dagger \leftrightarrow m\, .
\end{equation}

The Wightman function that encodes the particle distribution function and that is a solution of the Kadanov-Baym equations
\begin{equation}
    iS=\sum_{s=\pm1}iS_s \, ,
\end{equation}
is related to the
functions $g_L$ and $g_R$ via
\begin{equation}
\label{eq:LRdecomp}
iS_s=P_s\left[s \gamma^3 \left( g_R^{s} P_R - g_L^{s} P_L \right)
+ g_N^{s} P_R + g_N^{\dagger s} P_L \right] \, ,
\end{equation}
with $P_R = (1 + \gamma_5)/2$ and $P_L = (1 - \gamma_5)/2$ the projection operators on the right- and left-chiral fields and $P_s = (1 + s\, S_z)/2$, $s=\pm1$,  results from the spin projection operator along $z$
\begin{eqnarray}
\qquad S_z \equiv \frac{1}{\tilde{k}}\left(\gamma^0 k_0 - \gamma^1k_x - \gamma^2k_y\right)\gamma^3\gamma^5
\equiv \frac{\slashed{\tilde k}}{\tilde{k}}  \gamma^3\gamma^5, 
\qquad \tilde{k} = \textrm{sign}(k_0)(k_0^2-k_x^2-k_y^2)^{1/2} \, .
\end{eqnarray}
$g^s_N$ is defined in Eq.~(\ref {eq:defgRandgN}) of  Appendix \ref{sec:KB_eq_spin_projection}.

Equation~(\ref{eqn:right-density}) describes the evolution of the particle distribution functions of several flavors\footnote{We removed subscript $^s$  in (\ref{eqn:right-density}) to lighten notations.}. In particular the second term (involving the commutator with the masses)  induces flavor mixing. In order to simplify the discussion, we neglect all the off-diagonal components in the basis where the mass is diagonal. This is justified in the case where the different masses are not too degenerate and the flavor oscillations sufficiently damp any off-diagonal particle densities. This coincides with the adiabatic regime studied in Ref.~\cite{Cirigliano:2009yt}. In fact, in this limit the derivative expansion will break down for the off-diagonal entries in the Wightman function since the oscillation length $\sim k/\Delta m^2$ can be of the same order as the typical inverse momentum $\sim 1/T$. In effect, the fast oscillation will imply that the off-diagonal densities are never much populated and can be neglected.

These equations for the right and left handed density matrices can be disentangled for the following linear combinations
\begin{equation}
    \begin{split}
        g_{0d}^{s}=\frac{1}{2}\left(g_{Rd}^{s}+g_{Ld}^{s}\right) \qquad g_{3d}^{s}=\frac{1}{2}\left(g_{Rd}^{s}-g_{Ld}^{s}\right)\\
        g_{1d}^{s}=\frac{1}{2}\left(g_{Nd}^{s}+g_{Nd}^{s \dagger}\right) \qquad g_{2d}^{s}=\frac{-i}{2}\left(g_{Nd}^{s}-g_{Nd}^{s \dagger}\right)
    \end{split}
\end{equation}
that correspond to the decomposition
\begin{equation}
iS_s=-P_s\left[s\gamma^3\gamma^5g^s_0-s\gamma^3g_3^s+\mathbb{I}g_1^s-i\gamma^5g_2^s\right] \, .
\end{equation}
We use the subscript $d$ in order to identify matrices that are in their diagonal form. Equation~(\ref{eqn:right-density}) then admits a very simple form for the density $g_{0d}^{s}$ (again neglecting any off-diagonal elements)\footnote{We note that in \cite{Prokopec:2003pj} the CP-violating force in the  kinetic equation for $g_{0dii}^{s}$ is different and involves the combination 
\begin{equation}
 \Im\left[V^\dagger{m^\dagger}'mV\right]'_{ii}
\not=  \Im\left[V^\dagger{m^\dagger}''mV\right]_{ii} \, ,
\end{equation}
 The discrepancy comes from the fact that we work in the limit 
where flavor oscillations are relatively fast and one can neglect all off-diagonal in the basis where the masses are diagonal. 
On the other hand, Ref.~\cite{Prokopec:2003pj} works in the limit where oscillations are very slow and the derivative expansion even holds for the off-diagonal entries of the Wightman function.} 
\begin{equation}\label{eqn:our_kinetic_equation}
    \left(k_z\partial_{ḱ_z}-\frac{1}{2}\left(\left[V^\dagger \left({m^\dagger}m\right)' V\right] + \frac{s}{\tilde{k}_0}\Im\left[V^\dagger{m^\dagger}''mV\right]\right)_{ii}\partial_{k_z}\right)g^s_{0dii}\approx 0
\end{equation}
$V$ is the transformation matrix used in the mass diagonalization (see Eq. (\ref{eq:flavourrotations}) in Appendix \ref{sec:KB_eq_spin_projection}).
The subscript $ii$ refers to the diagonal entries and does not stand for the conventional summation. Note that in order to obtain this result we had to make use of the constraint equations (Eq.~(\ref{eqn:g0_as_function_of_g3}) in appendix \ref{sec:constraint}) to the lowest order in the gradient expansion. 

\subsection{CP-violating force from varying yukawas across the bubble wall}

 Under the hypothesis of diagonal entries as stated above,  the commutator terms (terms 2, 6 and 7) in equation~(\ref{eqn:right-density}) do not contribute to equation~(\ref{eqn:our_kinetic_equation}). From the derivative structure, we can see that the CP-conserving force $\propto [V^\dagger \left({m^\dagger}m\right)' V]$ in equation~(\ref{eqn:our_kinetic_equation}), has to come from a combination of the terms 3, 4 and 5, whereas the CP-violating force $({s}/{\tilde{k}_0})\Im [V^\dagger{m^\dagger}''mV ]$ follows from terms 8 and 9.
It is then easy to show that the CP-violating part vanishes for the SM which has constant Yukawa couplings. Indeed, for constant Yukawas $\Im\left[V^\dagger{m^\dagger}''mV\right]\propto\Im\left[V^\dagger{Y^\dagger}YV\right]\phi''\phi$ and since  $V^\dagger{Y^\dagger}YV$ is hermitian, the diagonal entries are real. 

In summary, the only relevant CP-violating terms in equation~(\ref{eqn:right-density}) are terms 8 and 9. These are second order terms in the derivative expansion of $\hat{E}$. In the Standard Model, these terms vanish since derivatives of the mass matrix are proportional to the mass matrix itself. 
In the models we will study, this is no longer true. 
The purpose of this work is to explore the possibility that the variation of the mass terms of Standard Model fermions across the bubble wall provide the only source of CP violation to explain the observed baryon asymmetry.
These new CP-violating sources can be sufficient for baryogenesis provided that the Yukawa coupling starts with a value of order one in the symmetric phase.  This is possible even with only one fermionic flavor as  long as the complex phase of this mass is changing during the electroweak phase transition, a CP-violating axial current being induced due to a semi-classical force \cite{Cline:2000nw}.

This source of CP violation is different from the standard CP violation from the CKM phase. In this case, CP-violating processes have to involve at least three flavors and accordingly are suppressed by the Jarlskog invariant $J_{CP}$ \cite{Jarlskog:1985ht,Jarlskog:1985cw}. In principle, the Standard Model CKM  CP violation  also enters in our analysis, but it will do so via higher loop contributions to the self-energy $\Sigma$ in (68) and be very much suppressed \cite{Gavela:1993ts,Huet:1994jb}. In practice, we neglect the self-energies and hence the standard CKM type of CP violation.

In order to interpret the kinetic equation~(\ref{eqn:our_kinetic_equation}) in terms of currents, notice that in equilibrium $S \propto (\slashed{k} + m)$ such that $g_3/g_0 = s k_z/\tilde k$. The kinetic equation is consistent with this relation to first order in derivatives. Since $g_0$ is even under $s$ in equilibrium, the CP-violating force drives only the $s$-odd component at leading order that is the coefficient of $\gamma_5 \gamma_3$ and hence the axial current~\footnote{Notice the unfortunate sign convention of $g_L$ in equation~(\ref{eq:LRdecomp}) that we adopted from~\cite{Konstandin:2004gy}. }. 
The corresponding equation for $g_3$ does not contain any CP-violating source.
The following Ansatz fulfills the requirements and makes the link with the quasi-particle picture as justified in Appendix \ref{sec:constraint}:
\begin{eqnarray}
g_0^s &\propto& \tilde k  \, (P_+ f_L + P_- f_R)  \, \delta(k_0^2 - \omega^2)  \, , \nn \\
g_3^s &\propto& k_3  \, (f_L + f_R)/2  \, \delta(k_0^2 - \omega^2) \, ,
\end{eqnarray}
with $P_\pm$ a projection on spin. In order to respect parity we define the spin projection as 
\begin{equation}
P_\pm = \frac12 \left( 1 \pm s \, \textrm{sign}[k_z] \right) \,.
\end{equation}
We obtain the following equations for the particle densities $f_L$ and $f_R$
\begin{eqnarray}\label{eqn:BoltzmannEqns}
\left(k_z\partial_{ḱ_z}-\frac{1}{2}\left(\left[V^\dagger \left({m^\dagger}m\right)' V\right] \right)_{ii}\partial_{k_z}\right)f_{L,i} &\approx& {\cal S}_i \, , \nn \\
\left(k_z\partial_{ḱ_z}-\frac{1}{2}\left(\left[V^\dagger \left({m^\dagger}m\right)' V\right] \right)_{ii}\partial_{k_z}\right)f_{R,i} &\approx& -{\cal S}_i \, , 
\end{eqnarray}
with the CP-violating source
\begin{equation}
\label{eq:CPsource}
{\cal S}_i \equiv 
\frac{\textrm{sign}[k_z]}{2 \tilde{k}}\Im\left[V^\dagger{m^\dagger}''mV\right]_{ii}\partial_{k_z} f_{eq,i} \, .
\end{equation}
which agrees with the dominant contribution to the equations used in \cite{Fromme:2006wx} derived in the one-flavor case from the semi-classical force. Since the CP-violating source is of higher order in derivatives, we replaced the particle distribution function in (\ref{eq:CPsource}) by the one at equilibrium. 

Finally, let us comment on the dispersion relation. A derivation of the dispersion relation along the lines of \cite{Prokopec:2003pj, Prokopec:2004ic} is given in Appendix \ref{sec:constraint} (see equation (\ref{eqn:dispersion_relation})). We find that also in the adiabatic limit, the dispersion relations obtain small CP-violating shifts that will alter the 
on-shell condition of the particles in the wall in a CP-violating way. This can in principle induce 
additional CP-violating deviations in the particle densities. However, in the above equations these deviations will occur only locally in the wall and not diffuse in the symmetric phase. Once interactions are included in the analysis, also these deviations will diffuse but as long as the typical interaction length is larger than the wall thickness (which is indeed true), these effects will be sub-leading compared to the source (\ref{eq:CPsource}) coming from the forces. This agrees with the findings in \cite{Fromme:2006wx} and we also checked this numerically in some cases.

\section{The Diffusion Network}\label{sec:diffusion_Network}

With the Boltzmann equations derived in the previous section we are now able to determine the diffusion network. First we note that the expanding bubble will drive the plasma slightly out of equilibrium. We can parametrize this departure from equilibrium with a fluid-type Ansatz as done in~\cite{Fromme:2006wx}
\begin{equation}\label{eqn:fluidDistribution}
 	f_i=\frac{1}{e^{\beta (\omega_i+v_w k_z-\mu_i)}\pm 1} + \delta f_i ,
\end{equation} 
where the plus (minus) sign refers to fermions (bosons). Here the (small) chemical potentials $\mu_i (z)$ model a departure from equilibrium of the particle densities whereas the fluctuations $\delta f_i (z, k)$ quantify the departure from the kinetic equilibrium. They do not contribute to the particle density (i.e. $\int d^3p \delta f_i=0$) and they can be linked to the local velocity of the fluid. In the wall frame and in lowest order, this approaches the equilibrium distribution  
\begin{equation}
f_i \to f_i^0=(\text{Exp}[\beta (\omega_i+v_w \, k_z)]\pm 1)^{-1} \, .
\end{equation}
In the following, we assume the chemical potentials and the deviation $\delta f$ to be small and linearize the equations with respect to these quantities. We also linearize the system with respect to the wall velocity $v_w$, which is well justified in most models.

\subsection{Moments of the kinetic equations}

To obtain the diffusion equations we take different momenta of the equations (\ref{eqn:BoltzmannEqns}). We define the momentum of a function $Y$ as:
\begin{eqnarray}
	\left\langle Y(k,x) \right\rangle &=&  \frac{1}{N} \int\dd^3k  \, Y(k,x) \nn \\
   N &=& \int \dd^3 k  \, \frac{d}{d\omega} f^0 \, .
\end{eqnarray}
where for later convenience we normalized to the integral over $\frac{d}{d\omega} f^0$ that is the energy derivative of the equilibrium distribution of a massless fermion. To normalize to the distribution of a massless particle avoids an additional $z$-dependence in $N$ that would make charge conservation less transparent.
 
Up to this point the deviation $\delta f(z,k)$ still has an unknown momentum dependence. One way to solve this problem is to assume that this $k-$dependence factorizes in the following sense: We define the relative bulk velocity according to 
\begin{equation}
u \equiv \left< \frac{k_z}{\omega_0} \delta f \right> \, ,
\end{equation}
and assume factorization according to 
\begin{equation}
\left< Y \, \frac{k_z}{\omega_0} \delta f \right>  \simeq \left[ Y \, f^0 \right] \, u \,  
\end{equation}
which enforces the normalization $\left[ f^0 \right] = 1$, and accordingly
\begin{eqnarray}
	\left[ Y(k,x) \, f^0 \right] &=&  \frac{1}{\bar N} \int\dd^3k \,  Y(k,x)  \, f^0  \nn \\
   \bar N &=&   \int \dd^3 k   \, f^0  \, .
\end{eqnarray}
Of course, this assumption on factorization is not unique. This approach follows the analysis in \cite{Fromme:2006wx} where no significant dependence on the factorization choice was reported. The important requirements are that $u$ is understood as a quantity that is odd in $k_z$ and hence describes the relative flow of the fluid. It is also important that it produces the correct sound velocity which is indeed given, as we will show below.

In order to calculate the generated baryon asymmetry, we are mainly interested in the CP-odd components of the deviations from equilibrium. By a slight abuse of notation we denote the CP-odd components of $\mu$ and $u$ also as $\mu$ and $u$, in order to avoid overboarding notation.
Taking the moments $\left< \textrm{eq. (\ref{eqn:BoltzmannEqns})} \right>$ and 
$\left< k_z/\omega_0 \times \textrm{eq. (\ref{eqn:BoltzmannEqns})} \right>$, one obtains for each particle species an equation of the form
\begin{equation}
	\begin{split}
 v_w \, K_1 \, \mu' + v_w (m^2)' \, K_2 \, \mu 
+ u' -\langle \mathbf{C} \rangle 
&=0 \\
 -K_4 \, \mu' 
+v_w \, \tilde{K}_5 \, u' + v_w (m^2)' \, \tilde{K}_6 \,  u 
-\left\langle \frac{k_z}{\omega_{0i}}\mathbf{C} \right\rangle 
&= \pm v_wK_8 \Im\left[V^\dagger {m^\dagger}'' m V\right]_{ii} \, ,
	\end{split}
\label{eq:diffusion_0}
\end{equation}
where the plus (minus) sign correspond to left- (right-) handed fermions and where we have restored the collision term.

The $K$-factors can be computed numerically and are given by
\begin{equation}
\begin{array}{l l}
K_1  = -\left\langle \frac{k_z^2}{\omega_0} \frac{d^2}{d^2\omega} f^0 \right\rangle 
 = \left\langle \frac{d}{d\omega} f^0 \right\rangle  \, , &  
K_2  = \left\langle \frac{1}{2\omega_0} \frac{d^2}{d^2\omega} f^0 \right \rangle \, , \\ 
K_4  = \left\langle \frac{k_z^2}{\omega_0^2} \frac{d}{d\omega} f^0 \right\rangle \, , &  
\tilde{K}_5  = \left[ \frac{k_z^2}{\omega_0} \frac{d}{d\omega} f^0 \right ]  = -1 \, ,\\ 
\tilde{K}_6  = \left [ \frac{\omega_0^2 - k_z^2}{2\omega_0^3} \frac{d}{d\omega} f^0 \right ] \, , &  
K_8  = \left\langle \frac{|k_z|}{2\omega_0^2\omega_{0z}} \frac{d}{d\omega} f^0 \right \rangle \, .
\end{array}
\end{equation}
The naming of the $K$-factors corresponds to the one introduced in \cite{Fromme:2006wx}.
The equalities are obtained by partial integration.
Furthermore, in the massless limit 
$K_4 \simeq -\frac13 K_1 \, \tilde K_5$. This is important to correctly reproduce the speed of sound of a relativistic plasma. 

Notice also that the first equation represents the charge conservation 
\begin{equation}
\partial_z J^z = \left<\mathbf{C} \right> \quad \textrm{ with } \quad 
J_z = v_w  \, K_1 \, \mu  + u \, ,
\end{equation}
and the observation that $K_2 = \frac{d}{dm^2} K_1$. The second equation could be promoted to energy-momentum conservation if instead of the choice above, a different moment had been used. However, energy-momentum is not only broken by the interactions but also by the inflow of energy due to the free energy release in the Higgs sector~\cite{Konstandin:2014zta}. Hence there is no added benefit in using the moments that correspond to energy momentum conservation in equation (\ref{eq:diffusion_0}).

\subsection{Collision terms}

The collision integrals can be related to the different scattering rates \cite{Huet:1995mm, Moore:1995si, Cline:2000nw}
\begin{equation}
\label{eq:coll_def}
\langle \mathbf{C} \rangle = \Gamma^{\text{inel}}\sum_i \mu_{i} \, , \qquad 
\left\langle \frac{k_z}{\omega_{0i}}\mathbf{C} \right\rangle = -\Gamma^{\text{tot}} u \, .
\end{equation}
The emerging pattern is that the local bulk fluid $u$ can be reduced due to  2-to-2 scatterings of the very abundant light elements in the plasma like the light quarks and the gluons. Any change in the chemical potentials on the other hand has to come from particle number changing processes that are far less 
abundant. 

The degrees of freedom we treat in the plasma are the Higgs and  top, bottom, charm and strange quarks, except in Section \ref{sec:fullsystem} where we include all quark flavours. In practise, when considering that only the Yukawa of the charm-top system vary, we assign to the quarks of the second family (charm and strange quarks) the doubled number of degrees of freedom to take into account the first family of light quarks (up and down) that have for simplicity the same deviations as the second family. This leaves us with 9 particle species and 18 fluctuation fields to work with.

For the first equation of (\ref{eq:diffusion_0}) which describes the dynamics of the charges, it is instructive to write the equations in matrix form. The fluctuation fields $\mu$ and $u$ are then 9-component vectors  while the $K$-factors can be thought of as diagonal matrices. Only the collision terms mix the different particle species. The collision terms are proportional to specific linear combinations of chemical potentials that participate in the different interactions, as indicated in (\ref{eq:coll_def}). Moreover, the system can be arranged such that the inelastic collision terms are of the form ($i$ and $j$ are species indices, $\alpha$ runs over the different interaction processes)
\begin{equation}
\langle \mathbf{C} \rangle_i  = \sum_{\alpha, j} \Gamma_\alpha^{\text{inel}} \, \gamma_\alpha^i  \, ( \gamma^j_\alpha \, \mu_j ) \, 
\end{equation}
where $\gamma^i_\alpha$ denotes a vector that indicates which species $i$ participate in the interaction $\alpha$. It is then transparent that if a vector $q_i$ is orthogonal to all interactions,  the corresponding charge is conserved, 
\begin{equation}
\sum q_i \gamma_\alpha^i = 0 \quad \forall \, \alpha, \implies
\partial_z \left(\sum_i q_i J^z_i \right) = 0 \,.
\end{equation}
In the present system, this is achieved by counting two degrees of freedom for the Higgs and three for each quark. The whole system of four quarks is spelled out explicitly in~\cite{Fromme:2006wx}.

The only baryon number changing process is the sphaleron rate $\Gamma_{ws}$. However, the sphaleron rate is so slow that it can be decoupled from the actual diffusion. 
Often in the literature, the sphaleron rate is assumed to be vanishing in the broken phase and unsuppressed in the symmetric phase ($z \gtrless 0$). This is actually a  poor assumption. Here we will integrate the sphaleron diffusion equation numerically. We found that a much better approximation is actually to switch off the sphaleron when the exponent of the suppression factor is of order unity, namely $(a \, \phi/T) \simeq 1$, which is deep in the symmetric phase. Interestingly, the produced asymmetry can either be smaller or larger when doing the exact integration, due to a sign change in the CP violating  source in the wall.

\begin{table}
\begin{center}
\begin{tabular}{ |c|c|} 
\hline
interaction & rate\\
\hline
\hline
$t_L \leftrightarrow t_R + h$ & \multirow{2}{1cm}{$\Gamma_{y,t}$} \\ 
$b_L \leftrightarrow t_R + h$ &  \\ 
\hline
$b_L \leftrightarrow b_R + h$ & \multirow{6}{1cm}{$\Gamma_{y,b}$} \\ 
$c_L \leftrightarrow c_R + h$ & \\ 
$s_L \leftrightarrow s_R + h$ & \\ 
$t_L \leftrightarrow b_R + h$ & \\ 
$s_L \leftrightarrow c_R + h$ & \\ 
$c_L \leftrightarrow s_R + h$ & \\ 
\hline
$t_L \leftrightarrow t_R$ & $2\Gamma_{m,t}$ \\ 
\hline
$b_L \leftrightarrow b_R$ & \multirow{3}{1cm}{$2\Gamma_{m,b}$} \\ 
$c_L \leftrightarrow c_R$ & \\ 
$s_L \leftrightarrow s_R$ & \\ 
\hline
$t_L \leftrightarrow b_L$ & \multirow{2}{1cm}{$\Gamma_{W}$} \\ 
$c_L \leftrightarrow s_L$ & \\ 
\hline
$h \leftrightarrow 0$ & $\Gamma_h$ \\ 
\hline
{all L} $\leftrightarrow$ {all R} & $\Gamma_{ss}$ \\ 
\hline
\end{tabular}
\hskip 2 cm
\begin{tabular}{ |c|} 
\hline
inelastic rates \\
\hline
$\Gamma_{y,t} = 4.2 \times 10^{-3}  \, y_t^2 \, T$\\ 
$\Gamma_{y,b} = 4.2 \times 10^{-3}  \, y_b^2 \, T$  \\ 
$\Gamma_{m,t} = \frac{m_t^2}{63 T}$ \\ 
$\Gamma_{m,b} = \frac{m_b^2}{63 T}$  \\ 
$\Gamma_{W} = \frac{T}{60}$ \\ 
$\Gamma_{h} = \frac{m_W^2}{50 T}$ \\ 
$\Gamma_{ss} = 4.9 \times 10^{-4} T$ \\ 
\hline
\hline
elastic rates \\
\hline
$\Gamma_{tot,q} = \frac{T}{18}$ \\ 
$\Gamma_{tot,h} = \frac{T}{60}$ \\ 
\hline
\end{tabular}
\caption{
Interaction processes and rates used in the diffusion system. The numerical values for the rates in the right table are adapted from \cite{Fromme:2006wx}.
\label{tab:interac_and_rates}}
\end{center}
\end{table}

All interactions that are relevant are summarized in table~\ref{tab:interac_and_rates}. Specifically, we use the Yukawa rates $\Gamma_{y,t}$ and $\Gamma_{y,b}$ for the top and the light quarks respectively, $\Gamma_{m,t}$ and $\Gamma_{m,b}$ for the rates of the helicity flips for the top and the light quarks, $\Gamma_W$ for the $W$-scattering rate, $\Gamma_h$ for the Higgs number violation rate, $\Gamma_{ss}$ for the strong sphaleron rate  and  $\Gamma_{tot,q}$ and $\Gamma_{tot,h}$ for the quarks and the Higgs total elastic interaction rates.
We use a common Yukawa coupling for the bottom and the light quarks. Notice that the charged Higgses also contribute and their fluctuations are assumed to be equal to their uncharged counterparts.
In some of the flavor models we discuss below, the Yukawa couplings of light quarks are of order unity in the symmetric phase which is why we keep chirality flipping processes for the light quarks in the network.

We prefer not to use baryon number conservation to remove one set of fluctuations. Instead, we keep all fluctuations and check at the end that baryon number is indeed conserved up to this point. In the symmetric phase, there are actually more conserved charges. There, the Higgs VEV vanishes and the only active rates are actually the Yukawa, W-bosons and strong sphaleron interactions. Besides, not all of those interactions $\gamma^i_\alpha$ are independent. After all, only 7 different linear independent vectors arise and there are necessarily two conserved combinations. The second conserved charge actually arises because baryon number is conserved in the the first and second family separately.  

In conclusion of this section, we can now determine the profile of $\mu_L$ that controls the final baryon asymmetry (\ref{eqn:totalBaryonAsymmetry}), by solving a system of equations of the type (\ref{eq:diffusion_0}), involving in principle all particle species in the plasma. This system is written in matrix form, with matrices listed in Appendix \ref{sec:diffusionSystem}. We will apply this procedure to various systems of varying Yukawas in the next sections.

\section{A show case: CP violation in the top sector only}\label{sec:showcase_oneFlavour}

\subsection{Varying Top Yukawa}

In this section, we present in detail a case that was discussed several times in the literature: Baryogenesis from CP violation in the top sector only~\cite{Bodeker:2004ws, Espinosa:2011eu}.
We assume that  during the electroweak phase transition the top mass exhibits a changing complex phase, $m_t = | m_t(z)| \exp(i \, \theta(z))$. In this  case, the CP violating  source in the Boltzmann equation is
\begin{equation}
 \Im\left[V^\dagger {m^\dagger}'' m V\right]_{tt}=\Im\left[{m_t^\dagger}'' m_t \right]
=- \left[|m_t|^2 \theta'\right]' 
\end{equation}
where the first equality simply comes from the fact that, for one flavour, $m$ and $V$ are not matrices but just complex numbers and hence they commute.

Since the Yukawa couplings of the light quarks are small, their only interaction is the strong sphaleron. Besides, the left-handed bottom interacts with the left-handed top via the $SU(2)$ gauge interactions~\footnote{The left-handed quarks of the first two families also interact via the W-boson, but this is numerically not very important.}. This means that the chemical potentials of the first two families of quarks are all the same as the right-handed bottom (with opposite signs for left and right chirality). Using baryon number conservation, all these chemical potentials can be removed leaving only the left- and right-handed top and the left-handed bottom as relevant species. The corresponding diffusion system is given in~\cite{Fromme:2006wx}. We chose however to work with our more complete system, tracking 9 different species individually as pointed out in section~\ref{sec:diffusion_Network}. For this reason our results can differ by some small factor from the results in~\cite{Fromme:2006wx}.

For the Higgs VEV profile we assume a tanh that is parametrized by a wall thickness $L_w$
\begin{equation}
\label{eq:vevprofile}
v_\phi(z) = \frac12 \, \phi_T \, \left[ 1 - \text{Tanh}(z/L_w) \right] \, . 
\end{equation}
and for the CP-violating phase we assume,
\begin{equation}
\label{eq:varyingtheta}
  \theta(z)=\text{ArcTan}\left\{\frac{1}{2}\Delta\theta \left[ 1 - \text{Tanh}(z/L_w) \right] \right\}
\end{equation}
\begin{figure}[!t]
\centering
\includegraphics[width=0.45\textwidth]{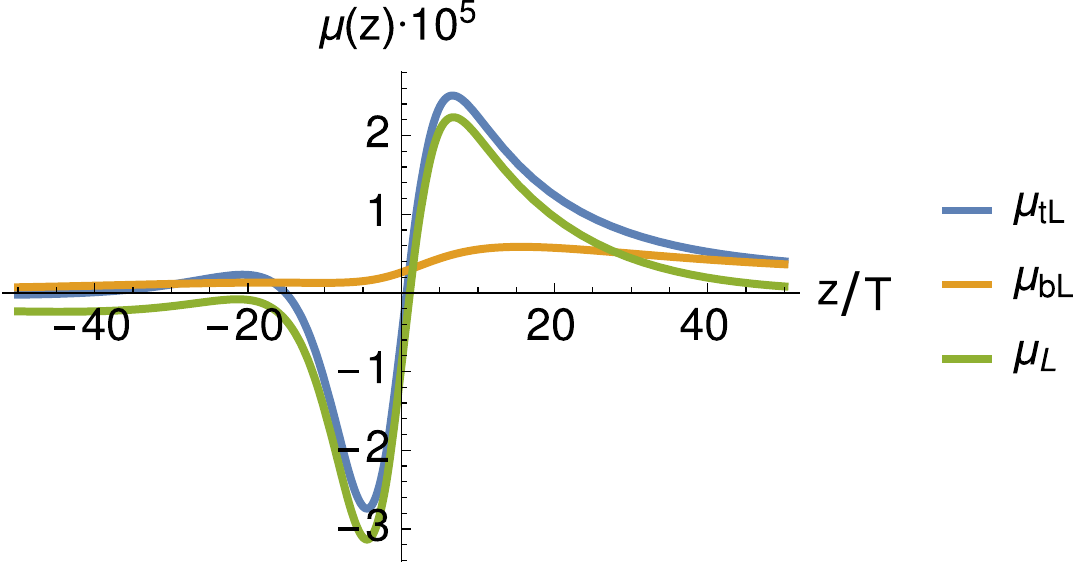}
\includegraphics[width=0.45\textwidth]{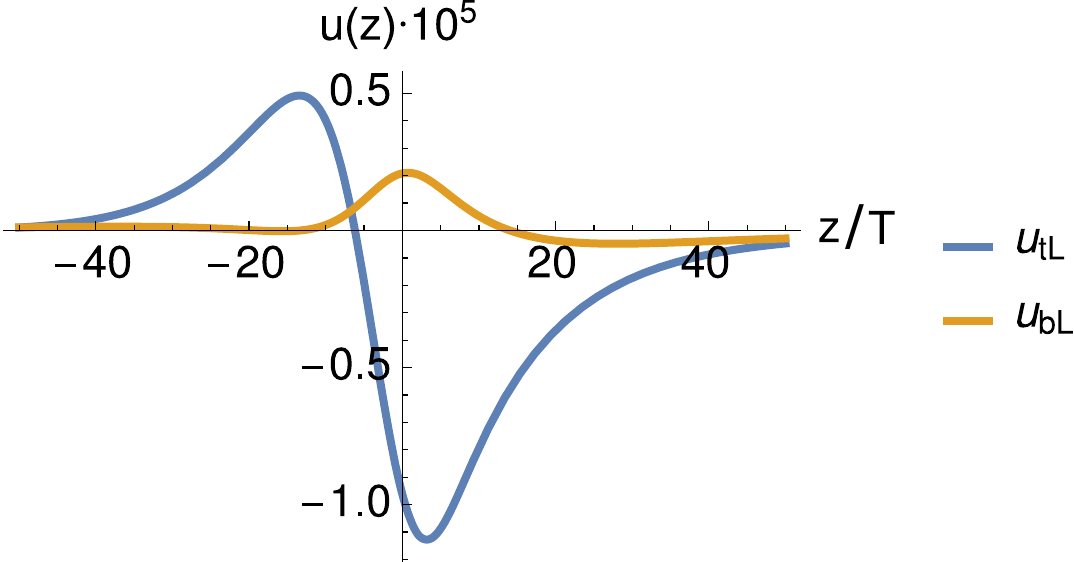}\\
\includegraphics[width=0.45\textwidth]{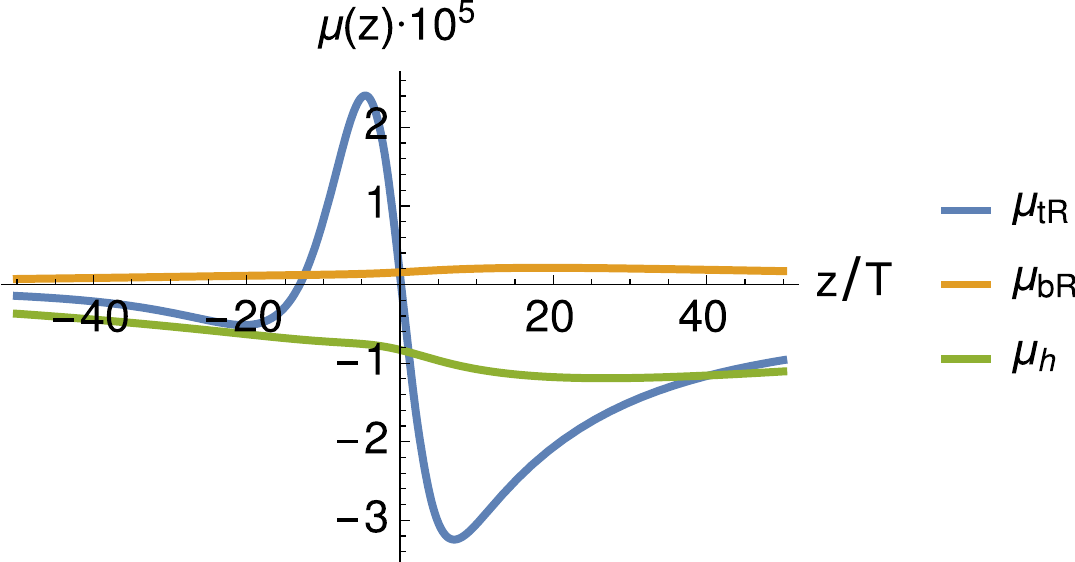}
\includegraphics[width=0.45\textwidth]{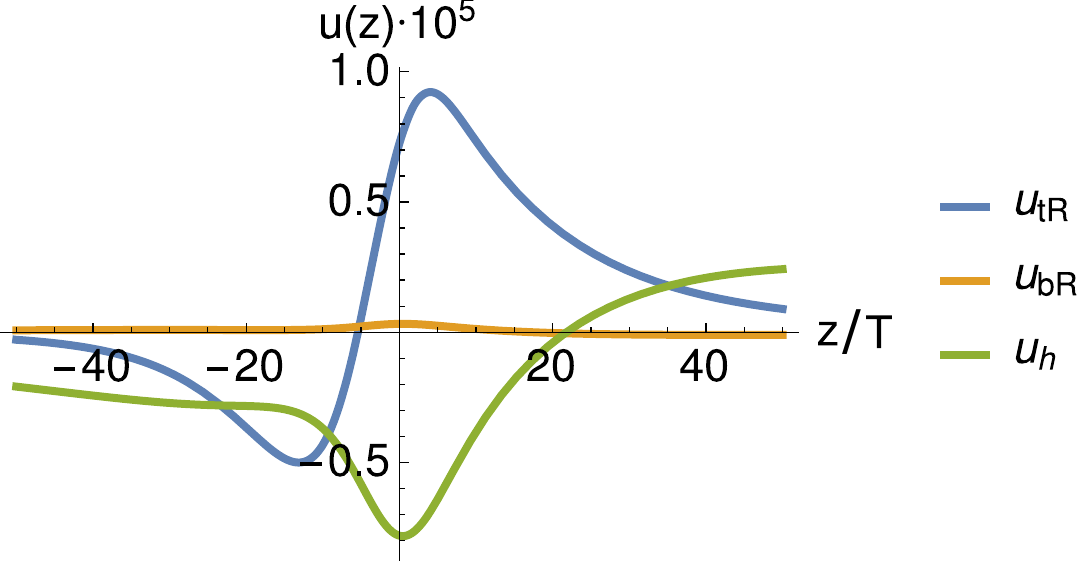}
\caption{\label{fig:top_mu_and_u} \em
Chemical potentials $\mu$ (left) and velocity fields $u$ (right) induced by a varying complex Top quark Yukawa coupling according to Eq. (\ref{eq:varyingtheta}). 
The upper graphs show the left-handed particle species (and the total left-handed chemical potential) while the lower graphs show the 
right-handed species and the Higgs. Note that the Higgs diffuses unhindered into the symmetric phase while all the quarks are damped. The parameters are $\phi/T = 1.5$, $L_w = 8 /T$, $v_w = 0.1$ and $\Delta\theta=0.1$.
}
\end{figure}
\begin{figure}[!h]
\centering
\includegraphics[width=0.45\textwidth]{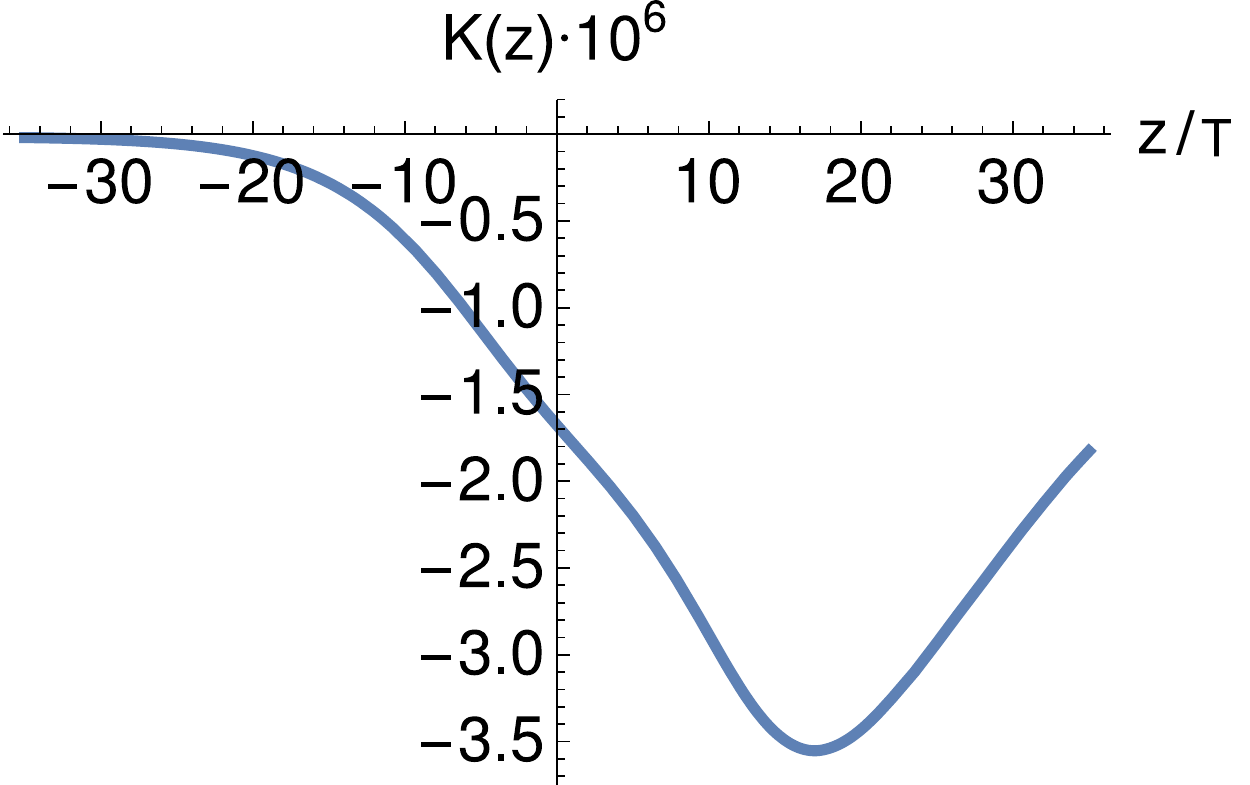}
\includegraphics[width=0.45\textwidth]{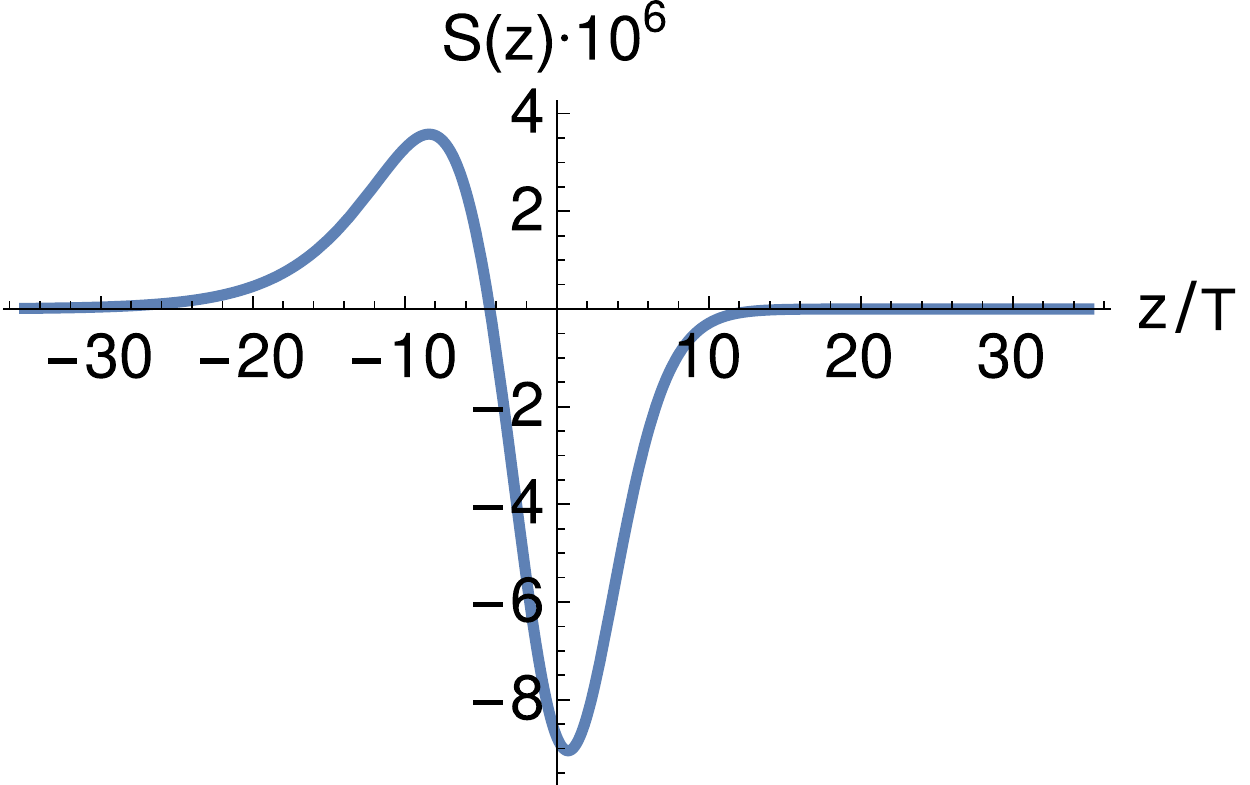}
\caption{\label{fig:top_G_and_S} \em
The kernel $K(z)$ and the source $\bar {\cal S}(z)$. The dip in $K(z)$ results from the competition between the suppression of sphaleron transitions inside the broken phase and the washout of the baryon asymmetry  far from the wall. The source $\bar {\cal S}(z)$ is peaked within the wall. The parameters are as in figure~\ref{fig:top_mu_and_u}. The resulting baryon asymmetry is $\eta_{B} =  1.01\cdot 10^{-10}$. 
}
\end{figure}
\begin{figure}[!h]
\centering
\includegraphics[width=0.5\textwidth]{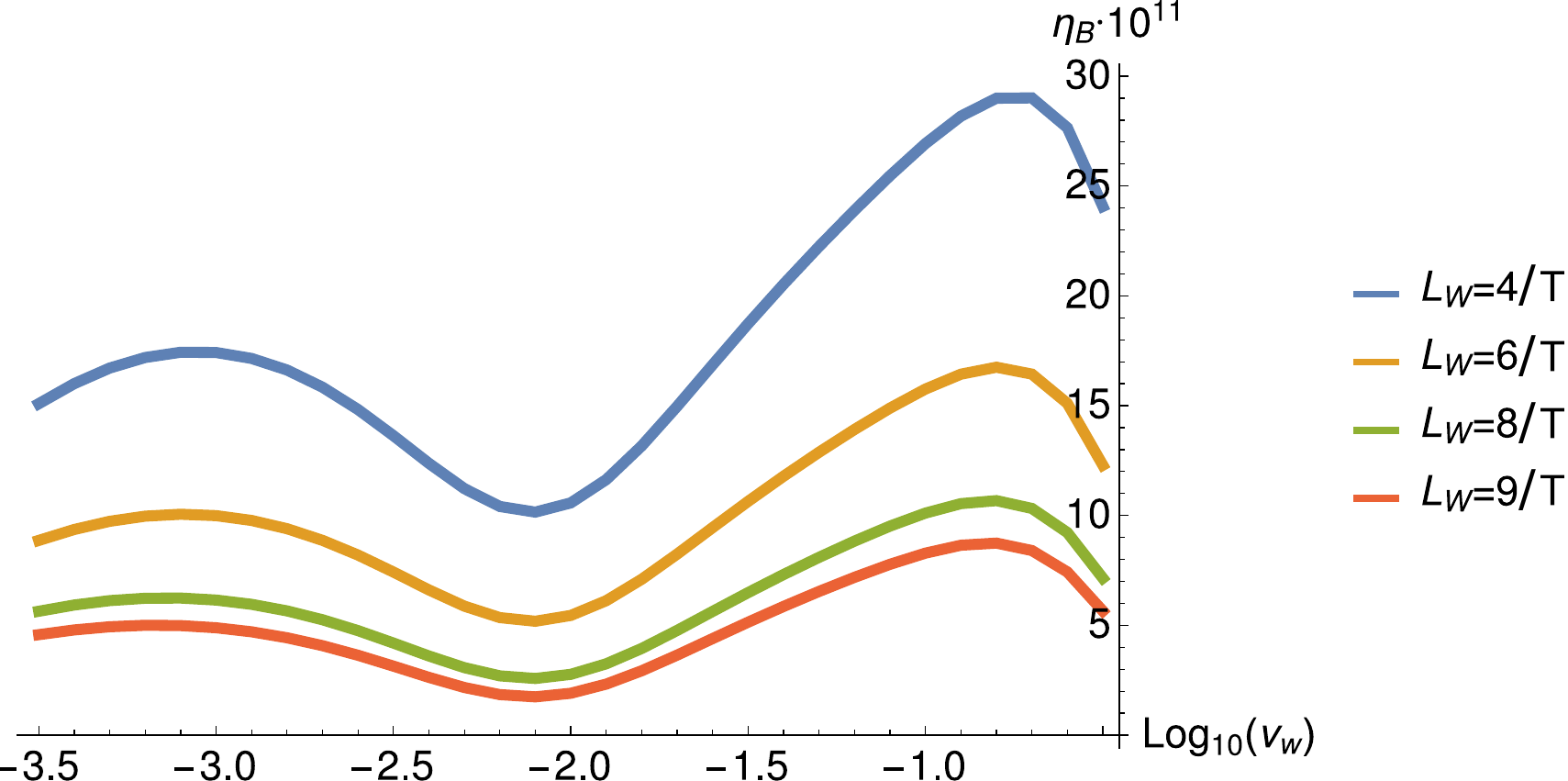}
\caption{\label{fig:top_eta_vs_v_and_l} \em
Dependence of the final baryon asymmetry on the wall velocity $v_w$ and the wall thickness $L_w$. The change in the complex phase of the top mass is $\Delta \theta = 0.1$. This plot is obtained using the 18D system.
}
\end{figure}
{Ultimately, this profile and the wall thickness
would have to be determined in a dedicated analysis. 
Since the CP violation is typically small, $\Delta\theta \ll 1$,  the CP phase $\theta$ is practically described by a $\text{Tanh}(z/L_w)$.
We  chose the parametrizations (\ref{eq:vevprofile}) and (\ref{eq:varyingtheta}) since this is the behavior that was found in the singlet extension of the Standard Model~\cite{Espinosa:2011eu} which is very close to our flavon model. The result depends strongly on the fact that the CP phase and the Higgs VEV change simultaneously during the phase transition. We introduce a more general parametrization in Section \ref{sec:generalParamTwoFlavour} in order to take this into account.
The wall thickness $L_w$ can be approximately inferred from the tunnel bounce that was calculated for the Froggatt-Nielsen models in~\cite{Baldes:2016gaf}. Typical values are between $3/T$ and $20/T$ and we use values between $L_w = 3/T$ and $L_w = 8/T$ for our analysis. }

We described in section~\ref{sec:baryonasymmetry} how the diffusion equations are solved numerically taking the mass dependence in the network into account. We confirm that the correct amount of baryon asymmetry can be obtained from a top quark Yukawa varying phase.
Numerical results are shown in  Figures~\ref{fig:top_mu_and_u} to~\ref{fig:top_eta_vs_v_and_l}. 
Figure~\ref{fig:top_mu_and_u} shows the profile of the chemical potentials and velocity fields across the bubble wall for the various relevant degrees of freedom in the plasma and some choice of parameters.
Figure \ref{fig:top_G_and_S} shows the kernel $K(z)$ and the Source $\bar {\cal S}(z)$. 

{Figure \ref{fig:top_eta_vs_v_and_l} shows the dependence of the final baryon asymmetry on the wall velocity $v_w$ and the wall thickness $L_w$. 
 The final asymmetry scales approximately as $1/L_w$ while it is rather insensitive to the wall velocity in the range $10^{-3} < v < c_s = 1/\sqrt{3}$. 
The wall velocity $v_w$ can be determined 
along the lines of~\cite{Moore:1995ua, Moore:1995si, John:2000zq, Kozaczuk:2015owa, Konstandin:2014zta}. It was calculated in the Standard Model in~\cite{Moore:1995si} and found to be of order $\sim 0.3$. The varying Yukawa couplings should somewhat enhance the friction and hence reduce the wall velocity and we use as a benchmark $v_w = 0.1$. }


\subsection{Comprehensive study of the one-flavour case}\label{sec:varyTopYukawaOneFlavour}
In order to estimate the possibility of achieving  baryogenesis with light flavours, we performed an analysis using the above system where the CP asymmetry is generated by one single flavour.
For this study we let the top Yukawa coupling take different values through the bubble wall. Concretely we parametrized this coupling as:
\begin{equation}
 Y_t(z)=Y_c+Y_v(z)e^{i\theta} \quad \text{where}\quad Y_v(z)=Y_v^b+(Y_v^s-Y_v^b)\left[1-\frac{\phi(z)}{v}\right]
 \label{eq:realisticansatz}
\end{equation}
In this case the change in the complex phase is induced by the fact that the Yukawa has two contributions, one that is constant during the phase transition and one that varies. If only one of the two contributions has a non-zero complex phase (or if both have non-zero but different phases) an effective $z$-dependence of the total phase is induced: $\Theta(z)=Arg(Y_t(z))$. This model has  four parameters and we present a scan over two of them. We fix $\theta=\pi/2$ where we expect a big CPV source. We also fix the ratio $Y_v^b/Y_c=f$, i.e. if $f=0$ the varying part of the Yukawa coupling goes to zero in the broken phase and if $f=1$ it will take the same value as the constant contribution. We vary  the values of the Yukawa couplings in the broken ($Y_t^b$) and in the symmetric ($Y_t^s$) phase, respectively.

\begin{figure}[t]
\centering
\includegraphics[width=0.45\textwidth]{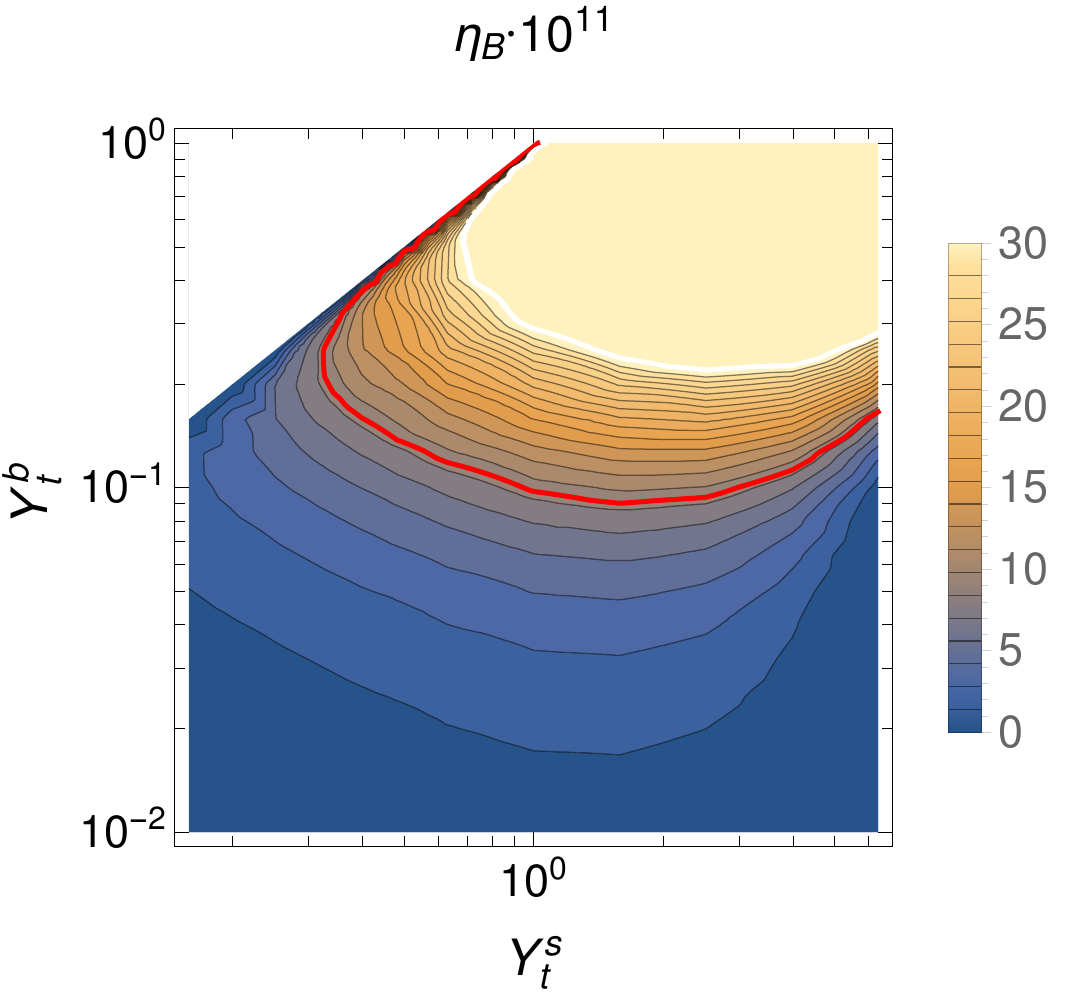}
\includegraphics[width=0.45\textwidth]{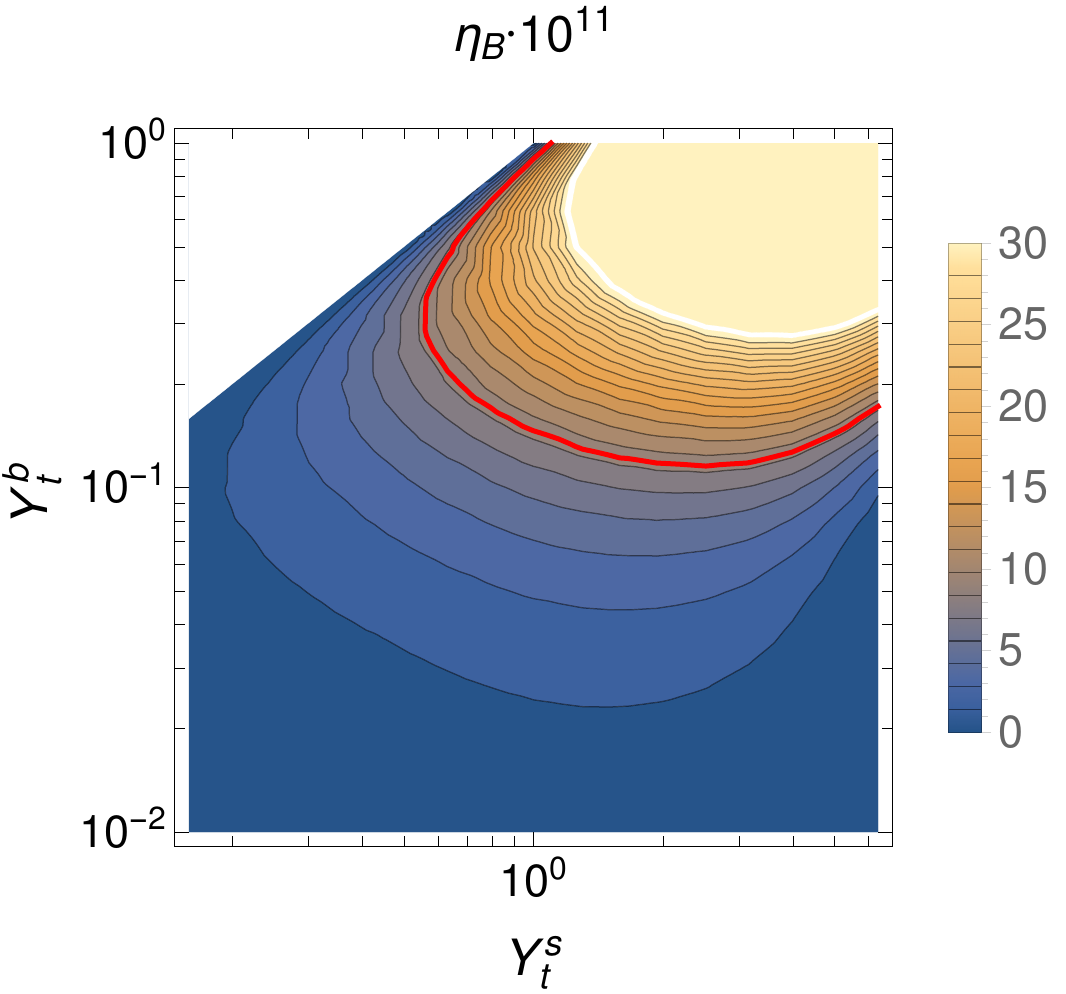}
\caption{\label{fig:scan_two_yukawas} \em
The case where a single flavour has a varying Yukawa: A scan over the values of the Yukawa coupling in the symmetric and the broken phases for two choices of $f=Y_v^b/Y_c$. $f=0$ on the left hand plot and $f=1$ on the right. For the choice $\theta=\pi/2$ the coupling in the symmetric phase is always larger than the coupling in the broken phase ($Y_t^b\leq Y_t^s$). In red we indicate the measured baryon asymmetry. The wall velocity and thickness were taken to be $v_w=0.1$ and $L_w=8/T$ respectively.
}
\end{figure}

Our main motivation is to investigate how low can the Yukawa coupling be in the broken phase if it is of order one in the symmetric phase, to account for the correct baryon asymmetry.
On figure~\ref{fig:scan_two_yukawas} we show a scan over the values for the Yukawas in the broken and symmetric phase for two choices of $f$, namely $f=0$ and $f=1$. 
The conclusion of this exercise is that, when considering only one flavour, we are forced to work with the top quark, as all the other fermions in the Standard Model would not be able to produce a large enough baryon asymmetry.

Note that if we were to remain completely agnostic about the physics behind the dynamical Yukawa coupling we could parametrize the Yukawa as:
\begin{equation}
	Y_t(z)=Y(z)e^{i\theta(z)}
\end{equation}
where $Y(z)$ and $\theta(z)$ are, a priori arbitrary, real functions. In this case, there is much more freedom and we can achieve successful baryogenesis for an appropriate choice of the functions $Y(z)$ and $\theta(z)$ even if $Y(z)$  takes very small values in the broken phase, i.e. for the light quarks. We confirmed numerically that the correct asymmetry can be generated with $Y(z)$ and $\theta(z)$ following hyperbolic tangents and $Y(z)$ taking very small values in the broken phase. Nevertheless there is little physical motivation for this rather academic exercise and  we do not discuss the results here.

In summary, in the one-flavour case, the CP-violating force comes from the term $\left(m^2 \theta^{\prime}\right)^{\prime}$ and therefore $\theta^{\prime}$ needs to be large. Using the realistic ansatz 
(\ref{eq:realisticansatz}), the phase tends to change in the region where the Yukawa coupling has reached its value in the broken phase, which suppresses the CP violation for light quarks other than the top quark. It will be different in the multi-flavour case where the flavour mixing (CKM matrix) changes across the bubble wall and this leads to additional CP violation independently of the variation of the phase of the Yukawa couplings, as we will see in Section \ref{sec:twoflavour}.

%

\subsection{Sensitivity to interaction rates}\label{sec:InteractionRates}
\begin{figure}[t]
\centering
\includegraphics[width=0.5\textwidth]{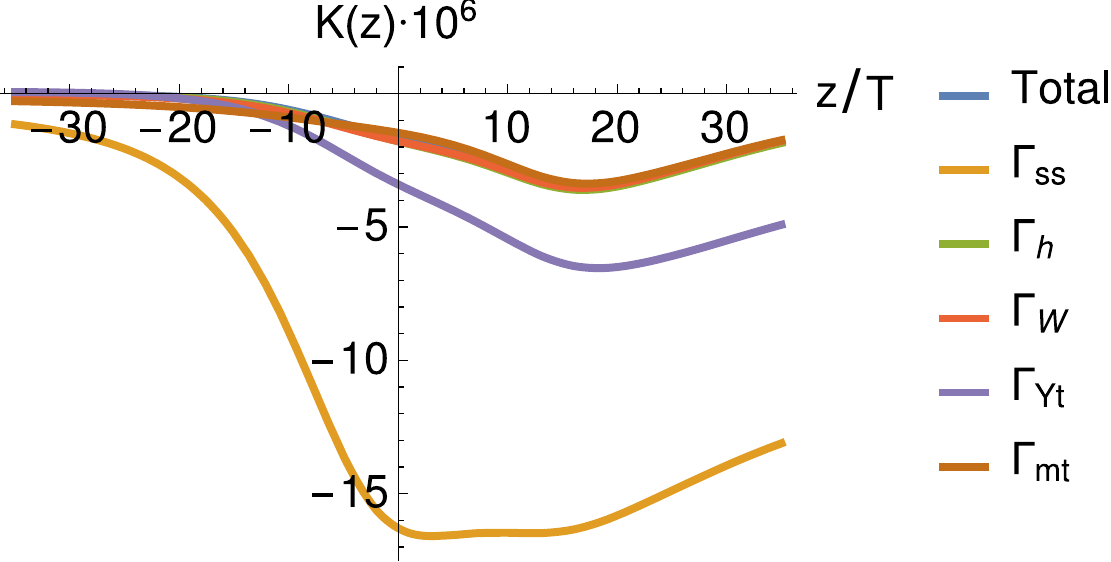}
\caption{\label{fig:interactionRates} \em
The kernels for the diffusion equation with different interaction rates turned off. The legend indicates the rates that have been turned off. All the kernels, with the exception of the strong sphaleron and the top Yukawa, are concentrated around the original kernel shown as Total. The change of the complex phase in the top mass is $\Delta \theta = 0.1$ and the wall velocity and thickness were taken to be $v_w=0.1$ and $L_w=8/ T$ respectively.
}
\end{figure}

In this subsection we study how the various interaction rates used in the diffusion equations influence the final result. This question is interesting for two main reasons. Some of the rates used in the diffusion equations are approximations (see Table \ref{tab:interac_and_rates}). It is therefore crucial to check the sensitivity  on the approximation made. The other reason is that we can adapt our system to study EW baryogenesis from  varying lepton Yukawa couplings. In this case, we will have to calculate the diffusion of the leptons through the wall. As leptons have different interaction rates (notably they do not couple to the strong sphaleron) we can estimate whether the absence of coupling with the strong sphalerons  \cite{Giudice:1993bb} can compensate for the fact that leptons have less degrees of freedom and smaller Yukawa couplings in the broken phase. We therefore analyze the kernels and the total baryon asymmetry in systems where  different interaction rates are switched off. The result is  shown in figure \ref{fig:interactionRates} where the kernels for different scenarios are shown. With the exception of the strong sphaleron and the top Yukawa rates, the rates do not have a strong influence on the kernel. The fact that the baryon asymmetry increases when the strong sphalerons are switched off is  encouraging in view of explaining the baryon asymmetry using leptons instead of quarks.

%
\section{Two-flavour and three-flavour cases}
\label{sec:twoflavour}

As we saw earlier, in models where only one flavour has a varying Yukawa, there is enough CP violation only if the Yukawa is large in the broken phase, namely it can only work for the top quark.
On the other hand, as we will show in this section, with multiple varying flavours, electroweak baryogenesis can be successful even using light flavours only. In fact,  in this case, in the CP-violating source  (\ref{eq:CPsource}), the variation of the flavour composition of the mass eigen states  $V(z)$ plays an important role.

\subsection{General parametrization}\label{sec:generalParamTwoFlavour}
\begin{figure}[t]
\centering
\includegraphics[width=0.45\textwidth]{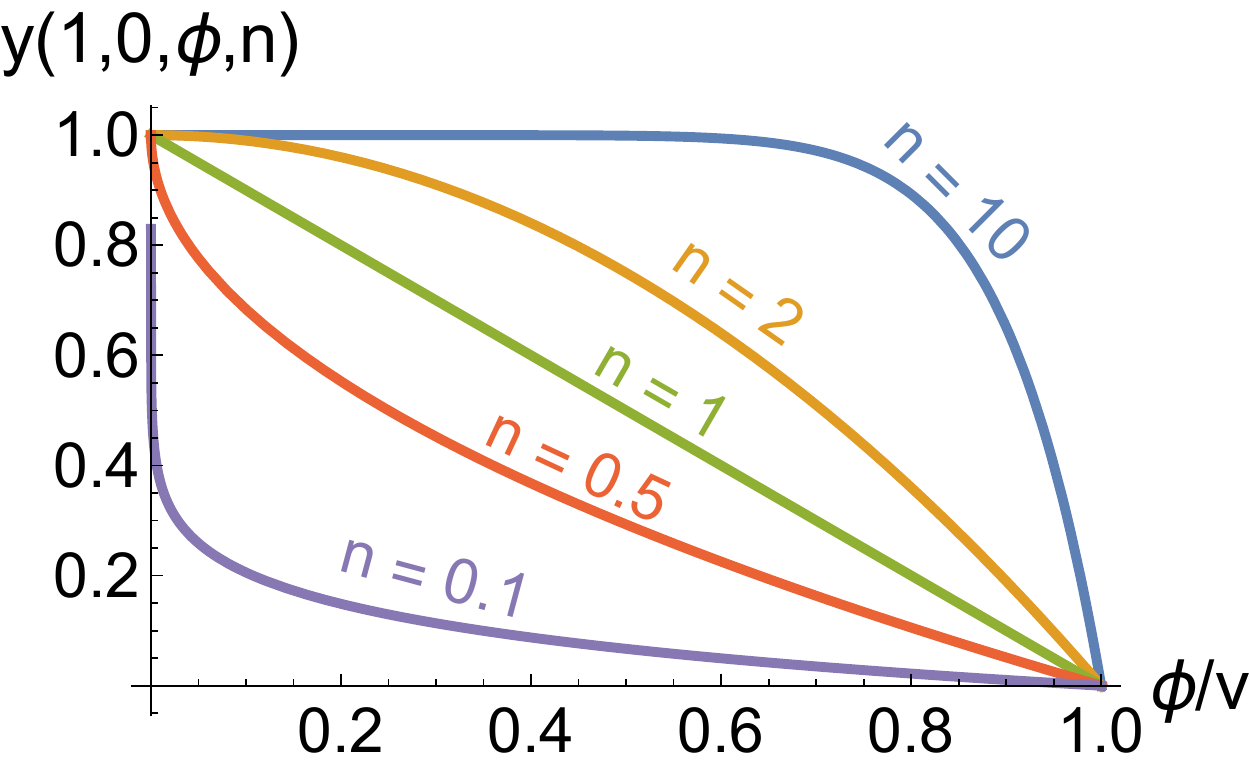}
\caption{\label{fig:yukawasGeneral} \em The general parametrization of the Yukawa variation across the bubble wall given in eq.~(\ref{eqn:generalParam}) for different values of $n$ (it does not apply for the top quark).}
\end{figure}

Before studying specific models of varying Yukawas, it is useful to consider some generic parametrization of the Yukawa variation to learn about the requirements on the flavour model  to explain the baryon asymmetry. A similar approach was presented in \cite{Baldes:2016rqn} to study the impact of varying Yukawas on the nature of the electroweak phase transition. In practise, one should include the degrees of freedom responsible for the dynamics of the Yukawas, typically the flavon field. However, we can  replace the flavon VEV by a function of the Higgs VEV once a field trajectory is assumed during the electroweak phase transition. 
 As the flavon is coupled to the Higgs, the Yukawas effectively depend on the value of the Higgs VEV. This can be parametrized as follows
\begin{equation}\label{eqn:generalParam}
  y(y_0,y_1,\phi,n)=(y_0-y_1)\left[1-\left(\frac{\phi}{v}\right)^n\right]+y_1
\end{equation}
where $y_0$ ($y_1$) is the value of the coupling in the symmetric (broken) phase, $\phi$ the Higgs VEV, $v$ the Higgs VEV in the broken phase (minimum of the potential) and $n$ is a free parameter that indicates where the change takes place. For a large value of $n$ the Yukawa coupling keeps the value of the symmetric phase until deep inside the broken phase, whereas for small values of $n$ the Yukawa changes rapidly in front of the wall (i.e. in the symmetric phase). This is illustrated on fig.~\ref{fig:yukawasGeneral} where the Yukawa coupling is shown as a function of $\phi$ for different choices of $n$. Somehow the choice of $n$ is a parametrisation of the trajectory in the (Higgs, Flavon) field space during the phase transition, which can be determined through a dedicated analysis of tunneling in a 2-field space \cite{CompositeEWPT}. Typically, the lighter the flavon, the larger $n$ can be. On the other hand, for a flavon much heavier than the Higgs, tunnelling will tend to happen along the flavon axis and therefore this will correspond to a small $n$ value. In this case, electroweak symmetry breaking happens after most of the flavour structure has already emerged, which will suppress CP-violation from Yukawa variation during the EW phase transition and therefore when sphalerons get out-of-equlibrium.

\begin{figure}[t]
\centering
\includegraphics[width=0.45\textwidth]{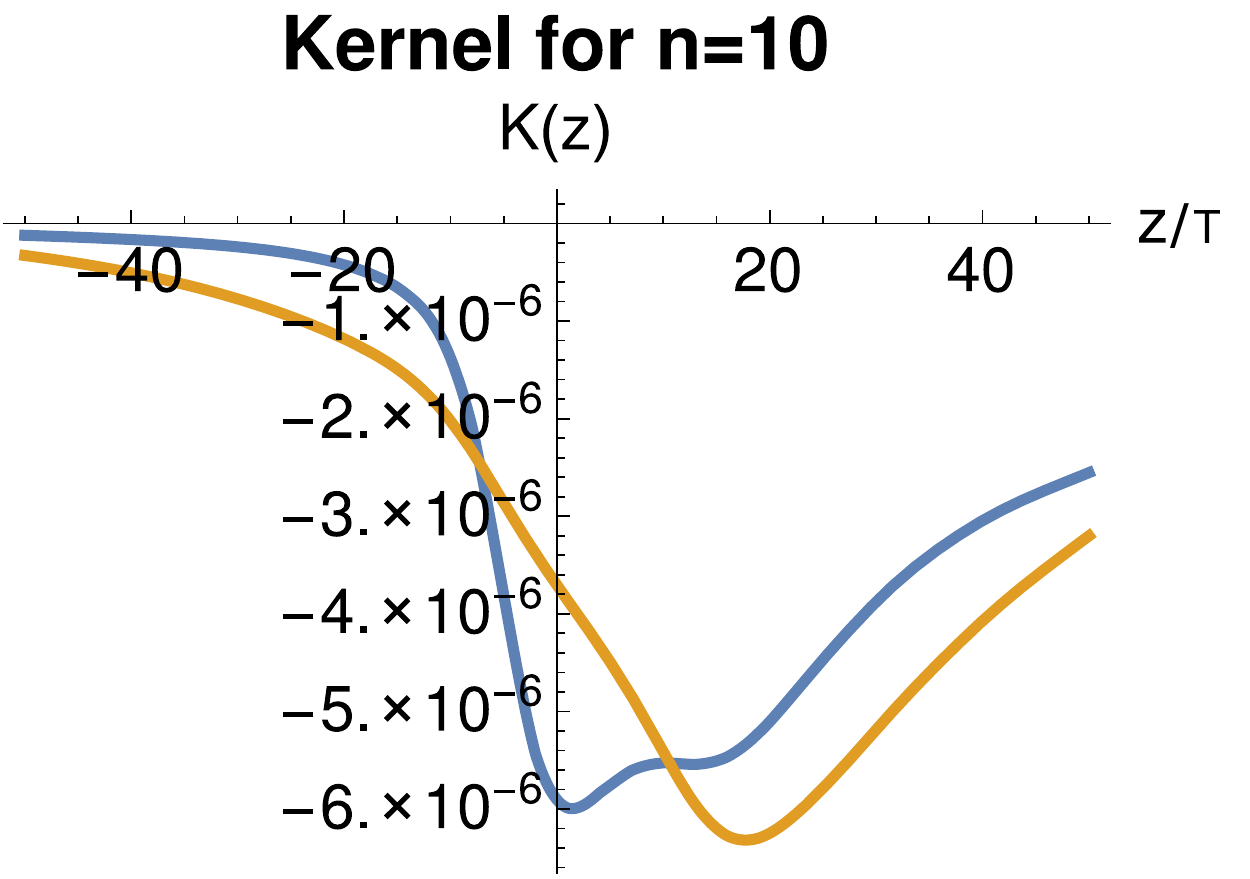}
\includegraphics[width=0.45\textwidth]{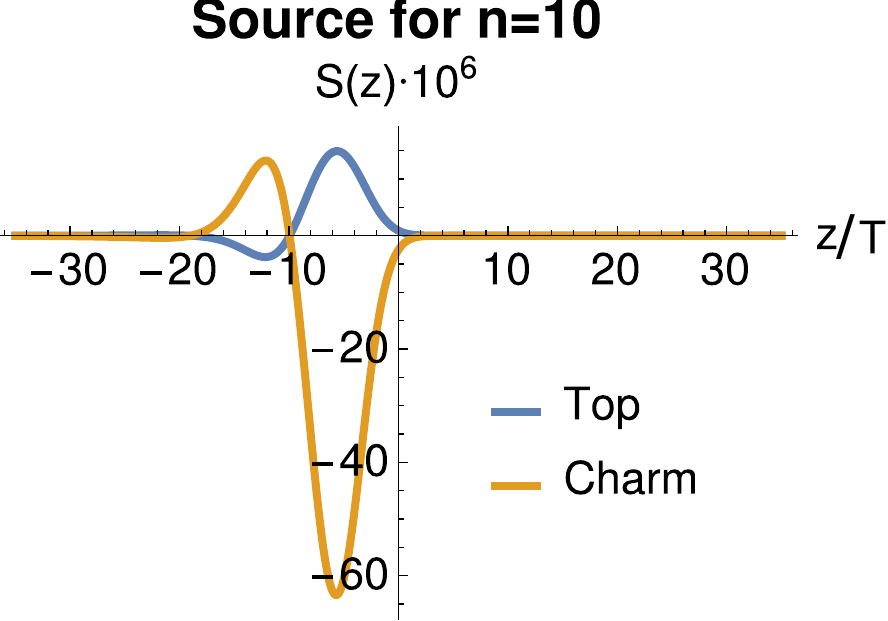}\\
\includegraphics[width=0.45\textwidth]{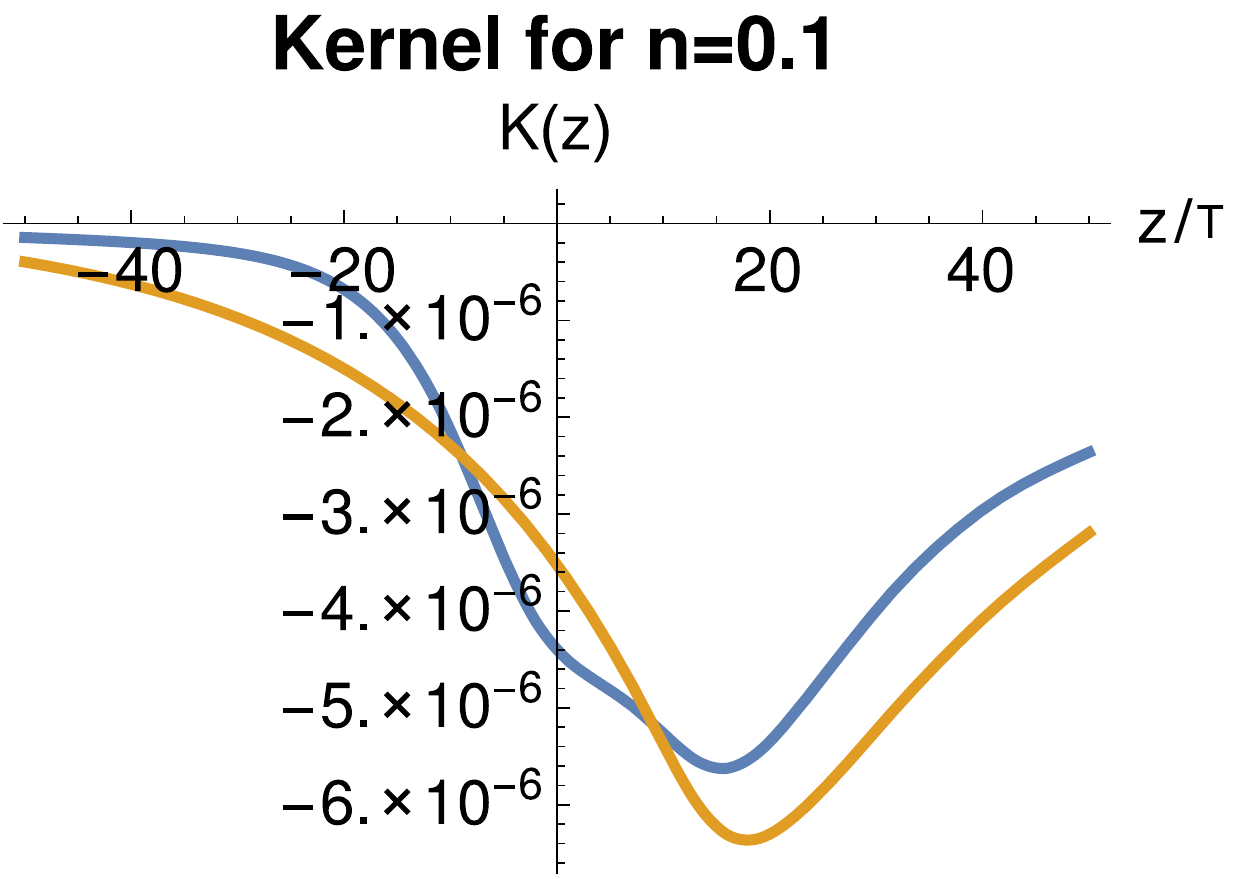}
\includegraphics[width=0.45\textwidth]{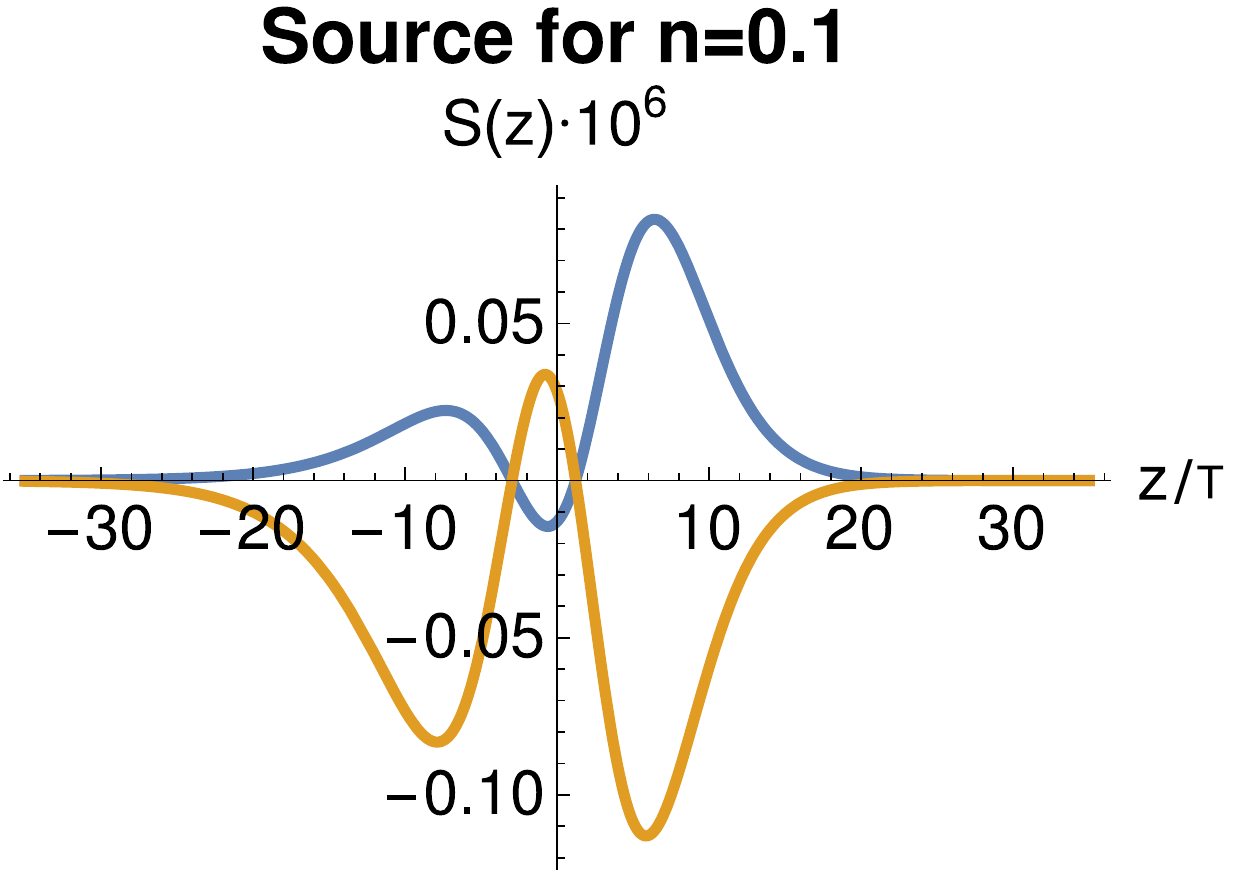}
\caption{\label{fig:GeneralKernelAndSource} \em
The kernel and source terms for the general parametrization assumed in this section. The kernels are not very dependent on the exact form of the Yukawa couplings as they only enter the calculation through the Yukawa interaction rates. The source term on the other hand gets the expected shift towards the symmetric  phase for small values of $n$ and it also gets a large suppression of the amplitude by $\sim 2$ orders of magnitude from the value of the Higgs VEV. The corresponding baryon asymmetries are: $\eta_B\approx 5.5 \cdot 10^{-10}$ for $n=10$ and $\eta_B\approx 2.1 \cdot 10^{-12}$ for $n=0.1$.
}
\end{figure}

As the CP violating source  is proportional to the second derivative of the mass parameter we expect the source for small (big) $n$ to be localized outside (inside) the broken phase. It would therefore seem beneficiary to have a model which would map onto a parametrization with a small $n$. As the weak sphaleron is most active in the symmetric phase, the baryon asymmetry can be augmented by creating the CP asymmetry in front of the bubble wall and therefore evading the suppression from the diffusion through the bubble wall. However, the CP violating source is also proportional to the mass parameter and hence every source deep inside the symmetric phase gets suppressed by the Higgs VEV which is rapidly approaching zero in the symmetric phase. This is shown on figure~\ref{fig:GeneralKernelAndSource}, where we report the results for the Yukawa matrix
\begin{equation}
	Y=\left(\begin{array}{cc}
		e^{i \theta}y(1,0.008,\phi,n) & y(1,0.04,\phi,n) \\
		y(1,0.2,\phi,n) & y(1,1,\phi,n)
	\end{array}
	\right) \qquad \text{with} \qquad \theta=1\, .
\end{equation}
We see a two orders of magnitude decrease in the amplitude of the source term. This cannot be compensated by the change in the kernel when going from the broken phase to the symmetric one,  which is less than one order of magnitude.
There is even an additional suppression in the case of a source term that is located in the symmetric phase. The source term for the particle $i$ is proportional to $\text{Im}\left[V^\dagger {m^\dagger}'' m V\right]_{ii}$. It is easy to show that the matrix $\text{Im}\left[V^\dagger {m^\dagger}'' m V\right]$ is traceless for constant complex phases in the Yukawa couplings. Therefore the source for the top always has the opposite sign of the source for the charm. If this were the full source term there would be an almost perfect (up to differences in the kernels) cancellation between the contribution from the top and the charm. The full source is also proportional to the factor $K_8$ which depends on the mass of the particle in question. For a source that is localized inside the broken phase, where the particles have very different masses, this factor $K_8$ prevents a big cancellation between the two contributions. On the other hand, for a source located in the symmetric phase, where both particles have a nearly vanishing mass, the cancellation is very much unsuppressed. 

This analysis shows that, even though we naively expect the mechanism to be more effective when the source is located in front of the bubble wall, models with a source located inside the broken phase are more effective in producing the baryon asymmetry. However, this is partially due to the fact that in this section we assume the same $n$ for all flavours in the parameterization (\ref{eqn:generalParam}). 
When studying specific models without this assumption, it is also possible to create a sizable baryon asymmetry using sources that are located inside the symmetric phase. This is for example the case for Randall-Sundrum models when focusing on the (top,charm) system which would then correspond to profiles with different $n$ for different flavours.

\subsection{Small Yukawas in the broken phase}

\begin{figure}[t]
\centering
\includegraphics[width=0.55\textwidth]{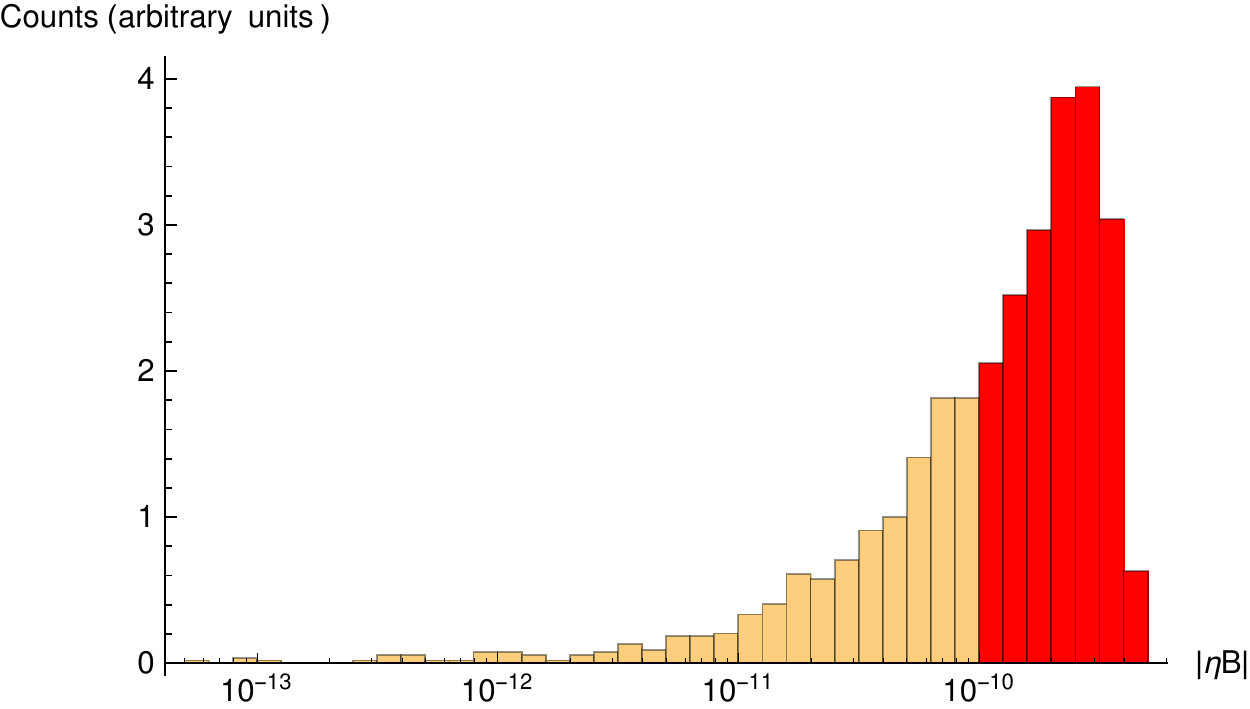}
\caption{\label{fig:histZeroYukawasMutliFlavour} \em The baryon asymmetry calculated for random values of the exponents $n_i$. In red is the fraction of the results that produced a larger baryon asymmetry than the one measured. The wall parameters were chosen to be $L_w=5/T$, $v_w=0.1$ and $\phi/T=1.5$.
}
\end{figure}

In the one flavour case we concluded that EW baryogenesis is possible only if we use the top quark as a source of CP-violation. On the other hand we will see that in the multi-flavour case the $\tau$-$\mu$ system is enough to generate the baryon asymmetry. It is interesting to see if we can also understand this behaviour using  the general parametrization discussed in this section. 
We need to take into account all entries in the Yukawa matrix. 
One approach would be  to scan the whole Yukawa matrix in the broken phase using a common parameter:
\begin{equation}
	Y(c)=\left(\begin{array}{cc}
		e^{i \theta}y(1,c~0.008,\phi,n) & y(1,c~0.04,\phi,n) \\
		y(1,c~0.2,\phi,n) & y(1,c~1,\phi,n)
	\end{array}
	\right)\qquad \text{with} \qquad c\in[0,1]\, .
\end{equation}
Where the function $y$ is given by Eq.~(\ref{eqn:generalParam}).
However, in the limit  $c \to 0$, the Yukawa matrix approaches
\begin{equation}
	Y\rightarrow\left(\begin{array}{cc}
		e^{i \theta}f(\phi) & f(\phi) \\
		f(\phi) & f(\phi)
	\end{array}
	\right)\, 
\end{equation}
which  will always produce a vanishing CP-violating source since $(Y^\dagger)''Y$ is a Hermitian matrix and hence the diagonal of $\Im[V^\dagger(m^\dagger)''mV]$ vanishes. In the end the total baryon asymmetry has to vanish for a vanishing $c$ and this parametrization does not help to understand the case of the leptons. This problem arises because all entries in the Yukawa matrix share the same exponent $n$. In the case of a Froggatt-Nielsen model this is however not the case. It is more realistic to give each of the four entries a different exponent. This rapidly increases the number of free parameters and makes a complete parameter scan intractable. Instead, we define a Yukawa matrix
\begin{equation}
	Y(n_1,n_2,n_3,n_4)=\left(\begin{array}{cc}
		e^{i 3.17}y(1,0,\phi,n_1) & e^{i 4.92}y(1,0,\phi,n_2) \\
		e^{i 5.29}y(1,0,\phi,n_3) & e^{i 2.04}y(1,0,\phi,n_4)
	\end{array}
	\right)
\end{equation}
and randomly chose values for $n_i$ 
from a uniform distribution in $[0,10]$. 
To assess how generic the production of the observed baryon asymmetry is in this model, we report the result of this study in a histogram (figure~\ref{fig:histZeroYukawasMutliFlavour}), where we show the absolute value of the baryon asymmetry.
The ballpark of results lies around $|\eta_B|\sim 1.47\cdot10^{-10}$, i.e. very close to the measured value ($\sim 62\%$ are above the measured value). This is a very promising result because there are still many parameters that we can change and adapt in order to increase the baryon asymmetry. For example, each entry in the Yukawa matrix can be multiplied by a numerical $\mathcal{O}(1)$ factor. We conclude that it is therefore not too difficult to find a set of parameters for which we can reproduce the correct baryon asymmetry, when using a 2-flavour system, even for light flavours (provided that they had large yukawas in the symmetric phase).

\subsection{Varying lepton Yukawas }
\begin{figure}[t]
\centering
\includegraphics[width=0.45\textwidth]{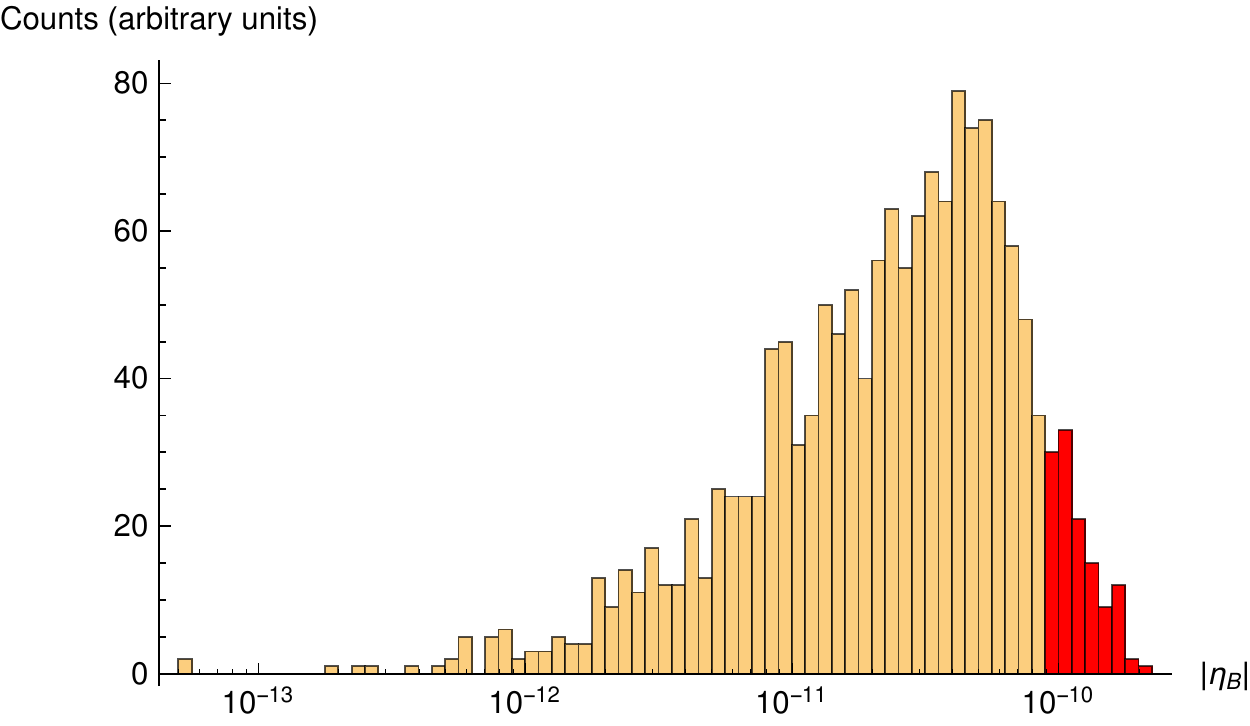}
\includegraphics[width=0.45\textwidth]{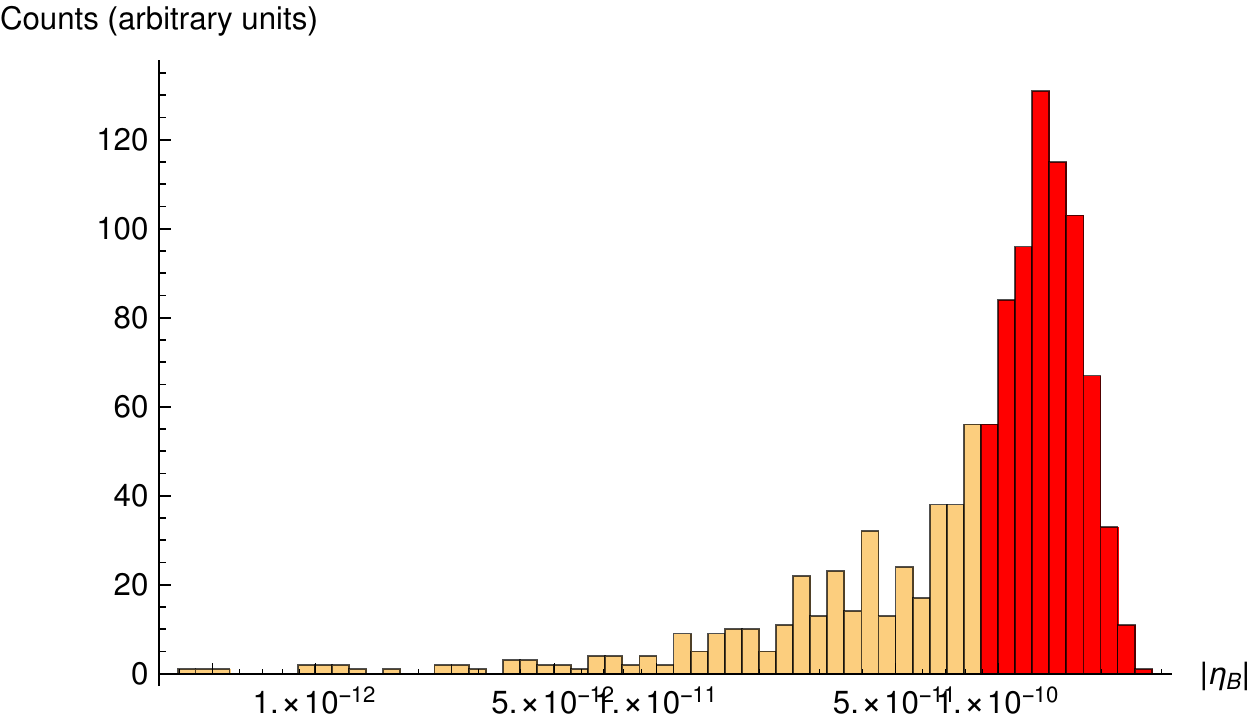}
\caption{\label{fig:histRandYukawasLept} \em The baryon asymmetry calculated in the case of the $\tau-\mu$ system for random values of the order one couplings and the phases. In red is the fraction of the results that produced a larger baryon asymmetry than the one measured. The left hand plot shows the result for $y_{ij}$ chosen from the {Rayleigh distribution} whereas on the right hand side we show the result for a uniform distribution of $y_{ij}$ in $[2,3.5]$. The wall parameters where chosen to be $L_w=3/T$, $v_w=0.1$ and $\phi/T=1.3$.
}
\end{figure}

In order to understand how to generically produce the right amount of baryon-asymmetry in the $\tau-\mu$ system we performed a random sampling of parameter space motivated by  Froggatt-Nielsen  presented in Section \ref{subsec:FN}. We sample randomly phases and $\mathcal{O}(1)$ couplings multiplying the flavon contribution in the Yukawa matrix: 

\begin{equation}\label{eqn:lepton_froggat_nielsen}
  Y_{\tau-\mu}=\left(\begin{array}{cc}
e^{i\theta_{\mu\mu}}y_{\mu\mu}\epsilon_\chi^4 & e^{i\theta_{\tau\mu}}y_{\tau\mu}\epsilon_\chi^5 \\
e^{i\theta_{\mu\tau}}y_{\mu\tau}\epsilon_\chi^2 &  e^{i\theta_{\tau\tau}}y_{\tau\tau}\epsilon_\chi^3
\end{array}
  \right)
  +
  \left(\begin{array}{cc}
6.1465\epsilon_\sigma^4 & -3.125\epsilon_\sigma^5 \\
0.5\epsilon_\sigma^2 & 7.3312\epsilon_\sigma^3
\end{array}
  \right) \, .
\end{equation}
Here $\theta_{ij}$ are the random phases chosen uniformly in $[0,2\pi]$.
{For the absolute values of the Yukawa couplings $y_{ij}$, we use two different distributions. The first one is the Rayleigh distribution (with average value $1.5$). This corresponds to Gaussian distributions for the real and imaginary parts of the couplings. The second distribution is uniform in $[2,3.5]$.} Finally, $\epsilon_\sigma$, $\epsilon_\chi$ are numbers smaller than unity controlled by the VEVs of flavon fields $\sigma$ and $\chi$ explaining  the hierarchy in Yukawa couplings (in Froggatt-Nielsen models they are ratios between the flavon VEVS and the scale $\Lambda$ of exotic vector-like  fermions, $\epsilon_\sigma=\langle \sigma \rangle / \Lambda_\sigma$, $\epsilon_\chi=\langle \chi \rangle / \Lambda_\chi$). We report the result of this sampling on figure \ref{fig:histRandYukawasLept} where the result for the non-flat distribution is shown on the left panel and the result for the uniform distribution is on the right. Roughly 10\% (65\%) of parameter space give a big enough baryon asymmetry when using the non-flat distribution (uniform distribution in $[2,3.5]$).

%

\subsection{Baryon asymmetry with varying Up and Charm quarks}
\begin{figure}[t]
\centering
\includegraphics[width=0.45\textwidth]{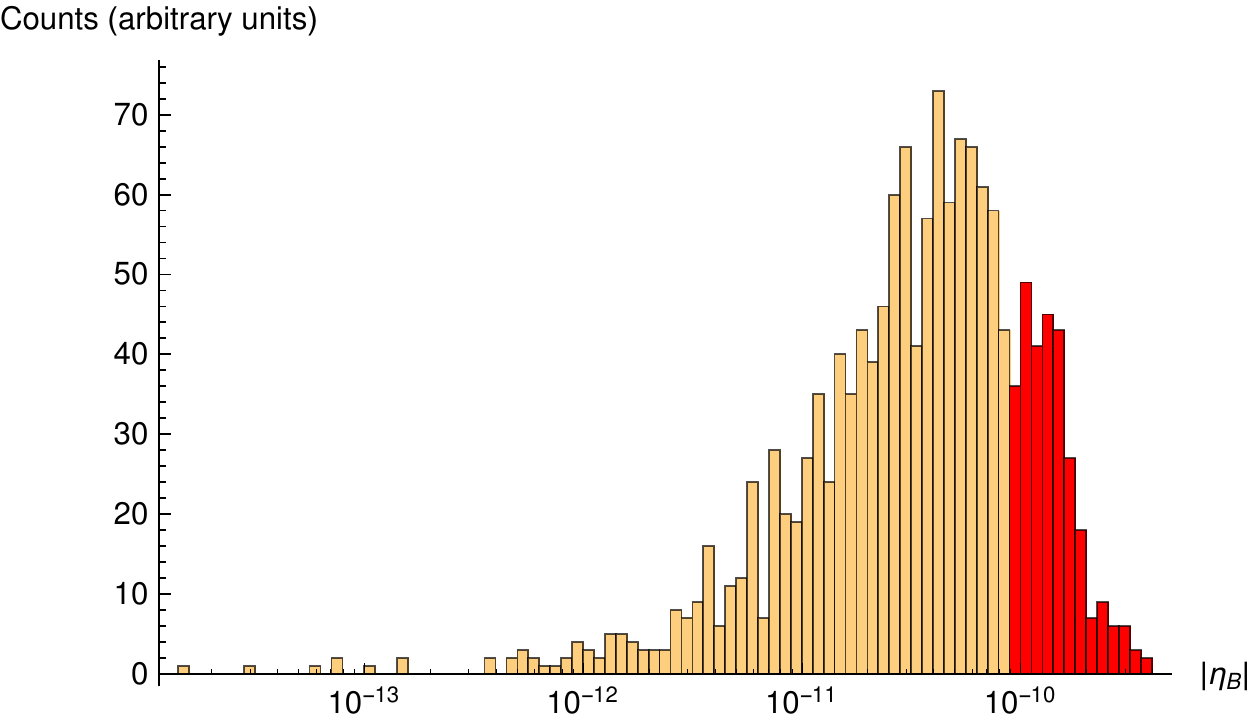}
\includegraphics[width=0.45\textwidth]{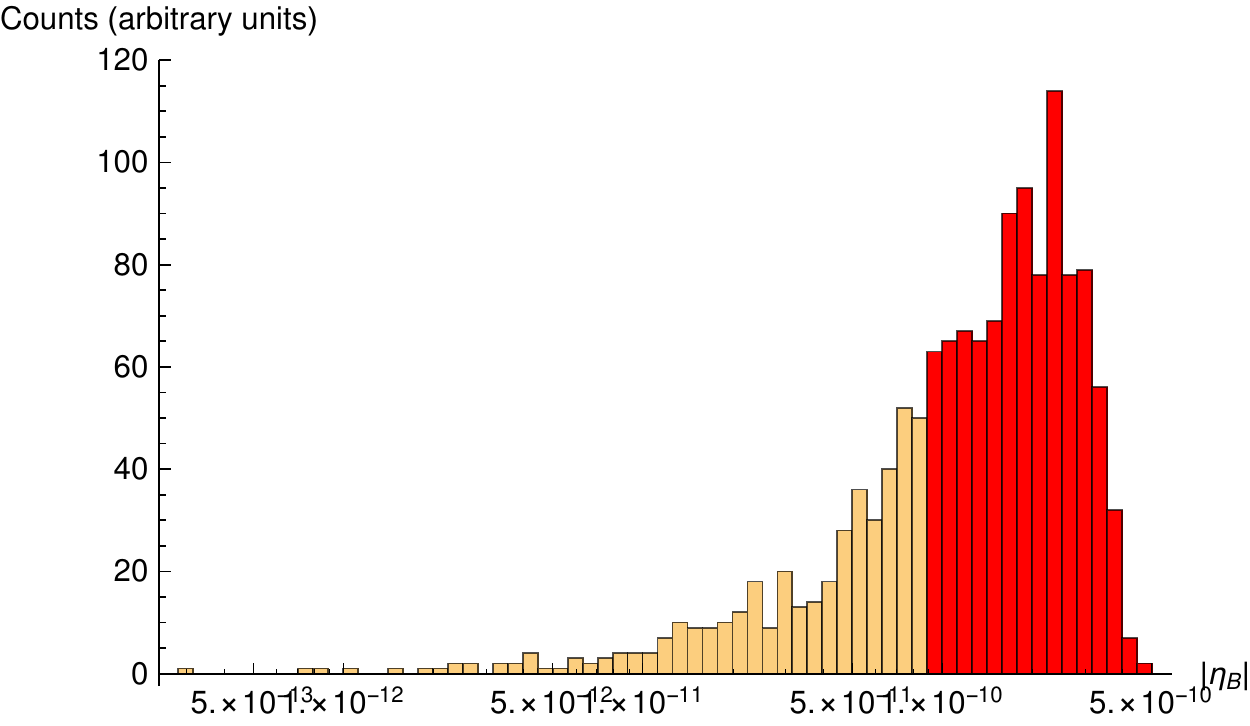}
\caption{\label{fig:histRandYukawasUC} \em The baryon asymmetry calculated in the up-charm system for random values of the order one couplings and the phases. In red is the fraction of the results that produced a larger baryon asymmetry than the one measured. The left hand plot shows the result for $y_{ij}$ chosen from the {Rayleigh distribution} whereas on the right hand side we show the result for a uniform distribution of $y_{ij}$ in $[2,3.5]$. The wall parameters where chosen to be $L_w=3/T$, $v_w=0.1$ and $\phi/T=1.3$.}
\end{figure}

We can perform a similar study to the one we performed for the leptons, using the up-charm system. In this case we use same system as the one discussed for the top-charm system where we adjust the exponents to match the FN-charges of the up-charm system:

\begin{equation}
  Y_{u-c}=\left(\begin{array}{cc}
e^{i\theta_{uu}}y_{uu}\epsilon_\chi^7+\tilde{y}_{uu}\epsilon_\sigma^7 &e^{i\theta_{uc}} {y}_{uc}\epsilon_\chi^4 +\tilde{y}_{uc}\epsilon_\sigma^4\\
e^{i\theta_{cu}}{y}_{cu}\epsilon_\chi^6+\tilde{y}_{cu}\epsilon_\sigma^6 &e^{i\theta_{cc}} {y}_{cc}\epsilon_\chi^3+\tilde{y}_{cc}\epsilon_\sigma^3
\end{array}
  \right)
\end{equation}

where $\theta_{ij}$ are the complex phases and ${y}_{ij}$ and $\tilde{y}_{ij}$ are order one real couplings. The same distributions have been chosen for the phases and the couplings as in the previous example. The result is also presented on a histogram (figure \ref{fig:histRandYukawasUC}). We observe that roughly 20\% (70\%) of parameter space can have successful baryogenesis when the couplings follow the non-flat (uniform) distribution.

\subsection{Full system}
\label{sec:fullsystem}
\begin{figure}[t]
\centering
\includegraphics[width=0.55\textwidth]{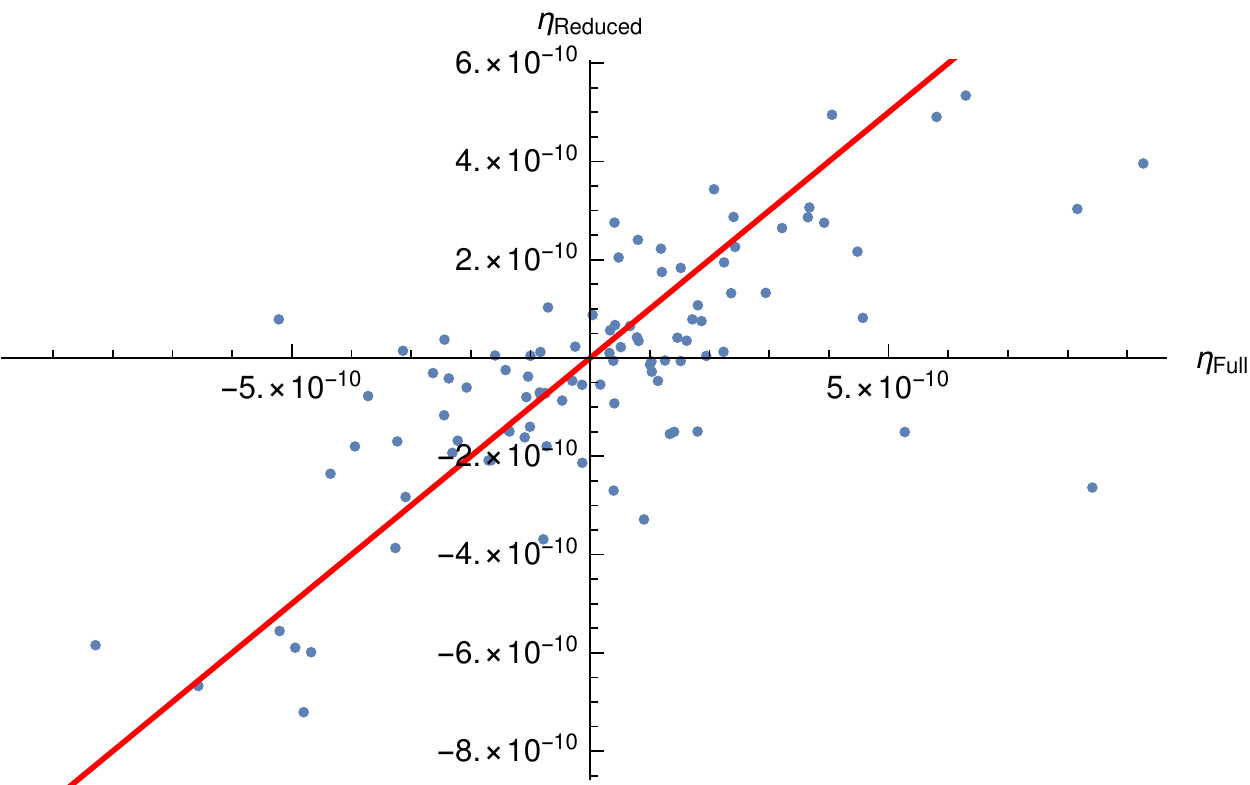}
\caption{\label{fig:FullVsCT} \em A direct comparison between the baryon asymmetries obtained using the full and the $c-t$ systems respectively. 
The red line correspond to equal results  for the full and the $c-t$ systems, $\eta_{Full}=\eta_{Reduced}$. The wall parameters where chosen to be $L_w=6.5/T$, $v_w=0.1$ and $\phi/T=1.3$.
}
\end{figure}

For completeness, we performed a numerical study of the full system including all 6 quarks and
compared with the analysis concentrating on the top-charm system where we neglected the other quarks. 
The size of the system quickly grows with the number ($N$) of quarks that we take into account ($2N\times 2N$) and the computational cost is much higher. In order to have a direct comparison between the top-charm system and the full system we chose to work in a Froggatt-Nielsen model with the following Yukawa couplings
\begin{eqnarray}
  Y_{u}&=\left(\begin{array}{ccc}
  e^{i\theta_{uu}}y_{uu}\left(\epsilon_\chi^7+\epsilon_\sigma^7\right) &e^{i\theta_{uc}}y_{uc}\left(\epsilon_\chi^4+\epsilon_\sigma^4\right) &e^{i\theta_{ut}} {y}_{ut}\left(\epsilon_\chi^3+\epsilon_\sigma^3\right)\\
e^{i\theta_{cu}}y_{cu}\left(\epsilon_\chi^6+\epsilon_\sigma^6\right) &e^{i\theta_{cc}}y_{cc}\left(\epsilon_\chi^3+\epsilon_\sigma^3\right) &e^{i\theta_{ct}} {y}_{ct}\left(\epsilon_\chi^2+\epsilon_\sigma^2\right)\\
e^{i\theta_{tu}}y_{tu}\left(\epsilon_\chi^4+\epsilon_\sigma^4\right) &e^{i\theta_{tc}}{y}_{tc}\left(\epsilon_\chi^1+\epsilon_\sigma^1\right) &e^{i\theta_{tt}} {y}_{tt}\left(1\right)
\end{array}
  \right) \\
  Y_{d}&=\left(\begin{array}{ccc}
  e^{i\theta_{dd}}y_{dd}\left(\epsilon_\chi^6+\epsilon_\sigma^6\right) &e^{i\theta_{ds}}y_{ds}\left(\epsilon_\chi^5+\epsilon_\sigma^5\right) &e^{i\theta_{db}} {y}_{db}\left(\epsilon_\chi^5+\epsilon_\sigma^5\right)\\
e^{i\theta_{sd}}y_{sd}\left(\epsilon_\chi^5+\epsilon_\sigma^5\right) &e^{i\theta_{ss}}y_{ss}\left(\epsilon_\chi^4+\epsilon_\sigma^4\right) &e^{i\theta_{sb}} {y}_{sb}\left(\epsilon_\chi^4+\epsilon_\sigma^4\right)\\
e^{i\theta_{bd}}y_{bd}\left(\epsilon_\chi^3+\epsilon_\sigma^3\right) &e^{i\theta_{bs}}{y}_{bs}\left(\epsilon_\chi^2+\epsilon_\sigma^2\right) &e^{i\theta_{bb}} {y}_{bb}\left(\epsilon_\chi^2+\epsilon_\sigma^2\right)
\end{array}
  \right)
\end{eqnarray}
for the full system and the $c-t$ submatrix of $Y_u$ for the $c-t$ system. This allows for a convenient comparison between the two systems. We still expect changes by a factor of $\mathcal{O}(1)$ because the full system also contains the down-type quarks which can give a non-negligible contribution. The result is shown on figure \ref{fig:FullVsCT} where we see that the ballpark of sample points are reasonably well described using only the $c-t$ system. Hence we can gain a lot of computing time when focussing on this sub-system. On the other hand, there are some points that are badly approximated using the small system only. In particular, the two systems can give different signs for the baryon asymmetry. Therefore,  in a very specific model, the full system should be used to calculate reliably the baryon asymmetry. 
Note that these differences come from the fact that we have CP violating sources in all families. In what has been done so far in the literature, CP violation was in the top sector only. In this case, the reduced system provides reliable results.

When concentrating on a sub-system with only two quarks we do not have to reconstruct the CKM-matrix in the broken phase as we can still use the remaining four quarks to get the correct CKM-matrix. In contrast, for the full system, we should choose our randomly sampled parameters in such a way that we correctly reproduce the CKM-matrix in the broken phase. Here, the purpose of the exercise is to show the validity and limits of the two-quark system and not to make a precise computation of the baryon asymmetry in a Froggatt-Nielsen model. Hence, we do not restrict the parameters to the hypersurface where the CKM-matrix is correctly recovered.

\section{Low-scale flavour models}\label{sec:flavour_models}

So far we have shown how a varying CKM matrix during the electroweak phase transition can lead to large sources of CP violation sufficient to produce the baryon asymmetry of the universe during a first-order electroweak phase transition, provided that Yukawa couplings are of order unity in the symmetric phase. 
To show this, we have relied on some general parametrisation of the Yukawa matrix.
In this section, we want to present some specific examples how this can happen in some benchmark flavour models.
The question of the compatibility with experimental bounds on the flavon fields is the topic of separate works, e.g. \cite{Baldes:2016gaf}.

We want to study different low-scale flavour models that are known to generate a strong first order phase transition and that induce varying Yukawa couplings. In models where the dynamics of the Yukawa coupling is linked to the mechanism of electroweak symmetry breaking,  the CP-violating source term discussed in chapter \ref{sec:kinetic_equations} are localized at or close to the bubble wall. This guarantees that the chiral asymmetry induced by the CP violating sources can be efficiently transformed into a baryon asymmetry by the weak sphalerons in front of the bubble wall and subsequently diffuse into the broken phase where the baryon asymmetry is frozen. For  the models studied here we calculate their kernels, their sources and the resulting baryon asymmetry generated for a natural choice of parameters. We could in principle calculate the width and velocity of the bubble wall, this is however beyond the scope of this paper and not necessary for what we want to stress, so we just take those two quantities as input parameters in our simulations (we show however the dependence of the results on these parameters in figure~\ref{fig:top_eta_vs_v_and_l}). {The strength of the phase transition, $\phi/T$ is in many cases known. For instance it was stressed in~\cite{Baldes:2016gaf} and~\cite{Baldes:2016rqn}  that Yukawa variation during EW symmetry breaking  can provide a strongly first order PT. An explicit example is given in Fig.~\ref{fig:FNpotential}  where the field trajectory in the Higgs-flavon plane during tunnelling clearly indicates that Yukawa couplings are varying at the same time as the Higgs VEV. We used in Sections \ref{subsec:FN} and \ref{subsec:leptons} the corresponding value of $\phi/T$.
 Other explicit examples are the Randall-Sundrum model studied in \cite{vonHarling:2016vhf} and composite Higgs models, presently under investigation (results suggest that a strength of the phase transition $\phi /T=2.5$ can be achieved)}

\subsection{Froggatt-Nielsen models}
\label{subsec:FN}
In Ref.~\cite{Baldes:2016gaf} the cosmology of Froggatt-Nielsen (FN) models was studied. In particular it was shown that models with a Higgs-Flavon mixing generally yield a strong first order phase transition. 
On the other hand, if we want in addition that the flavon VEV changes at the same time as the Higgs VEV for CP violation, this requires the flavon to be quite light, typically lighter than the Higgs, which is in tension with experimental bounds from flavour constraints \cite{Baldes:2016gaf,Bauer:2016rxs}.
Despite this, we present these models here for the purpose of illustration of how the Yukawa variation can arise from the flavon-Higgs dynamical interplay\footnote{A former related discussion was offered in \cite{Berkooz:2004kx}, where on the other hand, CP-violating and diffusion processes were not computed. The minimal model proposed in \cite{Berkooz:2004kx} was re-examined and ruled out in \cite{Baldes:2016gaf} which considered simple extensions, in particular added a second flavon $\chi$.} and then move to the more realistic Randall-Sundrum model.
The point is to expose the procedure to follow for any given model. One first need to determine the field path in the multi-field (at least two) scalar space during the phase transition, say $\sigma(\phi)$ where $\sigma$ is the new scalar  field controlling the  size of the Yukawa coupling,  which then is used to 
estimate the effective variation of the Yukawa coupling as a function of the Higgs field across the bubble wall:
\begin{itemize}
\item 1) Determine the field trajectory $\sigma(\phi)$ during the phase transition
\item 2) Plug $\sigma (\phi)$ into $Y(\sigma)$ to get $Y(\phi)$.
\end{itemize}

\begin{figure}[t]
\centering
\includegraphics[width=0.5\textwidth]{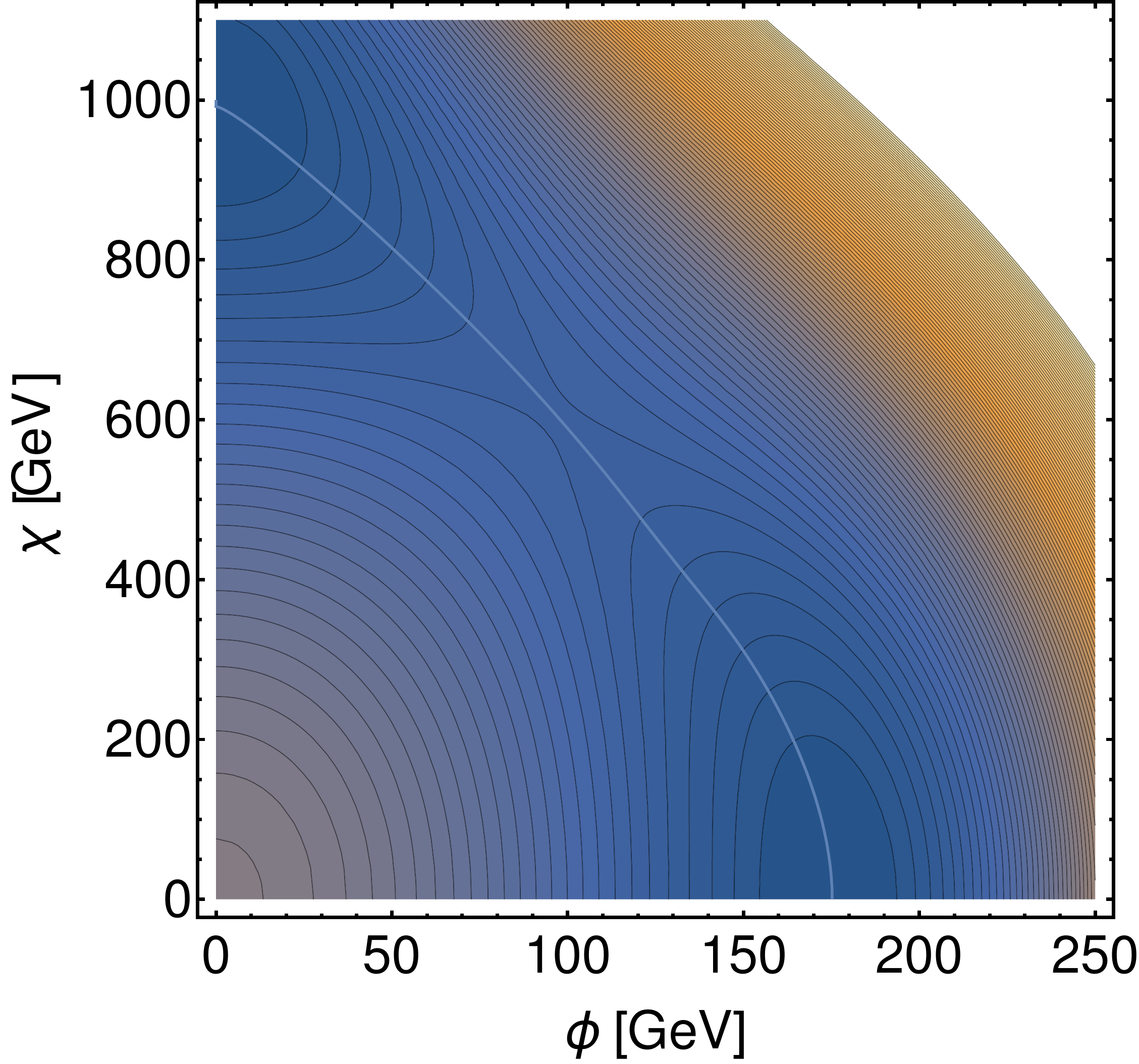}
\caption{\label{fig:FNpotential} \em
The two-field scalar potential of model B2 in ~\cite{Baldes:2016gaf} at the critical temperature. We show the approximate path of the phase transition in light blue. The $\chi$ self-coupling and the $\chi$-Higgs coupling are $\lambda_\chi=10^{-4}$ and $\lambda_{\chi\phi}=10^{-2}$ respectively.
}
\end{figure}

In the model B-2 of~\cite{Baldes:2016gaf}, two FN fields ($\sigma$ and $\chi$) with FN-charge $Q_{FN}(\sigma)=Q_{FN}(\chi)=-1$ are introduced. A typical FN-charge assignment for the SM fermions would be:
\begin{center}
\begin{equation}\label{eqn:FNCharges}
\begin{tabular}{lll}
$\bar{Q}_3 ~(0)$ & $\bar{Q}_2 ~(+2)$ & $\bar{Q}_1 ~(+3)$ \\
$U_3 ~(0)$ & $U_2 ~(+1)$ & $U_1 ~(+4)$ \\
$D_3 ~(+2)$ & $D_2 ~(+2)$ & $D_1 ~(+3)$
\end{tabular}
\end{equation}
\end{center}
Once the FN-symmetry $U(1)_{FN}$ is spontaneously broken the Yukawa interactions take the following form:
\begin{equation}
  \lag\supset \tilde{y}_{ij}\left(\frac{\langle\chi\rangle}{\Lambda_\chi}\right)^{\tilde{n}_{ij}}\bar{Q}_i\tilde{\phi} U_j+y_{ij}\left(\frac{\langle\chi\rangle}{\Lambda_\chi}\right)^{n_{ij}}\bar{Q}_i\phi D_j+\tilde{Y}_{ij}\left(\frac{\langle\sigma\rangle}{\Lambda_\sigma}\right)^{\tilde{n}_{ij}}\bar{Q}_i\tilde{\phi} U_j+Y_{ij}\left(\frac{\langle\sigma\rangle}{\Lambda_\sigma}\right)^{n_{ij}}\bar{Q}_i\phi D_j
\end{equation}
where $y_{ij}$, $\tilde{y}_{ij}$, $Y_{ij}$ and $\tilde{Y}_{ij}$ are (in principle complex) coupling constants that are all expected to be of order $|y_{ij}|\sim|Y_{ij}|\sim\mathcal{O}(1)$. $\phi$ is the SM Higgs and $\tilde{\phi}=i\sigma_2\phi^*$.  $\Lambda_\chi$ and $\Lambda_\sigma$ are associated with the mass scales of the vector-like FN fermions which are integrated out. The integers $n_{ij}$ and $\tilde{n}_{ij}$ are chosen to form singlets under $U(1)_{FN}$. Here the FN-symmetry is broken before the breaking of the electro-weak symmetry. During the electro-weak phase transition the VEVs of the FN-fields and the Higgs evolve roughly like:
\begin{equation}
  \phi: 0\rightarrow v_\phi \quad \sigma:\Lambda_\sigma/5\rightarrow\Lambda_\sigma/5 \quad  \chi:\Lambda_\chi\rightarrow 0
\end{equation}
where the dependence of the VEV of $\chi$ as a function of the VEV of $\phi$ can be found by following the minimal path in the $\chi-\phi$ potential at the critical temperature, as shown in figure~\ref{fig:FNpotential}.

In what follows we will concentrate on the Top-Charm system, where the CP-violation is expected to be biggest, and neglect the remaining quarks as sources of CP-violation. The Yukawa-matrix for the Top-Charm system is given by
\begin{equation}
  Y_{t-c}=\left(\begin{array}{cc}
\tilde{y}_{cc}\epsilon_\chi^3 & \tilde{y}_{ct}\epsilon_\chi^2 \\
\tilde{y}_{tc}\epsilon_\chi^1 & \tilde{y}_{tt}\epsilon_\chi^0
\end{array}
  \right)
  +
  \left(\begin{array}{cc}
\tilde{Y}_{cc}\epsilon_\sigma^3 & \tilde{Y}_{ct}\epsilon_\sigma^2 \\
\tilde{Y}_{tc}\epsilon_\sigma^1 & \tilde{Y}_{tt}\epsilon_\sigma^0
\end{array}
  \right),
\end{equation}
where $\epsilon_\chi=\langle \chi \rangle / \Lambda_\chi$ varies through the bubble wall, and $\epsilon_\sigma=\langle \sigma \rangle / \Lambda_\sigma\approx 0.2$ is constant during the EW phase transition. 
In this model, the flavon $\sigma$ which controls the size of Yukawa couplings today is  already settled during the EW phase transition. However, there is another scalar $\chi$ which evolves during the EW phase transition and is responsible for the variation of the Yukawa couplings, but the effect of  its VEV is negligible today.

\begin{figure}[t]
\centering
\includegraphics[width=0.45\textwidth]{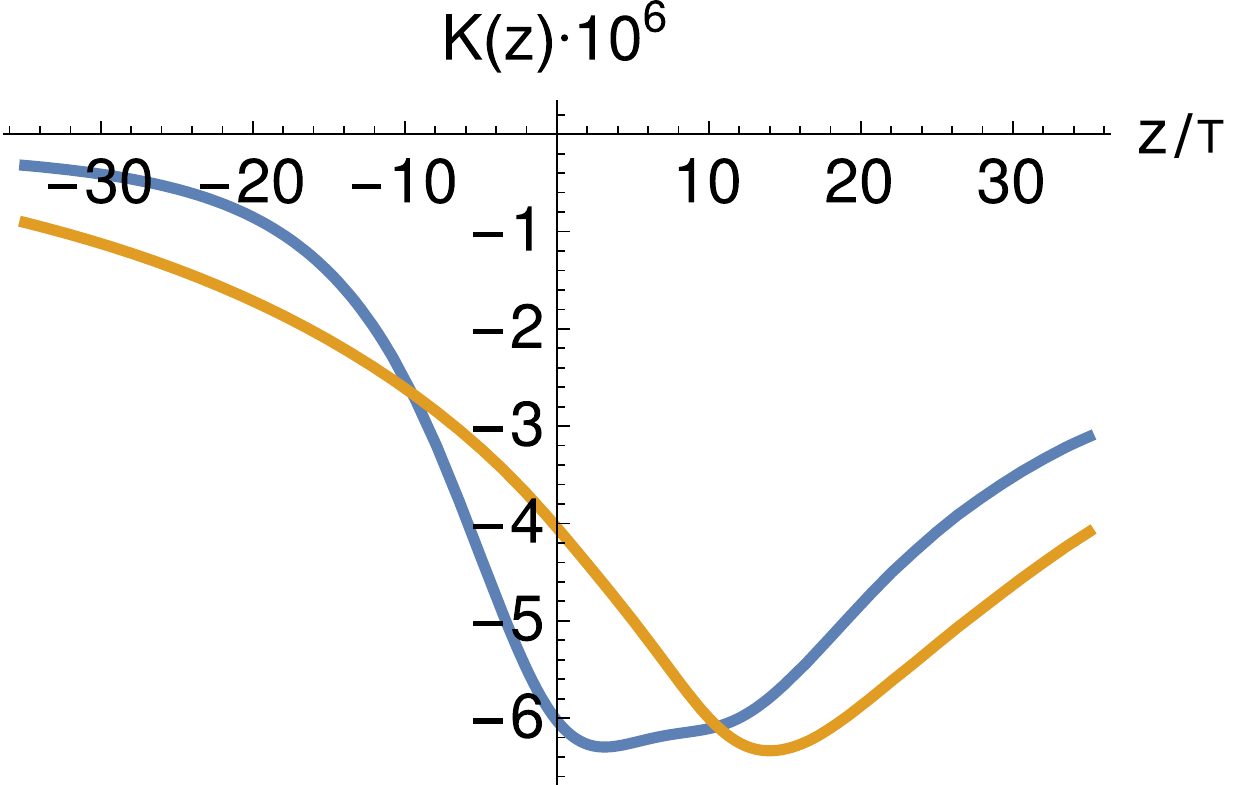}
\includegraphics[width=0.45\textwidth]{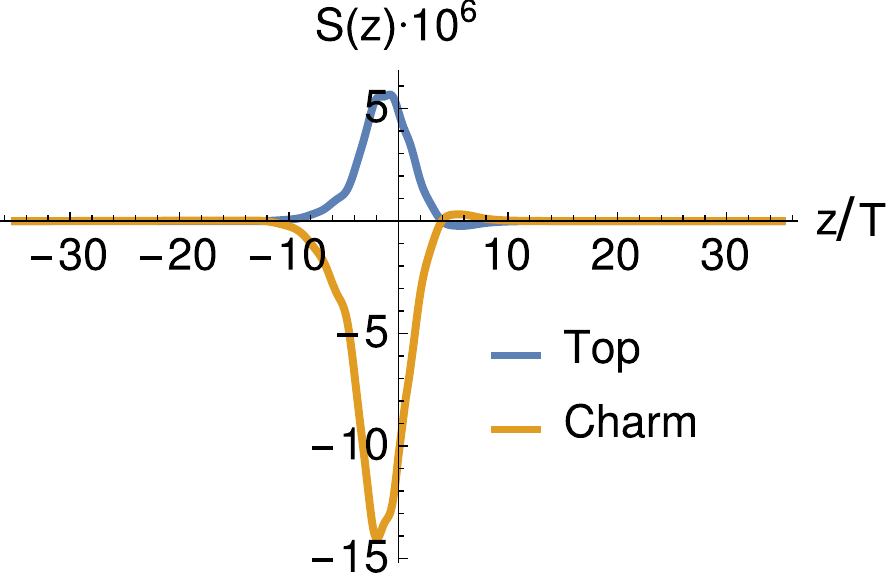}
\caption{\label{fig:FNKernelAndSource} \em
The kernel and the source for the Froggatt-Nielsen model discussed in this section. Here we chose $\theta=1$ in equation (\ref{eqn:FNBenchmark}), $L_w=6.5/T$ and $v_w=0.1$. The resulting baryon-asymmetry is $\eta_B\approx 11.46\cdot 10^{-11}$.
}
\end{figure}

Using equation (\ref{eq:diffusion_0}) with
\begin{equation}
  m=\frac{v_\phi(z)}{\sqrt{2}}Y_{t-c} \quad \text{\&} \quad v_\phi(z)=\frac{v_\phi(-\infty)}{2}\left(1-\text{Tanh}\left[\frac{z}{L_w}\right]\right)
\end{equation}
we can easily determine the CP-violating source term and compute the kernel for any given set of constants $\tilde{y}_{ij}$ and $\tilde{Y}_{ij}$. Figure~\ref{fig:FNKernelAndSource} shows the kernel and the source for our benchmark-point:
\begin{equation}\label{eqn:FNBenchmark}
  \tilde{y}_{i\neq j}=\tilde{Y}_{i\neq j}=1 \qquad \tilde{y}_{tt}=\tilde{Y}_{tt}=1/2 \qquad \tilde{y}_{cc}=\tilde{Y}_{cc}=e^{i\theta}\, .
\end{equation}
Note that this particular set of parameters seems quite unnatural as the constant and the varying contributions to the Yukawa coupling share common complex phases. There is a-priori no reason why this should happen. We made this choice to show that the mechanism works even in this case. In a more natural situation, we expect an enhanced baryon asymmetry as then $\Tr\left[V^\dagger (m^\dagger)''mV\right]\neq 0$ and the cancelation discussed in section~\ref{sec:generalParamTwoFlavour} is milder.
For more details on the model and the experimental constraints see \cite{Baldes:2016gaf}.

\subsection{Froggatt-Nielsen with leptons}
\label{subsec:leptons}
The results of the previous sections suggest that this mechanism may also work if the flavour model explains the flavour structure in the lepton sector only. To study this in more detail, we consider a Froggatt-Nielsen model for leptons, of the B2 type in ~\cite{Baldes:2016gaf} with the charge assignment of~\cite{Huitu:2016pwk}. We do not introduce right-handed neutrinos. The Yukawa matrix considered here is:
\begin{equation}
  Y_{\tau-\mu}=\left(\begin{array}{cc}
3.28e^{i\cdot1.33}\epsilon_\chi^4 &  2.34e^{i\cdot1.24}\epsilon_\chi^5 \\
1.67e^{i\cdot0.26}\epsilon_\chi^2 &  2.89e^{i\cdot2.53}\epsilon_\chi^3
\end{array}
  \right)
  +
  \left(\begin{array}{cc}
6.1465\epsilon_\sigma^4 & -3.125\epsilon_\sigma^5 \\
0.5\epsilon_\sigma^2 & 7.3312\epsilon_\sigma^3
\end{array}
  \right),
\end{equation}
where in this case the ratio between the flavon VEV and the FN fermion mass scale is $\epsilon_\sigma\approx 0.1$. 
Note that here, since the varying and the constant contributions to the Yukawas do not share common complex phases,  the phase transition will effectively induce a varying complex phase in the yukawas. Hence the cancelation due to the tracelessness of the source does not apply and we can gain a small factor in the final baryon asymmetry. To find $\epsilon_\chi$ we first have to determine the potential $V(\chi,\phi)$ and then follow its minimal path like in the case of quarks. For this estimate we took for the $\chi$-self coupling and the $\chi$-Higgs coupling the same values as in the  quark case  namely $\lambda_\chi=10^{-4}$ and $\lambda_{\chi\phi}=10^{-2}$. As a result, the phase transition is not  as strong as in the quark case (i.e. $\phi/T\sim 1$). Once again, our purpose is not to make a precise analysis of the phase transition in a specific model but  to illustrate the origin of the CP violation and show that it can be sufficient for electroweak baryogenesis.

\begin{figure}[t]
\centering
\includegraphics[width=0.45\textwidth]{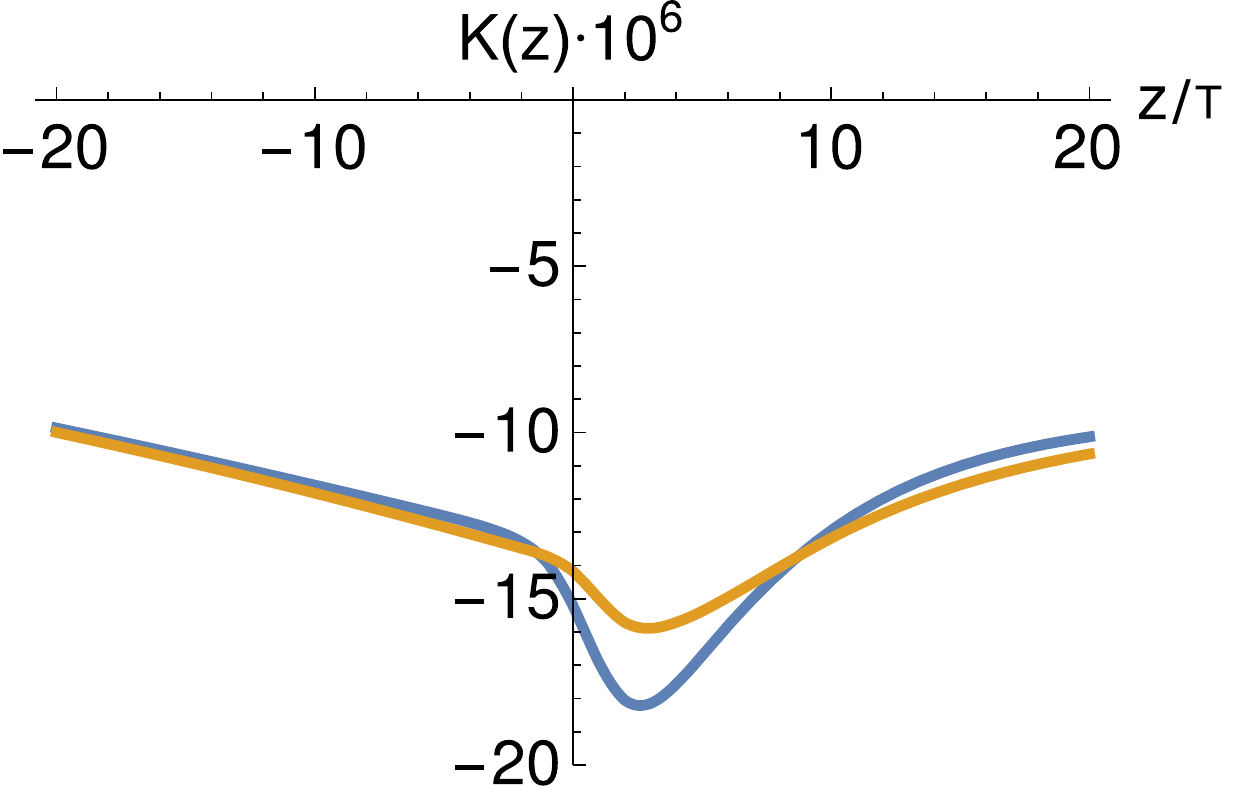}
\includegraphics[width=0.45\textwidth]{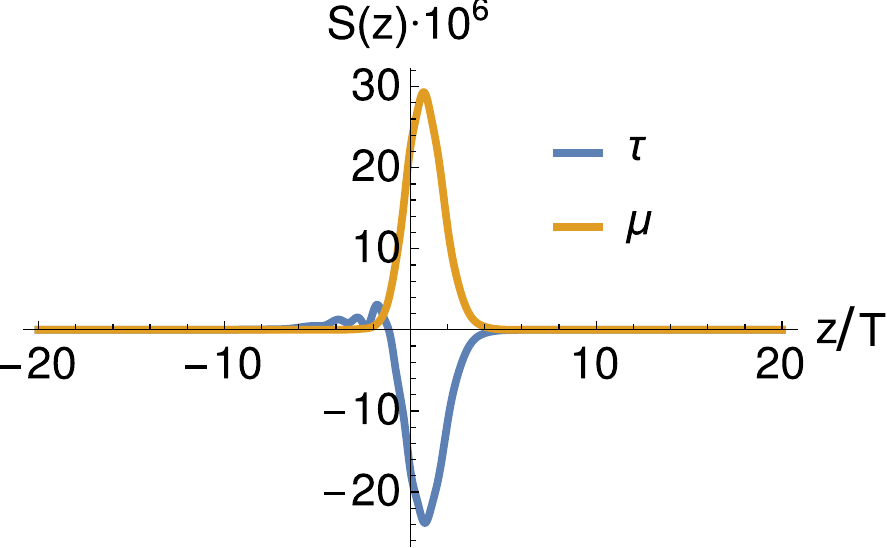}
\caption{\label{fig:FNKernelAndSourceLeptons} \em
The kernel and the source for the Froggatt-Nielsen model with leptons. The parameters for this evaluation where chosen as $L_w=3/T$ and $v_w=0.1$. The resulting baryon-asymmetry is $|\eta_B|\approx 1.1\cdot 10^{-10}$.
}
\end{figure}

The crucial difference with respect to the quark case is that leptons have a vanishing strong sphaleron interaction rate. As  seen in section~\ref{sec:InteractionRates}, switching off the strong sphaleron increases the baryon yield. This effect can counter balance  the suppression from the reduced number of degrees of freedom (by a factor of 3) compared to coloured particles and the fact that  all lepton Yukawa couplings  are very small in the broken phase.
As shown in figure \ref{fig:FNKernelAndSourceLeptons}, all suppressions for the leptons can be overcome and a large enough  baryon-asymmetry can be produced with our benchmark-point.

\subsection{Randall-Sundrum models}
Another and attractive possibility for coupling flavour physics to the Higgs sector comes from Randall-Sundrum models with a warped extra dimension. The model is a slice of AdS$_5$ geometry. The Higgs is localized on the IR brane whereas the SM fermions  propagate in the bulk. The 4-dimensional effective Yukawa interactions are determined by the wave function overlap of the fermions in the fifth dimension and the Higgs on the IR brane. Such a model can, in addition to solving the hierarchy problem, also address the flavour structure of the SM. If the wave-function of a fermion is peaked towards the UV brane, the corresponding 4D Yukawa is suppressed whereas if the fermion is rather peaked towards the IR brane, the effective Yukawa coupling is of order 1, like for the top quark.

The electroweak phase transition in these models has been studied  in~\cite{Creminelli:2001th,Randall:2006py,Nardini:2007me,Hassanain:2007js, Konstandin:2010cd,Konstandin:2011dr,Bunk:2017fic} and recent work~\cite{vonHarling:2016vhf} suggests that the dynamics of the Yukawas in the context of a first-order EW phase transition can be naturally implemented within this framework without conflicts with experimental bounds. In this case, the radion field which controls the interbrane distance plays the role of the flavon.

In the five-dimensional description, the symmetric phase is given by an AdS-Schwarzschild (AdS-S) solution with just a UV-brane where the graviton is peaked and no IR-brane is present whereas the broken phase is the usual RS1 model, i.e. a fifth dimension with an AdS$_5$ geometry bounded by  the UV- and the IR-branes. The phase transition happens via ``brane nucleation". During the phase transition the IR brane emerges from $y\to \infty$ and approaches towards  the UV brane until it is stabilized by the Goldberger-Wise  mechanism. In this context, we would therefore expect the Yukawa coupling of the fermions that are peaked close to the IR brane (e.g. the top quark) to increase as we go from the broken towards the symmetric phase whereas the 4D Yukawas of the light fermions, which are peaked at the UV brane, should decrease. This is clearly not what is needed for EW baryogenesis  where  Yukawas should be of order one in the symmetric phase (see section~\ref{sec:varyTopYukawaOneFlavour}). The solution studied in Model II of ~\cite{vonHarling:2016vhf} arises  from taking into account  the Yukawa coupling between SM fermions and  the Goldberger-Wise scalar. This causes the decreasing profiles of the fermions along the extra dimension to ``turn around" at some point and start to increase. This guarantees that the effective Yukawas grow as we go through the bubble wall towards the symmetric phase.
In appendix~\ref{sec:RSYukawas} we give the explicit formulae used for the Yukawa couplings.
More details can be found in ~\cite{vonHarling:2016vhf}. 
\begin{figure}[t]
\centering
\includegraphics[width=0.45\textwidth]{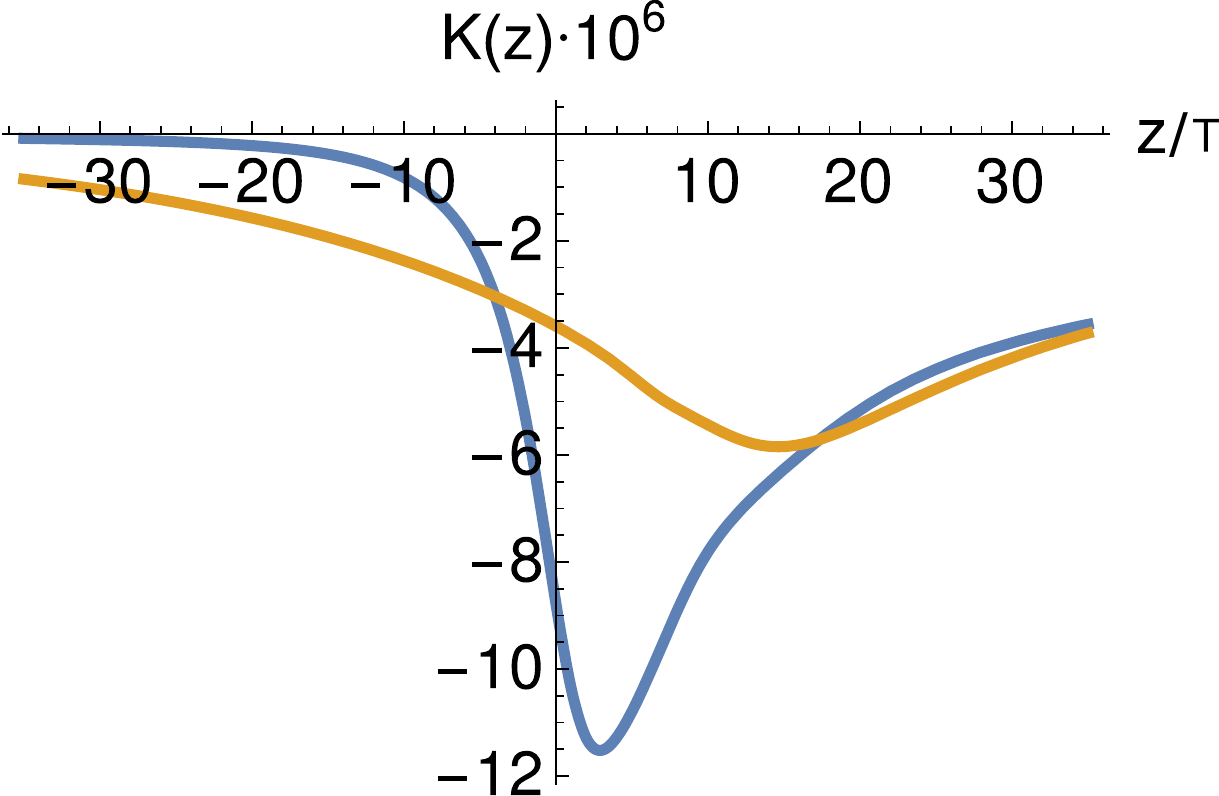}
\includegraphics[width=0.45\textwidth]{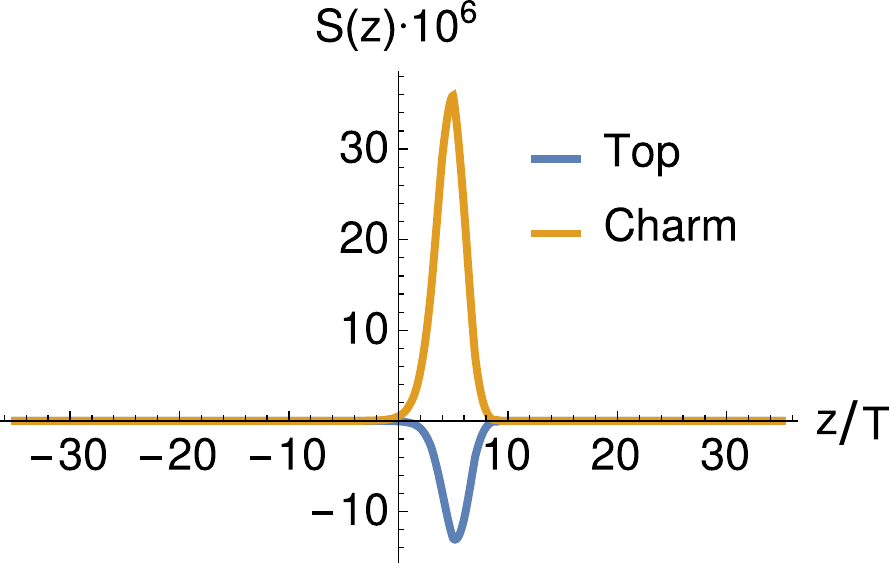}
\caption{\label{fig:RSKernelAndSource} \em
The kernel and the source in the Randall-Sundrum model II described in ~\cite{vonHarling:2016vhf}. The parameters for this evaluation where chosen as $L_w=6.5/T$, $v_w=0.1$ and $\phi/T=2.5$. The resulting baryon-asymmetry is $\eta_B\approx 9.89\cdot 10^{-11}$.
}
\end{figure}
On figure~\ref{fig:RSKernelAndSource} we show the kernels and the sources for a given point in parameter-space. This analysis shows  that the measured baryon asymmetry can be generated with a natural choice of parameters in the framework of Randall-Sundrum models.

\begin{figure}[t]
\centering
\includegraphics[width=0.45\textwidth]{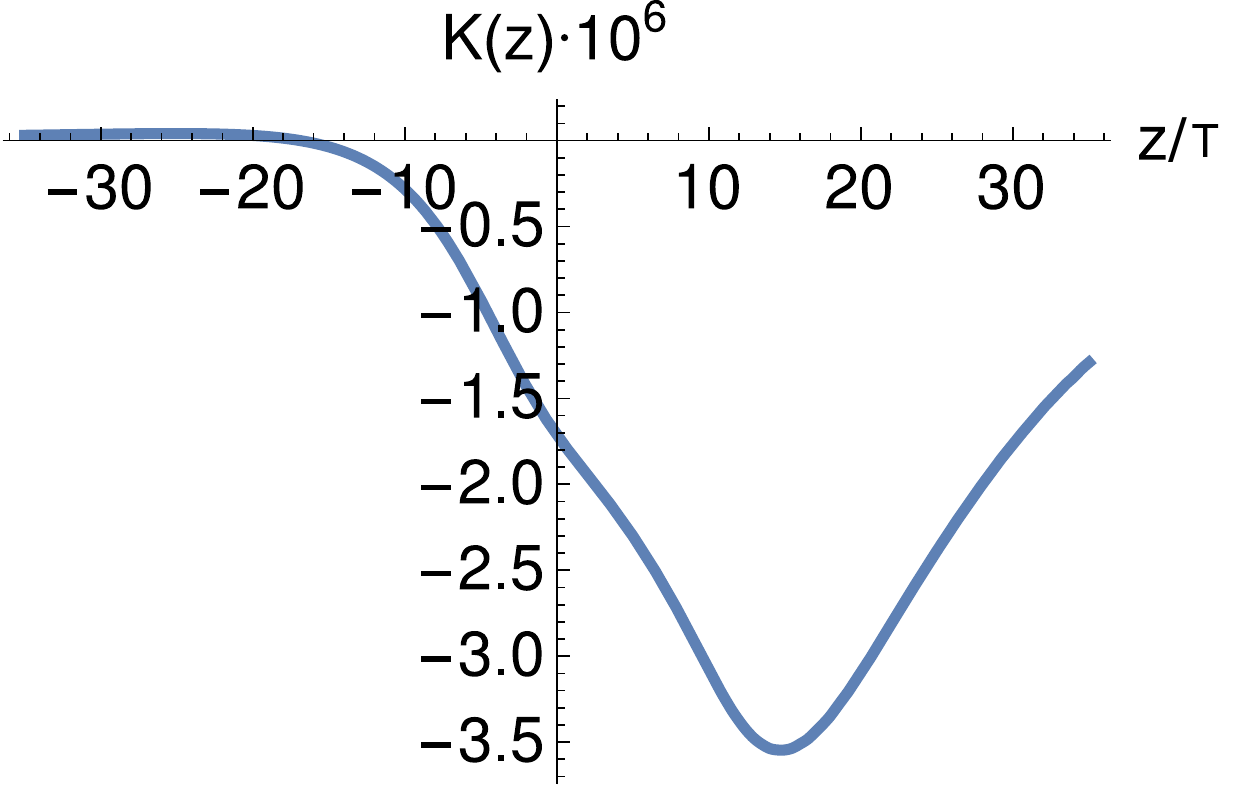}
\includegraphics[width=0.45\textwidth]{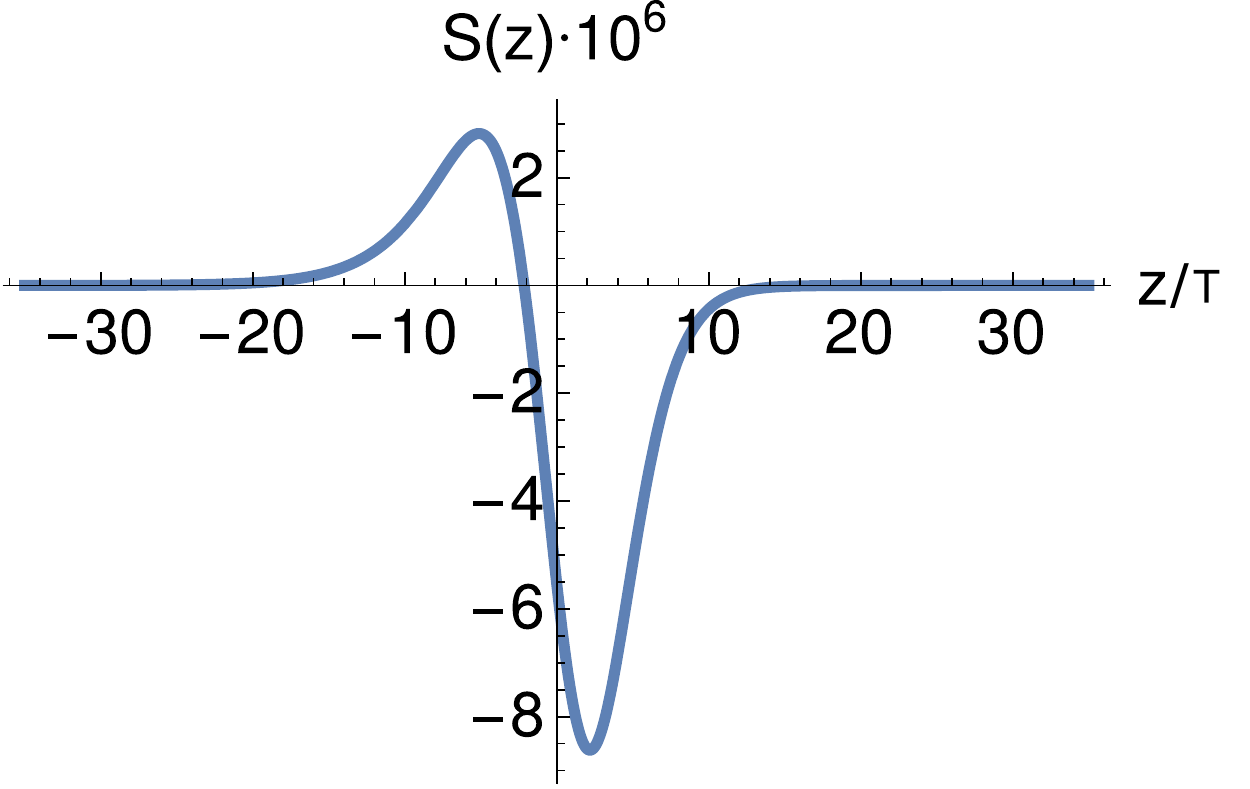}
\caption{\label{fig:RSOneFlavourKernelAndSource} \em
The kernel and the source in the Randall-Sundrum model I described in ~\cite{vonHarling:2016vhf} with only the top quark Yukawa varying. The parameters were chosen as $L_w=6/T$, $v_w=0.1$ and $\phi/T=2.5$. The resulting baryon-asymmetry is $\eta_B\approx 9.93\cdot 10^{-11}$.
}
\end{figure}

In ~\cite{vonHarling:2016vhf} the authors also point out a mechanism with which to generate varying Yukawas in the top sector alone (denoted Model I). In this case the top-Yukawa is given by:
\begin{equation}
	y_t(\sigma_{IR})\equiv \tilde{\lambda}_t k \sqrt{\frac{1-2c_Q}{1-\sigma_{IR}^{1-2c_Q}}}\sqrt{\frac{1-2c_t}{1-\sigma_{IR}^{1-2c_t}}}\, ,
\end{equation}
where
\begin{equation}
	\tilde{\lambda}_t(\sigma_{IR})\simeq \left[\lambda_t + 4~\kappa_t ~k~ v_ {IR}\left[1-\left(1+\sqrt{\frac{\epsilon}{4}}\right)\left(\frac{\sigma_{IR}}{\sigma_{IR}^{min}}\right)^\epsilon\right]\right]
\end{equation}
and $c_Q=0.278$, $c_t=-0.339$ and $\epsilon=1/20$ are constants. The combinations $\lambda_tk=d_\lambda \frac{l_5^{2/3}}{l_4^{1/2}}\frac{k}{M_5}$ and $4~\kappa_t ~k^2~ v_ {IR}=4d_\kappa\frac{l_5^{1/3}}{l_4^{1/2}}\frac{k^2v_{IR}}{M_5^{7/2}}$, with $l_d$ being the $d$-dimensional loop factor and $M_5$ the 5D Plank mass, are constants as well. They can be estimated using dimensional arguments (see~\cite{vonHarling:2016vhf}). For our example we chose $\lambda_tk\simeq 2.167e^{-i\cdot1.31}$ and $4~\kappa_t ~k^2~ v_ {IR}\simeq 5.769e^{-i\cdot \pi/2}$, which corresponds to the example in~\cite{vonHarling:2016vhf} with $d_\kappa=2e^{i\cdot 3\pi/2}$, $\tilde{\lambda}(\sigma_{IR}^{min})=0.929e^{-i\cdot 1.26}$, $k/M_5=1/2$ and $v_{IR}/M_5^{3/2}=1$. Finally $\sigma_{IR}=e^{-ky_{IR}}$ is the warp factor at the IR brane and can be linked to the Higgs VEV via $\phi=\xi\sigma_{IR}/\sigma_{IR}^{min}$ and where we took $\sigma_{IR}^{min}=2.5\cdot10^{-15}$ in agreement with~\cite{vonHarling:2016vhf}. 
With this set of parameters we can calculate the baryon asymmetry. The result is shown on figure~\ref{fig:RSOneFlavourKernelAndSource}.

To conclude this section, we point out that in \cite{vonHarling:2016vhf}, a simple linear relation was assumed between the Higgs VEV and the dilaton VEV. A complete analysis of the field path during tunnelling is being performed in a realistic dual framework of composite Higgs in Ref.~\cite{CompositeEWPT} and confirms that the Higgs and dilaton field can vary together and not subsequently during the phase transition. This is  encouraging and strongly motivates this class of models as natural models featuring varying Yukawas during the electroweak phase transition, in addition to being elegant solutions to the hierarchy problem.

\section{Summary and conclusion}

We  computed the baryon asymmetry generated in the charge transport mechanism of electroweak baryogenesis when the source of CP violation is induced by  the variation of Standard Model Yukawa couplings across the Higgs bubble wall.
We derived in details the kinetic equations and extracted the CP-violating force that arises from varying Yukawas across the bubble wall.

The final baryon asymmetry is given by Eq.~(\ref{eqn:totalBaryonAsymmetry}):
\begin{equation}
\nonumber
  \eta_B=\frac{135~N_c}{4\pi^2 v_w g_*}\int_{-\infty}^{+\infty} \dd z ~ \Gamma_{ws}~\mu_L~ e^{-\frac{3}{2} {\cal A} \frac{1}{v_w}\int_{-\infty}^{z} \dd z_0 \Gamma_{ws}}\, ,
\end{equation}
Its evaluation requires to know the profile along $z$ of  the chiral asymmetry $\mu_L$, which is obtained by solving the diffusion system Eq.~(\ref{eqn:diffusionSystemMatrix}):
\begin{equation}
\nonumber
 A(z)\cdot r'(z) +B(z)\cdot r(z) =\bar {\cal S}(z)
\end{equation}
that is a system of equations of the type (\ref{eq:diffusion_0}) for each fermionic flavour characterised by a chemical potential $\mu$ and velocity $u$:
\begin{equation}
\nonumber
	\begin{split}
 v_w \, K_1 \, \mu' + v_w (m^2)' \, K_2 \, \mu 
+ u' -\langle \mathbf{C} \rangle 
&=0 \\
 -K_4 \, \mu' 
+v_w \, \tilde{K}_5 \, u' + v_w (m^2)' \, \tilde{K}_6 \,  u 
-\left\langle \frac{k_z}{\omega_{0i}}\mathbf{C} \right\rangle 
&= \pm v_wK_8 \Im\left[V^\dagger {m^\dagger}'' m V\right]_{ii} \, ,
	\end{split}
\end{equation}
whose source term in the right-hand side originates from Eq.~(\ref{eq:CPsource})
\begin{equation}
\nonumber
{\cal S}_i \equiv 
\frac{\textrm{sign}[k_z]}{2 \tilde{k}}\Im\left[V^\dagger{m^\dagger}''mV\right]_{ii}\partial_{k_z} f_{eq,i} \, .
\end{equation}

The few equations above are the key master equations  which we applied in different cases. 
{We calculated the corresponding CP-violating source in the one-, two-, and three-flavour case, as well as in the fully leptonic case.}
We studied how the baryon asymmetry depends on the Yukawa profile, bubble wall width and bubble wall velocity. 

An important result is that  baryogenesis can be achieved with Standard Model light flavours, in particular leptons, provided that their  Yukawas vary from values of order 1 in the electroweak symmetric phase to their present values in the broken phase. 
In principle all flavours are playing a role. However, we showed that it is enough to consider the effect from two of them, which is also computationally much less demanding.

This opens new opportunities for baryogenesis as, so far, only the case of a varying top Yukawa had been used for electroweak baryogenesis (and the flavour-blind charginos or neutrinos in supersymmetric models). We  illustrated these results with specific models potentially  featuring natural Yukawa variation during the EW phase transition such as Froggatt-Nielsen or Randall-Sundrum models.

Given this success, the main next question is about the compatibility with experimental bounds.
The idea that the flavor structure of the Standard Model may have cosmologically emerged at the same time as electroweak symmetry breaking is quite appealing and requires a careful check of experimental constraints.

As studied in \cite{Baldes:2016gaf}, Yukawa coupling variation during  the electroweak phase transition typically requires a light additional scalar field below the TeV scale. The flavon in Frogatt-Nielsen models cannot be that light due to strong constraints from meson oscillations. On the other hand, in Randall-Sundrum models, a light dilaton is consistent with experimental bounds and naturally leads to 
Yukawa coupling variation during the electroweak phase transition  \cite{vonHarling:2016vhf}.
{The case of Composite Higgs models presents another very interesting and well-motivated possibility to realize this paradigm and their electroweak phase transition is currently investigated
\cite{CompositeEWPT}. }
These models are prototypes of models where flavour and EW symmetry breaking are intertwined as there is no Higgs potential in the absence of mixing between the elementary and composite sector which is  at the origin of Yukawa couplings. 
It will be interesting to look in more detail at the implications of models where the flavour structure of the Standard Model emerges at the same time as electroweak symmetry breaking.

\section*{Acknowledgements}
We thank Iason Baldes and Benedict Von Harling for collaboration on closely related topics.
We acknowledge support by the German Science Foundation (DFG) within the Collaborative Research Center (SFB) 676 Particles, Strings and the Early Universe.

\appendix
\section{Matrices of the diffusion equations}\label{sec:diffusionSystem}
Here we write out the matrices of equation (\ref{eqn:diffusionSystemMatrix}). Remember that the vector of unknown is given by:
\begin{equation}
	r=(\mu_{tL} ,  \mu_{tR} , \mu_{bL} ,  \mu_{bR} , \mu_{cL} ,  \mu_{cR} , \mu_{sL} ,  \mu_{sR} , \mu_{h} , u_{tL} ,  u_{tR} , u_{bL} ,  u_{bR} , u_{cL} ,  u_{cR} , u_{sL} ,  u_{sR} , u_{h} )^T
\end{equation}
This matrices follow directly from equation (\ref{eq:diffusion_0}) and the interactions specified in the text thereafter. We write them out for completeness but we insist that all the information needed to construct them is in the text around equation (\ref{eq:diffusion_0}). Remember that $N_c=3$ is the number of colours and $N_f=2$ accounts for the doubling of the number of degrees of freedom for the charm and strange quarks (see section \ref{sec:diffusion_Network} for more details on this).

For space reasons we display the matrix $A$ in four diagonal pieces:
\begin{equation}
	A=\left(\begin{array}{cc}A_1 & A_2 \\
A_3 & A_4
\end{array}\right)
\end{equation}
\begin{eqnarray}
	A_1&=&N_c v_w Diag\left(K_1^t,K_1^t,K_1^b,K_1^b,N_fK_1^c,N_fK_1^c,N_fK_1^s,N_fK_1^s,\frac{2K_1^h}{N_c}\right)\\
	A_2&=&N_c Diag\left(1,1,1,1,N_f,N_f,N_f,N_f,\frac{2}{N_c}\right)\\
	A_3&=&-N_c Diag\left(K_4^t,K_4^t,K_4^b,K_4^b,N_fK_4^c,N_fK_4^c,N_fK_4^s,N_fK_4^s,\frac{2K_4^h}{N_c}\right)\\
	A_4&=&N_c v_w Diag\left(\tilde{K}_5^t,\tilde{K}_5^t,\tilde{K}_5^b,\tilde{K}_5^b,N_f\tilde{K}_5^c,N_f\tilde{K}_5^c,N_f\tilde{K}_5^s,N_f\tilde{K}_5^s,\frac{2\tilde{K}_5^h}{N_c}\right)
\end{eqnarray}
and the matrix $B$ as follows:
\begin{equation}
 B=\left(\begin{array}{cc}
B_t^L | B_t^R | B_b^L | B_b^R | B_c^L | B_c^R | B_s^L | B_s^R | B_h & 0_{9\times 9} \\
0_{9\times 9} & B_d
\end{array}
\right)
\end{equation}
where $0_{9\times 9}$ is a ${9\times 9}$ zero matrix and $B_d$ is diagonal and we therefor only give the diagonal elements.

\begin{equation}
 B_t^L=N_c\left(\begin{array}{c}
-\Gamma_{ss}+v_w K_2^t (m_t^2)'-2\Gamma_{m_t}-\Gamma_W-(\Gamma_{y_b}+\Gamma_{y_t})\\
\Gamma_{ss}+2\Gamma_{m_t}+\Gamma_{y_t} \\
-\Gamma_{ss}+\Gamma_W\\
\Gamma_{ss}+\Gamma_{y_b} \\
-N_f \Gamma_{ss}\\
N_f  \Gamma_{ss}\\
-N_f \Gamma_{ss}\\
N_f \Gamma_{ss}\\
-(\Gamma_{y_b}+\Gamma_{y_t})
\end{array}
\right)
\end{equation}

\begin{equation}
 B_t^R=N_c\left(\begin{array}{c}
\Gamma_{ss}+2\Gamma_{m_t}+\Gamma_{y_t} \\
-\Gamma_{ss}+v_w K_2^t (m_t^2)'-2\Gamma_{m_t}-2 \Gamma_{y_t}\\
 \Gamma_{ss}+N_c \Gamma_{y_t} \\
- \Gamma_{ss}\\
N_f \Gamma_{ss}\\
-N_f \Gamma_{ss}\\
N_f  \Gamma_{ss}\\
-N_f \Gamma_{ss}\\
2\Gamma_{y_t}
\end{array}
\right)
\end{equation}

\begin{equation}
 B_b^L=N_c\left(\begin{array}{c}
 - \Gamma_{ss}+\Gamma_W \\
 \Gamma_{ss}+\Gamma_{y_t}\\
-\Gamma_{ss}+ v_w K_2^b (m_b^2)'-2 \Gamma_{m_b}-\Gamma_W-(\Gamma_{y_b}+\Gamma_{y_t})\\
 \Gamma_{ss}+2\Gamma_{m_b}+\Gamma_{y_b}\\
-N_f  \Gamma_{ss}\\
N_f  \Gamma_{ss}\\
-N_f  \Gamma_{ss}\\
N_f \Gamma_{ss}\\
- (\Gamma_{y_b}+\Gamma_{y_t})
\end{array}
\right)
\end{equation}

\begin{equation}
 B_b^R=N_c\left(\begin{array}{c}
\Gamma_{ss}+\Gamma_{y_b} \\
-\Gamma_{ss}\\
\Gamma_{ss}+2\Gamma_{m_b}+ \Gamma_{y_b} \\
- \Gamma_{ss}+ v_w K_2^b (m_b^2)'-2\Gamma_{m_b}-2 \Gamma_{y_b}\\
N_f  \Gamma_{ss}\\
-N_f \Gamma_{ss}\\
N_f  \Gamma_{ss}\\
-N_f  \Gamma_{ss}\\
2 \Gamma_{y_b}
\end{array}
\right)
\end{equation}

\begin{equation}
 B_c^L=N_f N_c \left(\begin{array}{c}
 - \Gamma_{ss} \\
 \Gamma_{ss} \\
- \Gamma_{ss}\\
  \Gamma_{ss} \\
-N_f  \Gamma_{ss}+ v_w K_2^c (m_c^2)'-2 \Gamma_{m_c}-\Gamma_W-(\Gamma_{y_c}+\Gamma_{y_s})\\
N_f  \Gamma_{ss}+2\Gamma_{m_c}+\Gamma_{y_c} \\
-N_f  \Gamma_{ss}+\Gamma_W \\
N_f  \Gamma_{ss}+\Gamma_{y_s} \\
- (\Gamma_{y_c}+\Gamma_{y_s}) 
\end{array}
\right)
\end{equation}

\begin{equation}
 B_c^R=N_f N_c\left(\begin{array}{c}
  \Gamma_{ss}\\
- \Gamma_{ss}\\
  \Gamma_{ss}\\
 -  \Gamma_{ss}\\
N_f  \Gamma_{ss}+2 \Gamma_{m_c}+ \Gamma_{y_c} \\
-N_f  \Gamma_{ss}+ v_w K_2^c (m_c^2)'-2 \Gamma_{m_c}-2  \Gamma_{y_c}\\
 N_f  \Gamma_{ss}+ \Gamma_{y_c} \\
 -N_f  \Gamma_{ss}\\
 2 \Gamma_{y_c}
\end{array}
\right)
\end{equation}

\begin{equation}
 B_s^L=N_f N_c \left(\begin{array}{c}
 - \Gamma_{ss}\\
\Gamma_{ss}\\
- \Gamma_{ss}\\
 \Gamma_{ss}\\
 -N_f  \Gamma_{ss}+ \Gamma_W \\
 N_f  \Gamma_{ss}+ \Gamma_{y_c}\\
-N_f  \Gamma_{ss}+ v_w K_2^s (m_s^2)'-2  \Gamma_{m_s}- \Gamma_W- (\Gamma_{y_c}+\Gamma_{y_s})\\
N_f  \Gamma_{ss}+2 \Gamma_{m_s}+ \Gamma_{y_s}\\
- (\Gamma_{y_c}+\Gamma_{y_s})
\end{array}
\right)
\end{equation}

\begin{equation}
 B_s^R=N_f N_c\left(\begin{array}{c}
   \Gamma_{ss}\\
- \Gamma_{ss}\\
  \Gamma_{ss}\\
 - \Gamma_{ss}\\
 N_f  \Gamma_{ss}+ \Gamma_{y_s} \\
 -N_f  \Gamma_{ss}\\
 N_f  \Gamma_{ss}+2 \Gamma_{m_s}+ \Gamma_{y_s} \\
 -N_f  \Gamma_{ss}+ v_w K_2^s (m_s^2)'-2 \Gamma_{m_s}-2  \Gamma_{y_s}\\
 2 \Gamma_{y_s}
\end{array}
\right)
\end{equation}

\begin{equation}
 B_h=\left(\begin{array}{c}
 -N_c(\Gamma_{y_b}+\Gamma_{y_t})\\
2N_c\Gamma_{y_t}\\
-N_c(\Gamma_{y_b}+\Gamma_{y_t})\\
2N_c\Gamma_{y_b}\\
-N_f N_c(\Gamma_{y_c}+\Gamma_{y_s})\\
2N_fN_c\Gamma_{y_c}\\
-N_f N_c(\Gamma_{y_c}+\Gamma_{y_s})\\
2N_fN_c\Gamma_{y_s}\\
2 v_w K_2^h (m_h^2)'-2 \Gamma_h-2N_c (\Gamma_{y_b}+N_f \Gamma_{y_c}+N_f \Gamma_{y_s}+\Gamma_{y_t})
\end{array}
\right)
\end{equation}

\begin{equation}
	B_d=Diag(B_d^t,B_d^t,B_d^b,B_d^b,B_d^c,B_d^c,B_d^s,B_d^s,B_d^h)
\end{equation}
where
\begin{eqnarray}
	B_d^t&=&N_c v_w \tilde{K}_6^t (m_t^2)'+N_c \Gamma_{tot,t}\\
	B_d^b&=&N_c v_w \tilde{K}_6^b (m_b^2)'+N_c \Gamma_{tot,b}\\
	B_d^c&=&N_fN_c v_w \tilde{K}_6^c (m_c^2)'+N_fN_c \Gamma_{tot,c}\\
	B_d^s&=&N_fN_c v_w \tilde{K}_6^s (m_s^2)'+N_fN_c \Gamma_{tot,s}\\
	B_d^h&=&2 v_w \tilde{K}_6^h (m_h^2)'+2 \Gamma_{tot,h}.
\end{eqnarray}
Finally we write the source as:
\begin{equation}
	\bar{\mathcal{S}}=\left(\begin{array}{c}
	0 \\
	0 \\
	0 \\
	0 \\
	0 \\
	0 \\
	0 \\
	0 \\
	0 \\
	-N_c v_w K_8^t ~ \Im\left[V^\dagger{m^\dagger}''mV\right]_{tt} \\
	N_c v_w K_8^t ~ \Im\left[V^\dagger{m^\dagger}''mV\right]_{tt} \\
	0 \\
	0 \\ 
	N_c v_w K_8^c ~ \Im\left[V^\dagger{m^\dagger}''mV\right]_{cc} \\
	-N_c v_w K_8^c ~ \Im\left[V^\dagger{m^\dagger}''mV\right]_{cc} \\
	0 \\
	0 \\
	0
	\end{array}\right).
\end{equation}

\section{The Kadanoff-Baym equations and spin projection} \label{sec:KB_eq_spin_projection}

In this appendix we derive the transport equations and the dispersion relations obtained in the Schwinger-Keldysh formalism. We present only a short sketch of the derivation of the different equations, a more complete derivation can be found in~\cite{Prokopec:2003pj, Prokopec:2004ic, Konstandin:2004gy}.

In the Schwinger-Keldysh formalism the Kadanoff-Baym (KB) equations for the Wightman functions can be written as
\begin{equation}
\label{eqn:kadanoff-baym1}
\left\{ \slashed{k} -m_h - i\gamma^5m_a - \Sigma_h \right\} e^{-i\Diamond} \left\{ S^{<,>} \right\}
- \left\{\Sigma^{<,>}\right\} e^{-i\Diamond} \left\{S_h\right\}=C_\psi,
\end{equation}
Here $\Sigma$ is the self energy and $m_h$ and $m_a$ are the hermitian and antihermitian part of the mass of the fermion respectively. The Wightman function that encodes the particle propagation and also the plasma is denoted as $S^{<,>}$.
The Moyal star product is denoted by 
\begin{equation}
\left\{\right\} e^{-i\Diamond} \left\{\right\} \qquad \text{with} 
\qquad \Diamond =\frac{1}{2}\left(\ola{\partial_p}\ora{\partial_x}-\ola{\partial_x}\ora{\partial_p}\right) \, .
\end{equation}
The subscript $h$ in $X_h=X^t-(X^>+X^<)/2$ is the hermitian part of $S$ and $\Sigma$ respectively and the superscript $t$ denotes time ordered quantities.
Finally,  the collision term is given by:
\begin{equation}\label{eqn:collisionInKB}
C_\psi=\frac{1}{2}e^{-i\Diamond}\left(\left\{\Sigma^>\right\}\left\{S^<\right\}-\left\{\Sigma^<\right\}\left\{S^>\right\}\right)
\end{equation}

The self-energy contributions of the left-hand-side of the KB equation leads mostly to a broadening of the propagators and we will neglect them here. The collision terms correspond to the gain- and lost-terms in the Boltzmann equation and will be 
detailed in the section on the diffusion network. 
For now we neglect the collision terms. 
They can be reintroduced finally without much extra work. 
In this case the equation (\ref{eqn:kadanoff-baym1}) reduces to:
\begin{equation}\label{eqn:tree-kadanoff-baym}
    \mathcal{D}iS^{<,>}=\left(\slashed{k}+\frac{i}{2}\slashed{\partial}-m_h e^{\frac{i}{2} \ola{\partial_z}\cdot\partial_k}-i\gamma^5m_a e^{\frac{i}{2} \ola{\partial_z}\cdot\partial_k}\right)iS^{<,>}=0
\end{equation}
where $\mathcal{D}$ is the Dirac kinetic operator which is found to commute with the spin operator 
\begin{eqnarray}
\qquad S_z \equiv \frac{1}{\tilde{k}}\left(\gamma^0 k_0 - \gamma^1k_x - \gamma^2k_y\right)\gamma^3\gamma^5
\equiv \frac{\slashed{\tilde k}}{\tilde{k}}  \gamma^3\gamma^5, 
\qquad \tilde{k} = \textrm{sign}(k_0)(k_0^2-k_x^2-k_y^2)^{1/2} \, .
\end{eqnarray}
In the wall frame where the masses and the Wightman function $S^{<,>}$ only depend on the $z$-coordinate, equation (\ref{eqn:tree-kadanoff-baym}) can be expanded in a block diagonal form in spin space
\begin{equation}
    iS^{<,>}=\sum_{s=\pm1}iS^{<,>}_s \qquad iS^{<,>}_s=-P_s\left[s\gamma^3\gamma^5g_0^{s<,>}-s\gamma^3g_3^{s<,>}+\mathbb{I}g_1^{s<,>}-i\gamma^5g_2^{s<,>}\right]
\end{equation}
where $P_s=\frac{1}{2}(\mathbb{I}+sS_z)$ is the spin projector. The result of this projection leads for the KB equation (\ref{eqn:kadanoff-baym1}):
\begin{eqnarray}
\label{eqn:spin_projected_kadanoff_baym}
2i\tilde{k} g_0-(2isk_z+s\partial_z)g_3-2im_h\hat{E}g_1-2im_a\hat{E}g_2&=0 \nn \\
2i\tilde{k} g_1-(2sk_z-is\partial_z)g_2-2im_h\hat{E}g_0+2m_a\hat{E}g_3&=0 \nn \\
2i\tilde{k} g_2+(2sk_z-is\partial_z)g_1-2m_h\hat{E}g_3-2im_a\hat{E}g_0&=0 \nn \\
2i\tilde{k} g_3-(2isk_z-s\partial_z)g_0+2m_h\hat{E}g_2-2m_a\hat{E}g_1&=0
\end{eqnarray}
where we defined $\hat{E}=e^{\frac{i}{2} \ola{\partial_z}\cdot\partial_k}$ and we have dropped the spin superscripts on the densities. In the case of several flavours this basis is not the most useful as the densities $g_i$ ($i=0,1,2,3$) do not have well defined transformation properties under flavour rotations (or chiralities). It is therefore useful to rewrite these equations in terms of linear combinations of the densities which will have nice properties under flavour rotations. The linear combinations considered here are
\begin{eqnarray}
\label{eq:defgRandgN}
    g_R=g_0+g_3 \qquad g_L=g_0-g_3 \nn \\
    g_N=g_1+ig_2 \qquad g_{N}^\dagger=g_1-ig_2 \ .
\end{eqnarray}
Their transformation properties under flavour rotation (mass diagonalization) are given by
\begin{equation}
    g_{Rd}=V^\dagger g_R V \, , \qquad   
    g_{Ld}=U^\dagger g_L U \, , \qquad
    g_{Nd}=U^\dagger g_N V \, , \qquad 
g_{Nd}^\dagger=V^\dagger g_N^\dagger U
\label{eq:flavourrotations}
\end{equation}
where the transformation matrices $V$ and $U$ are made of the eigenvectors of $m^\dagger m$ and $m m^\dagger$ respectively. Thus the masses transform as
\begin{equation}
    m_d=U^\dagger m V
\end{equation}
In this basis the KB equations (\ref{eqn:spin_projected_kadanoff_baym}) read
\begin{eqnarray}
\label{eqn:kadanoff_baym_in_flavour_rep}
2i\tilde{k} g_R-s(2ik_z+\partial_z)g_R-2im^\dagger \hat{E}g_N=0 \, , \nn \\
2i\tilde{k} g_L+s(2ik_z+\partial_z)g_L-2im \hat{E}g_N^\dagger=0 \, , \nn \\
2i\tilde{k} g_N+s(2ik_z+\partial_z)g_N-2im \hat{E}g_R=0 \, , \nn \\
2i\tilde{k} g_N^\dagger-s(2ik_z+\partial_z)g_N^\dagger-2im^\dagger \hat{E}g_L=0 \, .
\end{eqnarray}
From these equations we can obtain the constraint and kinetic equations. The constraint equations correspond to the antihermitian part of those equations and are given by
\begin{eqnarray}
\label{eqn:constraint_equations_exact1}
(2\tilde{k} - 2sk_z)g_R-m^\dagger \hat{E} g_N-g_N^\dagger \hat{E}^\dagger m &=&0 \, , \\
\label{eqn:constraint_equations_exact2}
(2\tilde{k} + 2sk_z)g_L-m \hat{E} g_N^\dagger - g_N \hat{E}^\dagger m^\dagger &=&0 \, , \\
\label{eqn:constraint_equations_exact3}
(2\tilde{k} - is\partial_z)g_N-m \hat{E} g_R - g_L \hat{E}^\dagger m &=&0 \, , \\
\label{eqn:constraint_equations_exact4}
(2\tilde{k} + is\partial_z)g_N^\dagger-m^\dagger \hat{E} g_L - g_R \hat{E}^\dagger m^\dagger &=&0 \, .
\end{eqnarray}
The kinetic equations on the other hand are given by the hermitian part of the equations (\ref{eqn:kadanoff_baym_in_flavour_rep})
\begin{eqnarray}
\label{eqn:kinetic_equations_exact1}
-s\partial_zg_R-im^\dagger\hat{E}g_N+ig_N^\dagger\hat{E}^\dagger m &=& 0\, , \\
\label{eqn:kinetic_equations_exact2}
s\partial_zg_L-im\hat{E}g_N^\dagger+ig_N^\dagger\hat{E}^\dagger m^\dagger &=& 0\, , \\
\label{eqn:kinetic_equations_exact3}
2isk_zg_N-im\hat{E}g_R+ig_L\hat{E}^\dagger m &=& 0\, , \\
\label{eqn:kinetic_equations_exact4}
-2isk_zg_N^\dagger-im^\dagger\hat{E}g_L+ig_R\hat{E}^\dagger m^\dagger&=& 0 \, .
\end{eqnarray}
Equations (\ref{eqn:kinetic_equations_exact1}--\ref{eqn:kinetic_equations_exact4}) are the ones which are used to eventually derive the Boltzmann equations (\ref{eqn:right-density}), after expansing them in a gradient expansion of $\hat{E}$. At this stage, they are, apart from neglecting the self energies, still exact relations. We will not derive (\ref{eqn:right-density}) (this is done in \cite{Konstandin:2004gy}), we will instead  illustrate the procedure and expand equations (\ref{eqn:constraint_equations_exact1}--\ref{eqn:constraint_equations_exact4})
to different orders in the gradient expansion $\hat{E}$ to determine the dispersion relation in the next appendix.

\section{Constraint equation and dispersion relation}\label{sec:constraint} 

The goal of this appendix is to eventually justify the quasi-particle picture.
We begin  with the exact constraint equations (\ref{eqn:constraint_equations_exact1}--\ref{eqn:constraint_equations_exact4}):
The densities $g_N$ and $g_N^\dagger$ can be eliminated using the relations
\begin{eqnarray}
    g_N&=&\frac{1}{2sk_z}(m\hat{E}g_R-g_L\hat{E}^\dagger m) \, , \\
    g_N^\dagger&=&\frac{1}{2sk_z}(g_R\hat{E}^\dagger m^\dagger - m^\dagger \hat{E} g_L) \, ,
\end{eqnarray}
obtained from the kinetic equations (eqns. \ref{eqn:kinetic_equations_exact1}-\ref{eqn:kinetic_equations_exact4}) in the stationary and perpendicular limit. 

The equations for $g_R$ and $g_L$ can therefore be expressed as:
\begin{eqnarray}
\nonumber (2\tilde{k} - 2sk_z)g_R-m^\dagger \hat{E} \left(\frac{m\hat{E}g_R}{2sk_z}\right)-\left(\frac{g_R\hat{E}^\dagger m^\dagger}{2sk_z}\right) \hat{E}^\dagger m\\
+m^\dagger \hat{E}\left(\frac{g_L\hat{E}^\dagger m}{2sk_z}\right) +\left(\frac{m^\dagger \hat{E} g_L}{2sk_z}\right)\hat{E}^\dagger m &=&0 \, , \\
\nonumber(2\tilde{k} + 2sk_z)g_L-m \hat{E} \left(\frac{g_R\hat{E}^\dagger m^\dagger}{2sk_z}\right)- \left(\frac{m\hat{E}g_R }{2sk_z}\right)\hat{E}^\dagger m^\dagger\\
 + m \hat{E} \left(\frac{m^\dagger \hat{E} g_L}{2sk_z}\right)  + \left(\frac{g_L\hat{E}^\dagger m}{2sk_z}\right)  \hat{E}^\dagger m^\dagger  &=&0 \, .
\end{eqnarray}

To first order in the derivative expansion of $\hat{E}$ the constraint equations for the left and right densities are given by
\begin{equation}\label{eqn:right_constraint}
\begin{split}
    2&(\tilde{k}-sk_z-\frac{1}{4sk_z}\left\{ m^\dagger m, \cdot \right\})g_R + \frac{1}{sk_z}m^\dagger g_L m  \\
    &+\frac{i}{4sk_z}\left( \left[ \partial_{k_z}g_R,m^\dagger m' \right] +\left[ \partial_{k_z}g_R,{m^\dagger}' m \right]+ 2 {m^\dagger}'\partial_{k_z}g_L m -2 m^\dagger \partial_{k_z}g_L m' \right)\\
    &+\frac{i}{4sk_z^2}\left( {m^\dagger}'m g_R - g_R m^\dagger m' + m^\dagger g_L m'- {m^\dagger}' g_L m \right) = 0 \, ,
\end{split}
\end{equation}
and 
\begin{equation}\label{eqn:left_constraint}
\begin{split}
    2&(\tilde{k}+sk_z+\frac{1}{4sk_z}\left\{ m m^\dagger, \cdot \right\})g_L - \frac{1}{sk_z}m g_R m^\dagger  \\
    &+\frac{i}{4sk_z}\left( \left[ m {m^\dagger}',\partial_{k_z}g_L \right] +\left[ m' m^\dagger,\partial_{k_z}g_L \right]+ 2 m\partial_{k_z}g_R{m^\dagger}' -2 m' \partial_{k_z}g_R m^\dagger \right)\\
    &+\frac{i}{4sk_z^2}\left( g_L m {m^\dagger}' -  m' m^\dagger g_L + m' g_R m^\dagger- m g_R {m^\dagger}' \right) = 0 \, .
\end{split}
\end{equation}

In the following we will diagonalize the two equations (\ref{eqn:right_constraint}) and 
(\ref{eqn:left_constraint}) and neglect off diagonal elements. This is 
justified in the adiabatic approximation as advocated in the main text. We obtain
\begin{equation}\label{eqn:right_constraint_diagonal}
\begin{split}
    2&\left(\tilde{k}-sk_z-\frac{1}{2sk_z}\left(|m_{di}|^2  + \frac{1}{2k_z}\Im\left[V^\dagger {m^\dagger}' m V\right]_{ii} \right) \right)g_{Rdii} + \frac{1}{sk_z} |m_{di}|^2 g_{Ldii}  \\
    &-\frac{1}{sk_z}\Im\left[V^\dagger {m^\dagger}' m V \right]_{ii}\partial_{k_z}g_{Ldii}+\frac{1}{2sk_z^2}\Im\left[V^\dagger {m^\dagger}' m V \right]_{ii}g_{Ldii} = 0 \, ,
\end{split}
\end{equation}
and
\begin{equation}\label{eqn:left_constraint_diagonal}
\begin{split}
    2&\left(\tilde{k}+sk_z+\frac{1}{2sk_z}\left(|m_{di}|^2  - \frac{1}{2k_z}\Im\left[U^\dagger m{m^\dagger}' U\right]_{ii} \right) \right)g_{Ldii} - \frac{1}{sk_z} |m_{di}|^2 g_{Rdii}  \\
    &-\frac{1}{sk_z}\Im\left[U^\dagger m{m^\dagger}' U \right]_{ii}\partial_{k_z}g_{Rdii}+\frac{1}{2sk_z^2}\Im\left[U^\dagger m{m^\dagger}' U \right]_{ii}g_{Rdii} = 0 \, .
\end{split}
\end{equation}
Equivalently for the densities $g_{0dii}$ and $g_{3dii}$, (obtained by adding and subtracting the previous two equations) the relations read
\begin{equation}\label{eqn:g0_constraint_diagonal}
    2\tilde{k} g_{0dii}-2\left(sk_z+\frac{1}{sk_z}|m_{di}|^2\right)g_{3dii}-\frac{1}{sk_z}\Im\left[V^\dagger {m^\dagger}' m V \right]_{ii}\partial_{k_z}g_{0dii} = 0 \, ,
\end{equation}
and
\begin{equation}\label{eqn:g3_constraint_diagonal}
    -2sk_zg_{0dii} +\left(2\tilde{k}-\frac{1}{sk_z^2}\Im\left[V^\dagger {m^\dagger}' m V\right]_{ii} \right)g_{3dii} +\frac{1}{sk_z}\Im\left[V^\dagger {m^\dagger}' m V \right]_{ii}\partial_{k_z}g_{3dii} = 0 \, .
\end{equation}
It should be noted that $\Im\left[V^\dagger {m^\dagger}' m V \right]_{ii}=\Im\left[U^\dagger m{m^\dagger}' U\right]_{ii}$. The latter equation can then be solved for $g_{0dii}$\footnote{Note that this relation between the $g_{0dii}$ and the $g_{3dii}$ is very useful when deriving the diagonalized version of the kinetic equations~(\ref{eqn:our_kinetic_equation}).}
\begin{equation}\label{eqn:g0_as_function_of_g3}
    g_{0dii}=\frac{\tilde{k}}{sk_z}g_{3dii}-\frac{1}{2k_z^3}\left(\Im\left[V^\dagger {m^\dagger}' m V\right]_{ii}\left(1-k_z\partial_{k_z}\right)\right)g_{3dii} \, .
\end{equation}
Injecting this result in equation (\ref{eqn:g0_constraint_diagonal}) gives
\begin{equation}
    \frac{2}{sk_z}\left(\tilde{k}^2-k_z^2-|m_{di}|^2-\frac{s\tilde{k}}{2k_z^2}\Im\left[V^\dagger{m^\dagger}'mV\right]_{ii}\right)g_{3dii}=0
\end{equation}
We can also proceed the other way round. From equation~(\ref{eqn:g0_constraint_diagonal}) we find
\begin{equation}\label{eqn:g3_as_function_of_g0}
    g_{3dii}=\left(sk_z+\frac{1}{sk_z}|m_{di}|^2\right)^{-1}\left(\tilde{k} g_{0dii}-\frac{1}{2sk_z}\Im\left[V^\dagger {m^\dagger}' m V \right]_{ii}\partial_{k_z}g_{0dii}\right)
\end{equation}
Upon multiplying equation~(\ref{eqn:g3_constraint_diagonal}) by $\left(sk_z+\frac{1}{sk_z}|m_{di}|^2\right)$ and using the relation found in equation~(\ref{eqn:g3_as_function_of_g0}) we find to first order in the gradient expansion
\begin{equation}
    2\left(\tilde{k}^2-k_z^2-|m_{di}|^2-\frac{\tilde{k}}{s(k_z^2+|m_{di}|^2)}\Im\left[V^\dagger{m^\dagger}'m V\right]_{ii}\right)g_{0dii}=0.
\end{equation}
%
Which justifies the following parametrization:
\begin{equation}\label{eqn:densities}
    g_{0dii}\propto\delta\left(\tilde{k}_0^2-k_z^2-|m_{di}|^2-\frac{\tilde{k}_0}{s(k_z^2+\Im_{di}|^2)}\Im\left[V^\dagger{m^\dagger}'m V\right]_{ii}\right)n_{i}(k_0,\vec{k},x)
\end{equation}
where $n_{i}(k_0,\vec{k},x)$ are the densities for the different particles.
This is again an algebraic relation and suggests the validity of the quasiparticle picture. For on shell particles we can determine the dispersion relation and find it to be identical to the one found in \cite{Prokopec:2003pj} in the limit where the CP-violating term is small, namely:
\begin{equation}\label{eqn:dispersion_relation}
\begin{split}
    \omega_i=&\sqrt{\omega_{0i}^2+\frac{1}{2\tilde{\omega}_{0i}^4}\left(\Im\left[V^\dagger{m^\dagger}'mV\right]_{ii}\right)^2+\frac{1}{2s\tilde{\omega}_{0i}^2}\Im\left[V^\dagger{m^\dagger}'mV\right]_{ii}\sqrt{4\tilde{\omega}_{0i}^2+\frac{1}{\tilde{\omega}_{0i}^4}\left(\Im\left[V^\dagger{m^\dagger}'mV\right]_{ii}\right)^2}}\\
    &\approx \omega_{0i}+\frac{s}{2\tilde{\omega}_{0i}\omega_{0i}}\Im\left[V^\dagger{m^\dagger}'mV\right]_{ii}
\end{split}
\end{equation}
where:
\begin{equation}\label{eqn:zero_energy}
    \omega_{0i}=\sqrt{\vec{k}^2+|m_d|^2_i} \qquad \tilde{\omega}_{0i}=\sqrt{k_z^2+|m_d|^2_i}
\end{equation}
The second term in (\ref{eqn:dispersion_relation}) is the small CP-violating contribution on which we comment in the last paragraph of Section \ref{sec:kinetic_equations}.
The group velocity can at this moment also be calculated and it is given by:
\begin{equation}\label{eqn:group_velocity}
    \vec{v}_g=\frac{\vec{k}}{\omega_{i}}\approx\frac{\vec{k}}{\omega_{0i}}\left(1-\frac{s}{2\tilde{\omega}_{0i}\omega_{0i}^2}\Im\left[V^\dagger{m^\dagger}'mV\right]_{ii}\right) 
\end{equation}
It is interesting to notice that those equations, determined in the case on multiple mixing fermion flavours, reproduce the result found in \cite{Fromme:2006wx} in the case of one flavour.

\section{Randall-Sundrum Yukawa couplings \label{sec:RSYukawas}} 
We adopt the model II using the Golberger-Wise scalar with a modified profile as suggested in~\cite{vonHarling:2016vhf}. Here we spell out the relevant equations without justification only for completeness. 
The formula for the Yukawa couplings is given by:
\begin{equation}
	y_u(\sigma_{IR})=\lambda_u k \mathcal{N}_{\tilde{c}_Q,c_Q}^{(0)}\mathcal{N}_{\tilde{c}_u,c_u}^{(0)}\sigma_{IR}^{c_Q+c_u-1}e^{\frac{(\tilde{c}_Q+\tilde{c}_u)\sigma_{IR}^\epsilon}{\epsilon}}\, ,
\end{equation}
where 
\begin{equation}
	\mathcal{N}_{\tilde{c}_Q,c_Q}^{(0)}=\sqrt{\epsilon}\left[\sigma_{IR}^{2c_Q -1}E_{1+\frac{1-2c_Q}{\epsilon}}\left(\frac{-2\tilde{c}_Q\sigma_{IR}^\epsilon}{\epsilon}\right) - E_{1+\frac{1-2c_Q}{\epsilon}}\left(\frac{-2\tilde{c}_Q}{\epsilon} \right) \right]^{-1/2}\, ,
\end{equation}
with $\epsilon=1/20$, $E_n(x)$ is the exponential integral function and
\begin{equation}
	\tilde{c}_{Q,u}=\rho_{Q,u}~\tilde{v}_{UV}/k \qquad c_{Q,u}=-\rho_{Q,u}~\beta/k\, ,
\end{equation}
where we took $\tilde{v}_{UV}=4k^{3/2}$ and $\beta=1.5k^{3/2}$ and
\begin{equation}
	\rho_{Q_2}=1.35~k^{-1/2} \quad \rho_{u_2}=1.43~k^{-1/2} \quad \rho_{Q_3}=1.22~k^{-1/2} \quad \rho_{u_2}=1.15~k^{-1/2}\, .
\end{equation}
The 5D Yukawa couplings are taken to be (in accordance with~\cite{vonHarling:2016vhf}):
\begin{equation}
	\lambda_u=\frac{1}{k}\left(
\begin{array}{cc}
0.76 \cdot e^{-1.46 i} & 0.74 \cdot e^{-2.13 i} \\
0.28 \cdot e^{0.39 i} & 0.93 \cdot e^{-1.26 i}
\end{array}
	\right)\, .
\end{equation}
Finally the warp factor at the IR brane $\sigma_{IR}=e^{-k y_{IR}}$ can be linked to the Higgs VEV via $\phi=\xi\sigma_{IR}/\sigma_{IR}^{min}$ where we took $\sigma_{IR}^{min}=2.5\cdot 10^{-25}$ consistently  with~\cite{vonHarling:2016vhf}.

\bibliographystyle{JHEP}  
\bibliography{biblio}

\end{document}